\documentclass[a4paper, 11pt]{article}
\usepackage[utf8]{inputenc}
\usepackage[OT1]{fontenc}
\usepackage[textwidth=16cm, bottom=3.5cm]{geometry}

\usepackage{graphicx}
\usepackage{amsmath}
\usepackage{amsfonts}
\usepackage{amssymb}
\usepackage{amsbsy}
\usepackage{dsfont}

\usepackage{latexsym}
\usepackage{array}
\usepackage{hyperref}
\usepackage{color}
\definecolor{bluegray}{rgb}{0.0, 0.0, 0.7}
\definecolor{darkbluegray}{rgb}{0.0, 0.0, 0.3}
\definecolor{gray}{rgb}{0.2, 0.2, 0.2}
\hypersetup{
    colorlinks=true,
    linkcolor=darkbluegray,
    citecolor=darkbluegray,
    urlcolor=darkbluegray,
}
\usepackage{float}
\usepackage{textcomp}
\usepackage{times}
\usepackage{authblk}
\usepackage{natbib}
\setcitestyle{authoryear,open={(},close={)}} 

\newcommand{\mb}{\mathbf}
\newcommand{\mbb}{\mathbb}
\newcommand{\mc}{\mathcal}
\newcommand{\bs}{\boldsymbol}
\newcommand{\bc}{\begin{center}}
\newcommand{\ec}{\end{center}}
\newcommand{\nit}{\noindent}

\newcommand{\beq}{\begin{equation}}
\newcommand{\eeq}{\end{equation}}
\newcommand{\beqa}{\begin{eqnarray}}
\newcommand{\eeqa}{\end{eqnarray}}
\newcommand{\beqan}{\begin{eqnarray*}}
\newcommand{\eeqan}{\end{eqnarray*}}
\newcommand{\bit}{\begin{itemize}}
\newcommand{\eit}{\end{itemize}}

\DeclareFontShape{OT1}{cmss}{b}{n}{<->ssub * cmss/bx/n}{}
\begin{document}


\title{Forecasting Brain Activity Based on Models of \\ Spatio-Temporal Brain Dynamics: \\ A Comparison of Graph Neural Network Architectures}

\author[1,2]{S. Wein \thanks{simon.wein$@$ur.de}}
\author[1]{A. Schüller}
\author[3]{A.~M.~Tom\'e}
\author[2]{W. M. Malloni}
\author[2]{M. W. Greenlee}
\author[1]{E. W. Lang}

\affil[1]{\normalsize CIML, Biophysics, University of Regensburg, Regensburg, Germany}
\affil[2]{Experimental Psychology, University of Regensburg, Regensburg, Germany}
\affil[3]{IEETA/DETI, Universidade de Aveiro, Aveiro, Portugal}

\maketitle


\begin{abstract}
Comprehending the interplay between spatial and temporal characteristics of neural dynamics can contribute to our understanding of information processing in the human brain. Graph neural networks (GNNs) provide a new possibility to interpret graph structured signals like those observed in complex brain networks. In our study we compare different spatio-temporal GNN architectures and study their ability to model neural activity distributions obtained in functional MRI (fMRI) studies. We evaluate the performance of the GNN models on a variety of scenarios in MRI studies and also compare it to a VAR model, which is currently often used for directed functional connectivity analysis. We show that by learning localized functional interactions on the anatomical substrate, GNN based approaches are able to robustly scale to large network studies, even when available data are scarce. By including anatomical connectivity as the physical substrate for information propagation, such GNNs also provide a multi-modal perspective on directed connectivity analysis, offering a novel possibility to investigate the spatio-temporal dynamics in brain networks.
\end{abstract}


\section{Introduction}

Distinct concepts of brain connectivity can provide different but complementary aspects of information processing in the brain \citep{Lang2012}. On the one hand, imaging modalities like functional magnetic resonance imaging (fMRI) allow us to temporally resolve dynamic neural activity patterns in distinct spatial locations in the human brain. Different statistical approaches that describe the coherency of activity profiles in brain networks have been proposed based on the notion of functional connectivity (FC). On the other hand, diffusion tensor imaging (DTI) can provide insights into the structural and relatively static aspects of the brain. By reconstructing white matter tracks from DTI data, the anatomical or structural connectivity (SC) between different brain areas can be estimated. Also directed and potentially causal relationships between regions become of interest in fMRI and are studied with respect to directed functional or effective connectivity. The latter is most often inferred from Granger causality or dynamic causal modeling \citep{Friston2013, Bielczyk2019}. 

Based on these concepts, spatial and temporal relationships between brain areas can be represented by graphical models, which have recently received increasing attention in the field of machine learning \citep{Wu2020, Bronstein2017}. So-called graph neural networks (GNNs) allow us to effectively process signals in the non-Euclidean geometry of graphs, providing also novel possibilities for applications in brain connectivity research \citep{Rosenthal2018, Wein2021, Ktena2018, Arslan2018, LiX2019, Kim2020}. Given a decomposition of the brain into specified areas, their spatio-temporal neural activity patterns can be interpreted as graph structured signal distributions. Nodes in brain networks can be associated with variables like the temporal neuronal activity of neuron pools, while edges in such networks reflect the strength of interactions between neural populations \citep{Bullmore2011}. As proposed in our recent study \citep{Wein2021}, these complex signals exhibited in non-Euclidean geometries can be processed with a variant of GNN denoted as spatio-temporal graph neural network (STGNN). Such STGNNs can allow us to simultaneously models spatial and temporal dependencies in such graph structured signals and thereby provide a new possibility to combine DTI with fMRI data. Activity propagation based approaches made already various interesting contributions to brain connectivity research, and could for example explain how resting-state FC \citep{Cole2016} or SC \citep{Yan2021} are related to cognitive task activations observed in task-based fMRI. Moreover, they could give us insights in what way the hierarchical organization of the brain is related to the propagation of information along structural connections \citep{Rodriguez2020}. In our study we compare different approaches based on GNNs to study the spatio-temporal propagation of information in brain networks from a multi-modal and data-driven perspective.

Recently, several different GNN architectures have been proposed to model the flow of information across time and space in graphical signals \citep{Wu2020}. Convolution operations, often applied in deep learning, have recently been extended successfully to graphical models and allow us to capture inherent spatial dependencies on non-Euclidean network structures \citep{Defferrard2016}. They were later combined with recurrent neural networks (RNNs) \citep{Rumelhart1986}, which can detect sequential relations in signals. This combined spatio-temporal GNN framework was proposed in the notion of diffusion convolution recurrent neural network (DCRNN) \citep{Li2018}. However RNNs can have problems with long time series and, when combined with graph convolution operations, their gradients are more likely to explode or vanish \citep{Seo2018, Li2018}. This was the motivation for introducing approaches that combine spatial graph convolutions with standard one-dimensional temporal convolutions \citep{Wu2020}. But their receptive field size can only grow if many hidden layers are used (linear growth) or global pooling is applied. To alleviate such shortcomings so-called WaveNets (WNs) have been introduced that employ stacked dilated temporal convolutions, which provide a long-term temporal memory \citep{Oord2016}. They have recently been combined with graph convolution operations in an architecture denoted as graph WaveNet (GWN) \citep{Wu2019}. As an alternative method for the temporal processing, also attention mechanisms have been recently included in STGNN architectures \citep{Zheng2020}. Attention mechanisms select, from all inputs, information that is critical to the task at hand and modify edge connection strengths accordingly. They have been applied already to natural language processing, speech recognition and image processing \citep{Xu2015, Vaswani2017, Liang2019}, but applications to analyze the dynamics in neural signals are still missing. In this study we compare these different STGNN architectures with each other and evaluate their effectiveness in replicating functional activity distributions observed in brain networks. In addition to these distinct temporal models, we study different approaches to model the spatial information exchange between brain regions. At first we employ the structural connectivity as the substrate for information propagation between brain regions. Further we evaluate the effectiveness of employing connectome embeddings (CEs) of the anatomical network to characterize the node relations. In a recent study \cite{Rosenthal2018} have shown that embeddings of nodes in the anatomical network can inherently capture higher order topological relations between different structurally connected nodes in this network. Finally we compare it to the case when we incorporate no predefined spatial layout into the GNN models, trying to learn the spatial structure by gradient descent based optimization during model training. We demonstrate that by modeling the functional information exchange between regions in STGNNs based on structural connectivity, we can significantly increase the accuracy of predicting future neural signals. This points out that STGNN models are capable of learning informative functional interactions between areas in such brain networks. Based on these different comparisons, we at first try to identify the most effective STGNN architectures to investigate such spatial and temporal dynamics in brain networks.

In a subsequent step we then compare this STGNN based approach to a currently popular data-driven model for characterizing directed functional information exchange in brain networks. Until now methods for the inference of directed functional or effective connectivity are often based on Granger causality \citep{Barnett2013} or dynamic causal modeling \citep{Friston2013} and its recent extensions \citep{Frassle2018, Prando2020}. In addition various algorithmic and information theory based methods have been developed meanwhile in this field \citep{Bielczyk2019, Ramsey2011}. In the following, we compare the performance of the STGNN prediction models to the forecasting model most often used in Granger causality analysis \citep{Barnett2013}. Granger causality is based on the idea that if one event $ A $ would cause another event $ B $, then $ A $ should precede $ B $ and the occurrence of event $ A $ should contain information about the occurrence of event $ B $ \citep{Friston2013}. In the context of neuroimaging this is realized in a predictive framework, by testing if adding information on activity in a region $ A $ improves the prediction of activity in region $ B $. For practical applications of this idea, the underlying predictive model in Granger causality is usually based on a vector auto regression (VAR) for multivariate timeseries inference \citep{Friston2013, Barnett2013, Bielczyk2019}. In a brain network with $ N $ regions of interest (ROIs) the parameters in a VAR model grow with $ N^2 $, so for larger brain networks it can be challenging to accurately fit the model if only limited data are available. This can be problematic in fMRI, where the temporal sampling rate is relatively low, while its good spatial resolution would allow for a detailed, high-resolution network analysis. Therefore it would be desirable to have a predictive model that can learn interactions between all brain areas of interest, and in addition naturally scales to larger brain networks. In our study we compare the STGNN approaches to a classical VAR model and test their accuracy on a variety of network sizes and data set sizes. We show that by learning localized functional interactions based on the anatomical network, the prediction accuracies of STGNN models remain significantly more accurate, even when brain networks become very complex and only limited data are available to fit the models. This points out that the STGNN approaches are robust among a large variety of MRI study scenarios, and are therefore also suitable for the analysis of smaller subject cohorts, like in studies of patients with rare neurological diseases.

Finally we analyze the spatial interactions between brain regions, which are learned by the STGNN models. By integrating prior knowledge on the brain anatomy in form of structural connectivity or based on connectome embeddings, such models can provide multi-modal perspective on directed relations between brain areas. So far, a variety of approaches were proposed to study the structure-function relation in the human brain \citep{Suarez2020}, which are based on computational modeling \citep{Honey2009, Deco2012, Deco2013, Messe2015, ChenX2018, Messe2014}, graph theory \citep{Liang2017, Abdelnour2018, Becker2018, Lim2019} and machine learning \citep{Sarwar2021, Deligianni2016, Amico2018, Rosenthal2018}. Biophysically inspired models for example could describe how functional connectivity patterns emerge from neural dynamics with static couplings characterized by anatomical connections \citep{Deco2013, Messe2014, Honey2009}, and were recently used to also study spatial heterogeneity of local circuit properties across the cortex \citep{Demirtas2019, Wang2019}. Methods from graph theory can supplement such computational frameworks by specifically pointing out how indirect structural connections contribute to the inference of FC strength \citep{Liang2017, Becker2018}. Also hybrid methods could demonstrate that the typology of structural brain networks supports in neuromorphic networks their memory capacity \citep{Suarez2021}. Such insights into structural and functional connectivity can then provide a basis to better understand the cognitive information processing in the human brain \citep{Ito2020}. The vast majority studies on structure-function relations focuses currently on inferring overall FC patterns from their SC, although static coherency based measures of FC might have limitations in their ability to capture the rich nature of dynamic brain activity \citep{Wein2021b}. To the contrary, STGNNs are able to directly predict the measured BOLD dynamics, and their interactions between brain regions, without relying on the indirect representation of functional dynamics based on coherency. This characteristic of STGNNs allows us to additionally investigate the structure-function coupling in brain networks from a novel perspective. To study this relation further on the individual brain region level, we demonstrate how a perturbation based approach can be utilized to reconstruct how dynamic functional interactions are mediated by their structural substrate in STGNN models. By inferring how information is propagated between individual regions in STGNNs, these predictive models have the potential to reveal directed relationships between individual areas in brain networks from a multi-modal perspective.  
In general due to the low temporal sampling rate and physiological artifacts, fMRI can have several limitations in detecting directed relations in brain networks \citep{Smith2011, Webb2013}. Still some recent fMRI studies and computational simulations could demonstrate that also lag-based methods like Granger causality can be useful for detecting such directed dependencies in fMRI data \citep{Seth2012, Wang2014, Mill2017, Duggento2018}.

The possibility to combine structural and functional imaging data in STGNNs can make these models as well interesting for several practical applications in brain connectivity research. For instance they can be used to investigate differences in the structure-function relationship between resting-state and task-based fMRI. Furthermore, in clinical applications these models could be employed to study how lesions in the structural connectome affect the functional organization of the brain network. In our current study we compare therefore such mechanisms for spatial and temporal modeling in STGNNs, with the objective to establish their methological foundation for brain connectivity research, and thereby providing a basis for future applications of STGNNs in multi-modal neuroimaging studies.



\section{Results}\label{sec:results}

\subsection{Graph neural network models}

\begin{figure}[!htb]
\bc
\includegraphics[width=\textwidth]{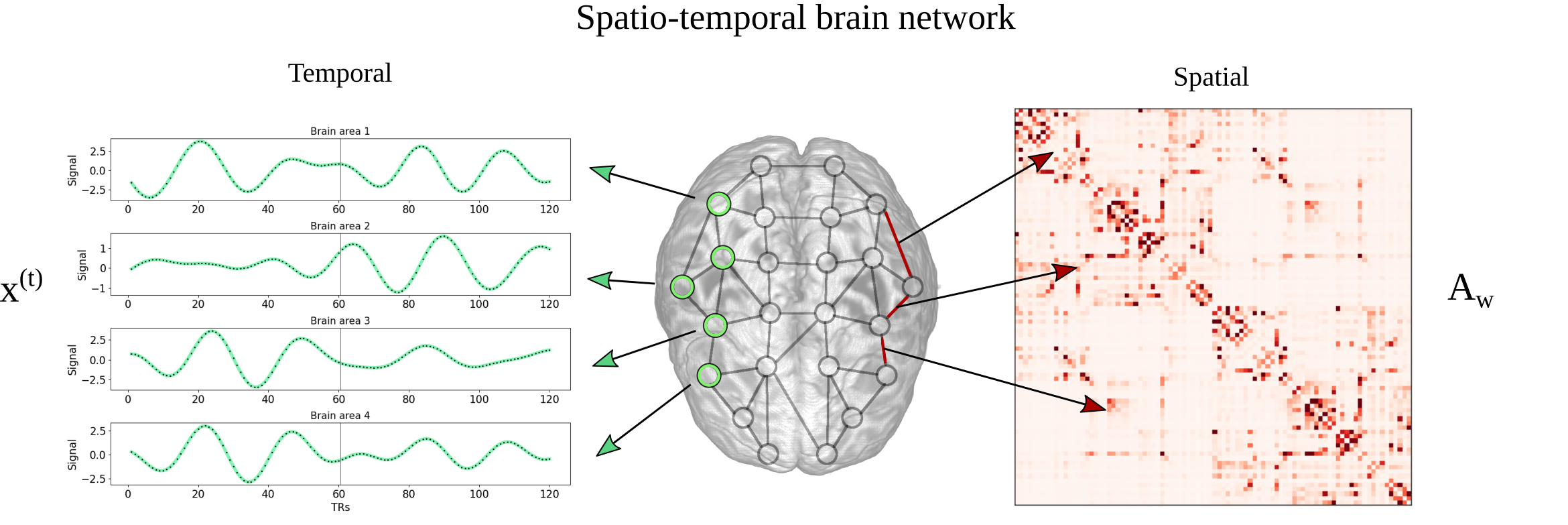}
\ec 
\caption{The spatio-temporal representation of a signal in a brain network is illustrated. The temporal component of the graph-like signal is given by the BOLD signal $ \mb{x}^{(t)} \in \mbb{R}^{N} $ in $ N $ brain regions sampled at different timesteps $ t $, as shown on the left side. The strength of the edges in the brain network are defined by a weighted adjacency matrix $ \mb{A}_w \in \mbb{R}^{N \times N} $, as illustrated on the right side. One entry $ w_{nn'} $ of the matrix  $ \mb{A}_w $  characterizes thereby the spatial relation between brain region $ n $ and $ n' $ in the network. 
}
\label{fig:overview_graph_signal}
\end{figure}

In our context of MRI the goal of the spatio-temporal GNNs will be to forecast the observed BOLD signal as accurately as possible in order to precisely capture the spatio-temporal dynamics of the underlying mechanisms in the brain. The learning objective can be formalized by introducing a graph signal $ \mb{x}^{(t)} \in \mbb{R}^{N} $, representing the BOLD signal measured at timestep $ t $ in $ N $ different brain regions. The goal of the models is to predict from an input sequence of $ T_p $ past neural activity states $ t = 1, \ldots, T_p $ a sequences of future states $ t = 1, \ldots, T_f $. In addition to the temporal information, also spatial dependencies are included in the GNN architectures. The spatial relations between the $ N $ brain regions can be represented in the notion of a graph $ \mc{G} = (\mc{V},\mc{E},\mb{A}_w) $, containing vertices (nodes) $ \mc{V} $, with $ |\mc{V}| = N $, and edges $ \mc{E} $. The structure of the graph is characterized by a weighted adjacency matrix $\mb{A}_w \in \mbb{R}^{N \times N}$, where an entry $ w_{nn'} $ describes the connection strength between brain region $ n $ and $ n' $. An illustration of such a graphical representation of a dynamic brain state is provided in figure \ref{fig:overview_graph_signal}. Based on this concept, the task of the GNN models is to derive a function $ h(\cdot) $ that best predicts $ T_f $ future activity states from an input sequence of $ T_p $ past states: 

\beq \label{eqn:graph_signal_prediction}
[\mb{x}^{(1)}, \ldots, \mb{x}^{(T_p)}; \mc{G}] \xrightarrow{h(\cdot)} [\mb{x}^{(T_p+1)}, \ldots, \mb{x}^{(T_p+T_f)}]
\eeq

\begin{figure}[!htb]
\bc
\includegraphics[width=0.87\textwidth]{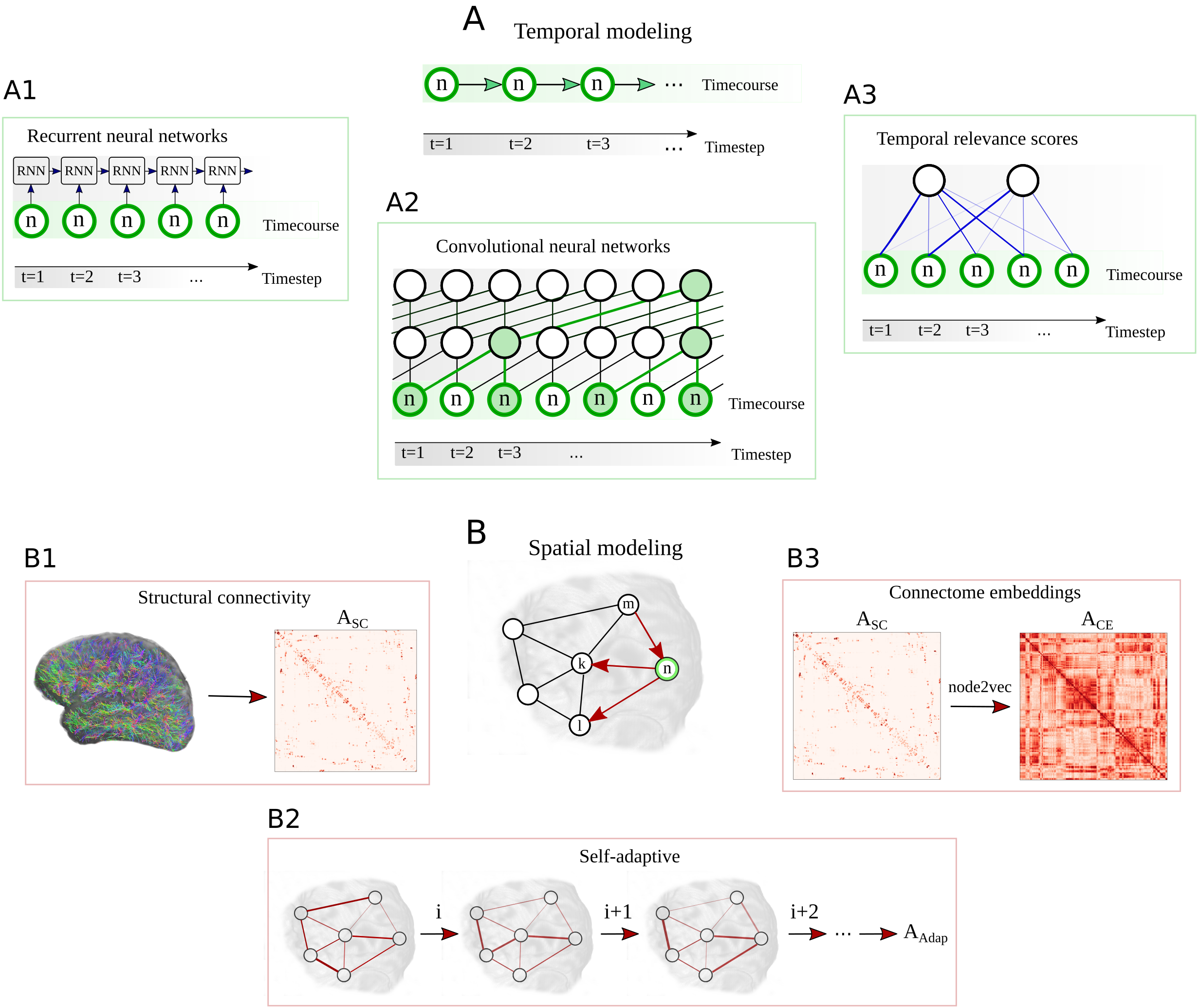}
\ec 
\caption{Overview on the different techniques used by spatio-temporal GNNs to learn the spatial and temporal dynamics in brain networks. The temporal component of the STGNN tries to infer dependencies between the activity at different timesteps $ t = 1, 2, 3, \ldots $ in a certain brain region $ n $, as illustrated in (A). A recurrent neural network (RNN) architecture processes in a recursive manner the activity states at the different timesteps $ t $ (A1). In our study this principle will be implemented in a RNN based sequence-to-sequence architecture for timeseries forecasting. As an alternative, convolutional neural network (CNN) architectures employ one-dimensional convolutions in the time-domain (A2). This idea is picked up in the WaveNet (WN) architecture, which introduces dilated convolutions to obtain exponentially growing receptive fields, as highlighted here in green. Following the idea of an attention mechanism, temporal relevance scores can be dynamically computed to weight the importance of a temporal feature observed at timestep $ t $ (A3).  A temporal attention (TAtt) based architecture is thereby composed of a multitude of stacked attention layers. Based on these temporal modeling approaches, STGNNs additionally detect spatial dependencies between a node $ n $ and other interconnected nodes in the brain network, as illustrated in (B). The substrate for the spatial interactions can be based on the structural connectivity matrix $ A_{SC} $, as reconstructed from DTI data (B1). Alternatively, the node2vec algorithm can be used to detect higher order relationships between  brain regions in the structural connectome, for characterizing the spatial similarity based on connectome embeddings $ A_{CE} $ (B3). Finally, spatial connections can be tried to be learned by the model itself, by freely adapting the edges in an adjacency matrix  $ A_{Adap} $ at different iterations $ i, i+1, i+1 \ldots $ during the model training (B2).  
}
\label{fig:overview_GNNs}
\end{figure}

Until now various spatio-temporal GNN architectures have been proposed to account for spatial and temporal dependencies of such graph structured signals \citep{Wu2020}. An overview of different possibilities for the temporal modeling is given in figure \ref{fig:overview_GNNs} (A). In timeseries analysis, recurrent neural networks (RNNs) \citep{Rumelhart1986} provide one efficient way to detect patterns in sequential data structures, like in our context the BOLD signal, subsequently sampled at different timesteps $ t $ (Figure \ref{fig:overview_GNNs} A1). This approach can be extended to a RNN based sequence-to-sequence architecture, whereby an encoder recursively processes an input sequence of $ T_p $ past neural activity states $ \mb{x}^{(t)} $ and encodes the temporal information into a hidden state $ \mb{H}^{(T_p)} $ \citep{Sutskever2014}. Next a decoder network uses the information in $ \mb{H}^{(T_p)} $ to generate a prediction for $ T_f $ future activity states. To account for vanishing gradients during training, the encoder and decoder consist of gated recurrent unit (GRU) cells \citep{Chung2014}. An alternative for detecting repetitive patterns in sequential data is provided by convolutional neural networks (CNNs) (Figure \ref{fig:overview_GNNs} A2). By employing one-dimensional convolutions in the time domain they are used in our context to process temporal dynamics of neural activity. To more efficiently capture long-term dependencies in temporal data the WaveNet (WN) architecture has been proposed \citep{Oord2016}. This model introduces dilated causal convolution operations to generate a large receptive field when using only relatively few network layers, which alleviates the processing of long temporal input horizons. The growth of the receptive field of a neuron (marked in green) in a WN layer is illustrated in figure \ref{fig:overview_GNNs} (A2). More recently, also attention mechanisms have been proposed to detect underlying hidden correlations in sequential data structures \citep{Vaswani2017}. In timeseries analysis, the idea of a temporal attention (TAtt) architecture is thereby to adaptively focus on the most important temporal features in a sequence (Figure \ref{fig:overview_GNNs} A3).

These different fundamental approaches for temporal dependency modeling have been recently combined with techniques to additionally capture spatial relationships in graph structured signals \citep{Li2018, Wu2019, Zheng2020}. Graph convolutional neural networks can be incorporated to model the propagation of information between adjacent nodes in the graphical representation of the signal \citep{Defferrard2016}. The neighborhoods of the vertices/nodes $ \mc{V} $ in the network are characterized by the adjacency matrix $ \mb{A}_w $. In our study we additionally investigate different possibilities for defining the spatial layout for the information propagation between brain regions, as illustrated in figure \ref{fig:overview_GNNs} (B). As a first choice for the adjacency matrix we will employ the structural connectivity $ \mb{A}_{SC} $ between the $ N $ brain areas, as it could be reconstructed from DTI data (Figure \ref{fig:overview_GNNs} B1). This choice is motivated by the idea that white matter connections obtained from this modality would establish the anatomical backbone for information exchange between brain areas. In a recent study \cite{Rosenthal2018} demonstrated that connectome embeddings (CE) can be utilized for projecting the structural connectome into a continuous vector space, which captures meaningful correspondences between different brain areas. This technique can thereby allow us to additionally account for long range and inter-hemispheric homotopic connections, which are usually only weakly expressed in DTI based anatomical connectivity \citep{Thomas2014}. In our study we utilized this technique to represent the edge weight $ w_{nn'} $ in the adjacency matrix as the similarity between the vector representations of two nodes $ n $ and $ n' $, which will be denoted as $ \mb{A}_{CE} $ (Figure \ref{fig:overview_GNNs} B3). The information is accordingly propagated between brain regions which possess high similarity based on their neighborhood role within the anatomical network. Finally we compare these techniques to the case when the model is given the freedom to learn spatial dependencies between the $ N $ regions itself. In this setup the adjacency matrix is represented by a self-adaptive matrix $ \mb{A}_{Adap} \in \mbb{R}^{N \times N} $, which is learned during the model optimization (Figure \ref{fig:overview_GNNs} B2). A detailed formal description of the model architectures and the training involved is outlined in the \nameref{sec:materials_methods} section. In the following we will assess the effectiveness of the different spatial and temporal modeling approaches by comparing their predictive performance on a MRI dataset from the Human Connectome Project (HCP) \citep{VanEssen2013}.


\subsection{Data description}\label{sec:data_preparation}

For the different evaluations in this study, resting-state fMRI data provided by the HCP \textit{S1200 release} was incorporated \citep{Smith2013}. To define the nodes of the brain network, the multi-modal parcellation proposed by \cite{Glasser2016} was applied, which is composed of $ 180 $ segregated regions within each hemisphere. The average of the BOLD signal was computed within each brain region, so for each resting-state session, $ N = 360 $ time courses were obtained (180 per hemisphere). During one session $ T = 1200 $ fMRI images were collected, so that the ROI timeseries can be represented by a data matrix $ \mb{X} \in \mbb{R}^{N \times T} $. We filtered the resting-state fMRI timeseries data with a $0.04-0.07 Hz$ Butterworth bandpass filter, because this frequency band has shown to be most reliable and functionally relevant for gray matter activity \citep{Glerean2012, Bruckner2009, Deco2017, Biswal1995, Achard2006}. 

For learning the predictions of the BOLD signal, samples of input and output sequences were generated from the timeseries data in $ \mb{X} $ \citep{Wein2021}. This was achieved by selecting windows of length $ T_p $ to obtain input sequences of neural activity states $ [\mb{x}^{(1)}, \ldots, \mb{x}^{(T_p)}] $, and respective target sequences of length $ T_f $ denoted as $[\mb{x}^{(T_p+1)}, \ldots, \mb{x}^{(T_p+T_f)}] $. The time index $ t $ was propagated through each fMRI session, where in total $ T - T_p - T_f + 1 $ input-output pairs were generated per session. For each fMRI session the first $ 80\% $ of those time window samples were used as the training data for the models, the subsequent $ 10\% $ as a validation set, and the last $ 10\% $ were employed for testing. For the following comparisons, the length of the input and output sequences were selected to be $ T_p = T_f = 60 $, which corresponds to a time span of roughly $ 43\ s $, based on a sampling interval of $ TR = 0.72\ s $ \citep{Ugurbil2013}. This time window has been shown to be long enough to be sufficiently challenging for the models and to make clear the differences in their performance. Likewise, the time window of $ 60 $ timepoints is short enough for them to make reasonable non-random forecasts of the BOLD signal.

In addition to the functional dynamics in the different brain regions derived from fMRI, the structural connectivity between those regions was reconstructed from DTI data. For this purpose the DTI dataset in the HCP \textit{S1200 release} was processed using the multi-shell, multi-tissue constrained spherical deconvolution model \citep{Jeurissen2014}, made available in the \textit{MRtrix3} software package \citep{Tournier2019}. White matter tractography was performed to estimate the anatomical connection strength between the regions defined by the multi-modal parcellation atlas \citep{Glasser2016}. The number of the streamlines which connect two atlas regions was used to determine the structural connectivity values between the $ N $ brain regions, which were then collected in a structural connectivity matrix $ \mb{A}_{SC} \in \mbb{R}^{N \times N} $. A detailed description of the MRI datasets and their preprocessing is provided in the \nameref{sec:dataset} section. In addition the embeddings of the nodes within the structural network $ \mb{A}_{SC} \in \mbb{R}^{N \times N} $ were generated using the node2vec algorithm \citep{Grover2016}. The parameters for this algorithm are outlined in detail in the \nameref{sec:spatial_dependencies} section and further Pearson correlation was used to quantify the degree of similarity of structural nodes in their connectome embedding space. The pairwise similarities between the $ N $ nodes were then collected in the matrix $ \mb{A}_{CE} \in \mbb{R}^{N \times N} $.


\subsection{Comparison of GNN architectures}\label{sec:comparison_GNNs}

\begin{figure}[!htb]
\bc

\makebox[\textwidth][c]
{ 
  \begin{minipage}[c]{0.30\textwidth}
  \includegraphics[width=1\textwidth]{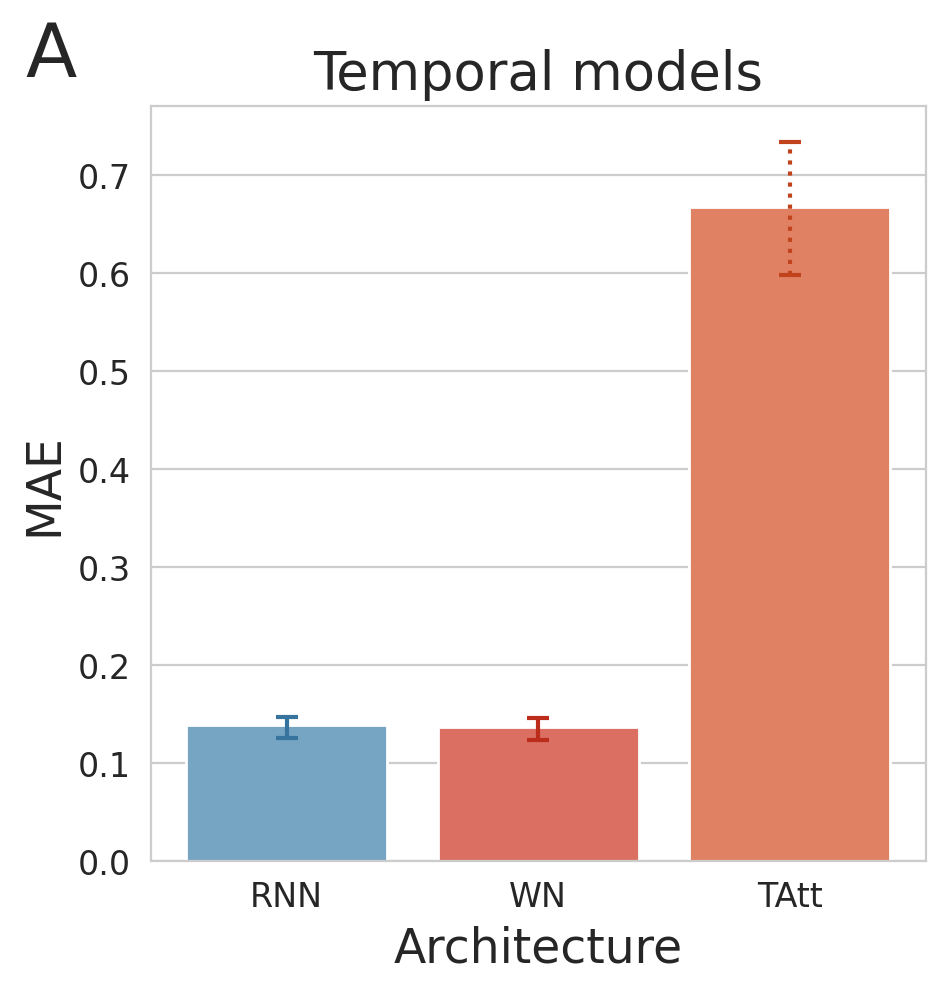}  
  \end{minipage}
    
  \hspace{7mm}    
    
  \begin{minipage}[c]{0.52\textwidth}
  \includegraphics[width=1\textwidth]{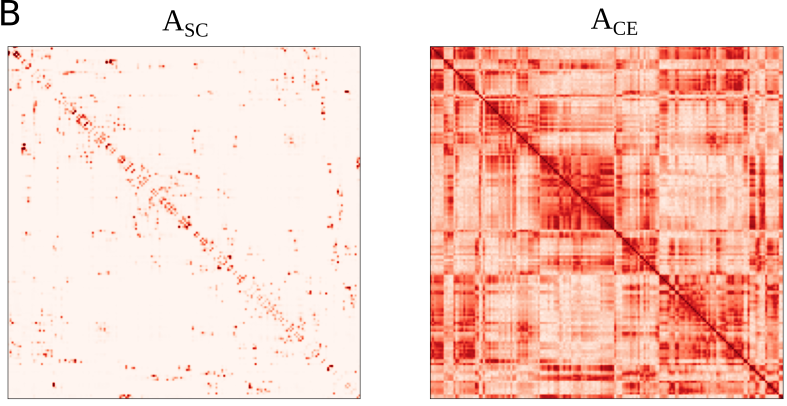}
  \vspace{0mm}
  \end{minipage}    
}

\vspace{5mm}

\makebox[\textwidth][c]
{
  \begin{minipage}[b]{0.30\textwidth}
  \includegraphics[width=1\textwidth]{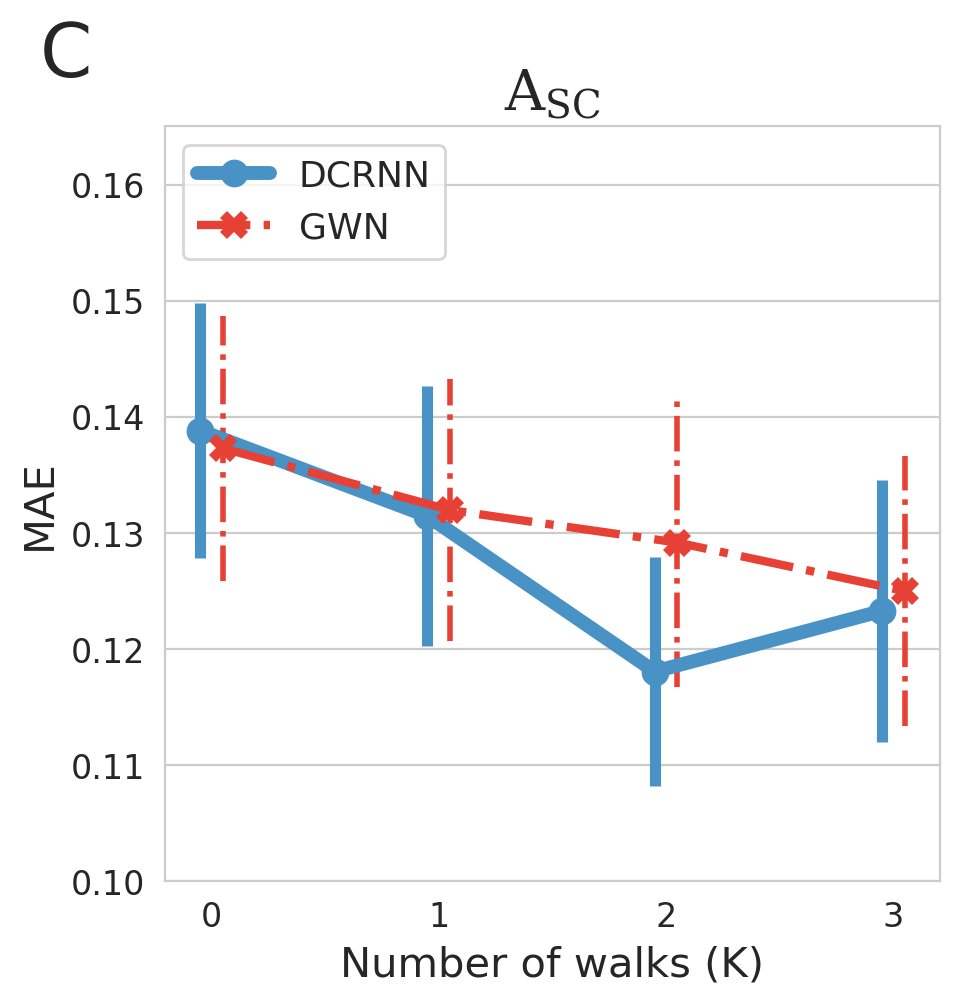}  
  \end{minipage}
  
  \hspace{1mm}
  
  \begin{minipage}[b]{0.30\textwidth}
  \includegraphics[width=1\textwidth]{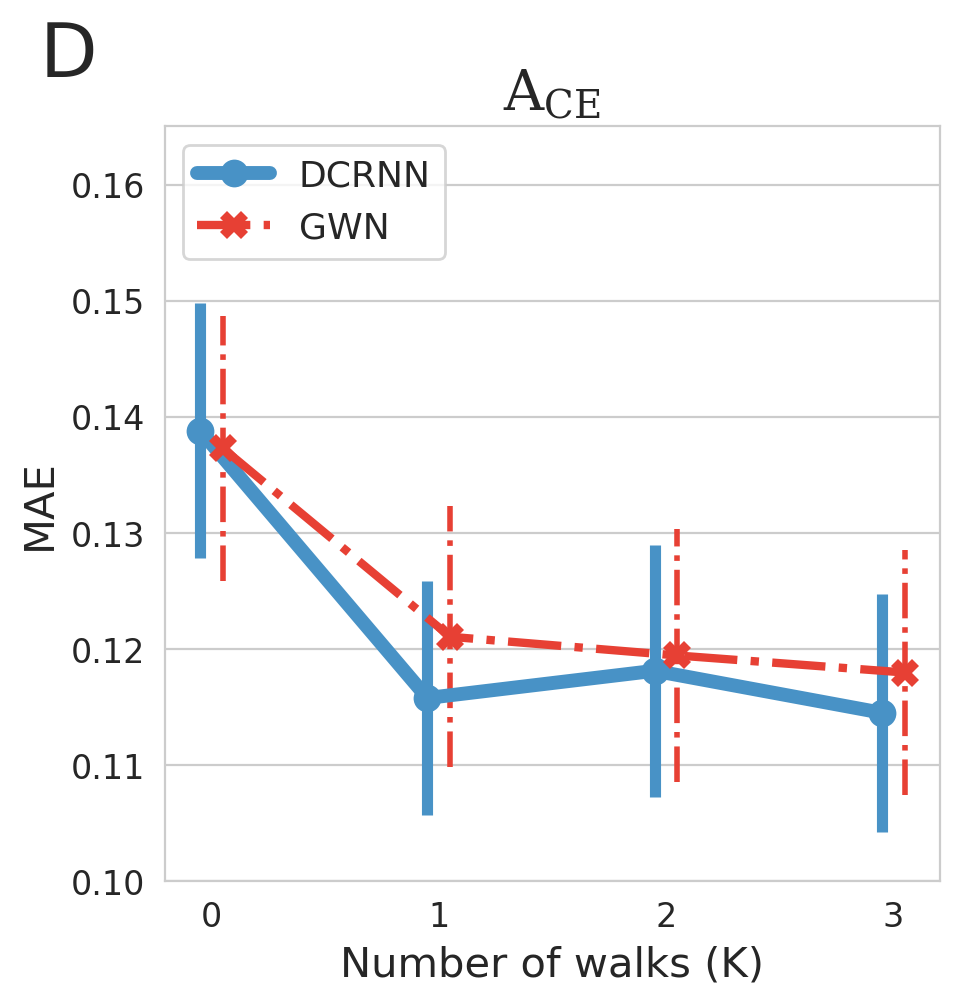}
  \end{minipage}  
  
  \hspace{1mm}
  
  \begin{minipage}[b]{0.30\textwidth}
  \includegraphics[width=1\textwidth]{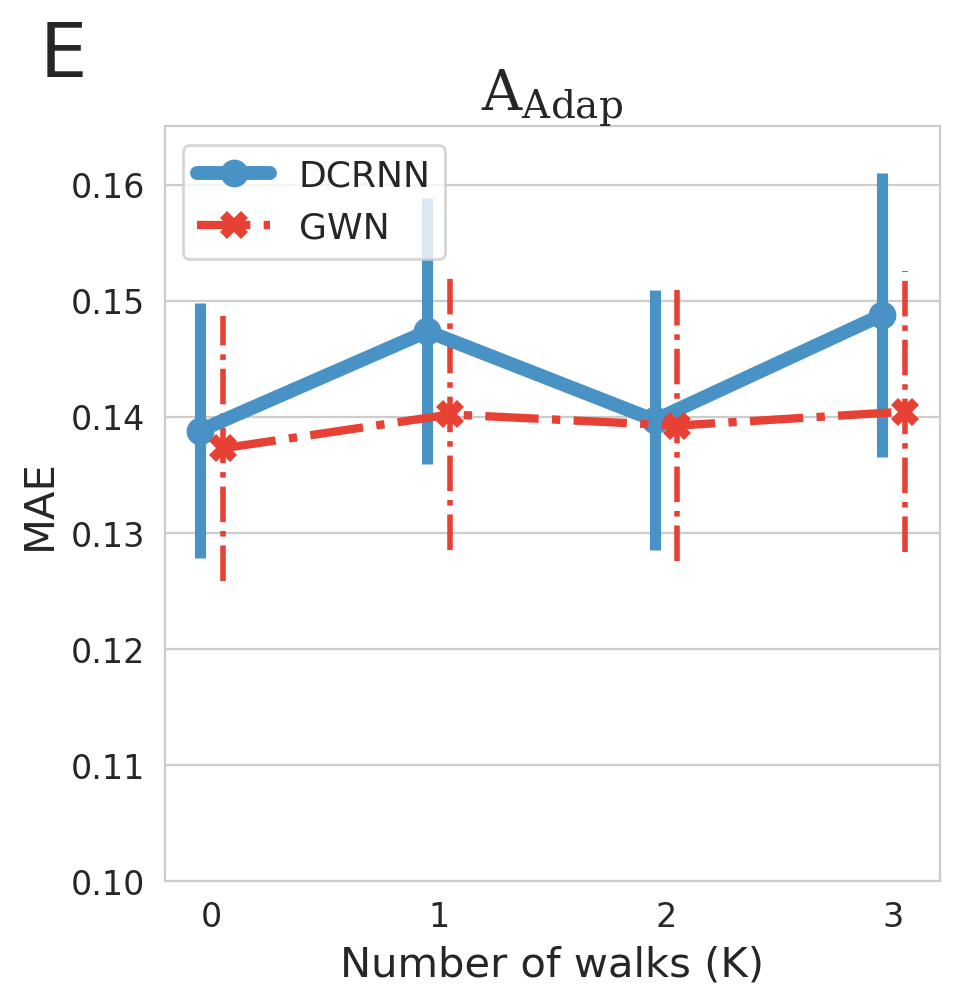}
  \end{minipage}  
}

\ec 
\caption{Figure (A) shows a comparison of different modeling strategies for temporal dynamics in the BOLD signal, comparing the test MAE of the recurrent neural network (RNN), the WaveNet (WN) and the temporal attention (TAtt) architecture. The overall test error was computed as an average across samples, brain regions and subject sessions. The errorbars represent the standard deviation of the test MAE across subjects. Due to their high accuracy in the temporal domain, we focus on RNN and WN based approaches for forecasting the spatio-temporal dynamics in the following. Spatial relations are added to the temporal models in form of graph convolutions, and the spatio-temporal extension of the RNN and WN models are respectively denoted as diffusion convolution recurrent neural network (DCRNN) and graph WaveNet (GWN) \citep{Li2018, Wu2020}. Spatial transitions are based on the relations of network nodes captured in a weighted adjacency matrix, which is either based on structural connectivity ($ \mb{A}_{SC} $), connectome embedding similiarity ($ \mb{A}_{CE} $), or adapted during model training ($ \mb{A}_{Adap} $). In (B) the adjacency matrix $ \mb{A}_{SC} $ based on structural connectivity within the $ 180 $ regions of the right hemisphere is illustrated, together with the adjacency matrix $ \mb{A}_{CE} $ derived from structural connectome embedding similarities. The regions in this illustration are ordered according to the atlas proposed by \cite{Glasser2016}. Figure (C), (D) and (E) show the prediction accuracies of the DCRNN and GWN model in dependence of the walk order $ K $. In figure (C) the overall test MAE is shown when incorporating the SC as an adjacency matrix $ \mb{A}_{SC} $, figure (D) illustrates the test MAE when employing CEs in an adjacency matrix $ \mb{A}_{CE} $ to define spatial relationships, and (E) displays the case when using a self-adaptive weight matrix $ \mb{A}_{Adap} $.}
\label{fig:comparison_temporal_spatial}
\end{figure}

Before evaluating the performance of different models on a larger variety of MRI study scenarios, we will first focus on the effects of different temporal and spatial modeling techniques. For this purpose a dataset with a sample size of a medium sized fMRI study including $ 25 $ subjects will be incorporated. Each resting-state fMRI session was decomposed into pairs of input and output samples, as discussed in the \nameref{sec:data_preparation} section, and the generated training, validation and test samples were then aggregated across the $ 25 $ fMRI sessions. The neural signal in regions within the right hemisphere \citep{Glasser2013}, consisting of $ N = 180 $ ROIs, will be included in the following comparison. At first we evaluate the prediction accuracy of the different temporal modeling strategies. For this purpose we compare the recurrent neural network (RNN) model, with the WaveNet (WN) model and the temporal attention (TAtt) model. The influence of the model hyperparameters, which are used in the following comparisons are described in the \nameref{sec:model_training} section and discussed in detail in \nameref{sec:supp_hyperparameters}.  The BOLD signal data was scaled to zero mean and unit variance for the evaluations, to obtain values of a magnitude that is easier to interpret. Figure \ref{fig:comparison_temporal_spatial} (A) shows the test mean absolute error (MAE) between the predicted and the true neural activity. By generating windowed input-output pairs of activity values from the fMRI data, the last $ 10 \% $ of samples from each session correspond to $ 108 $ of such input-output pairs per session for testing, each containing $ 60 $ output time points (corresponding to roughly $ 43 s $ of activity). The overall test errors were computed as the average across all these test samples from the $ 25 $ subjects and across all $ 180 $ brain regions. The comparison shows that RNN and WN have very similar capabilities in predicting the BOLD signal, while the TAtt model exhibits a worse performance. 
To test the significance of this difference between the models, we further computed the test MAE of each individual subject as an average across predicted timepoints, brain regions and test samples per subject. By applying a paired t-test, the differences between the WN and RNN model to the TAtt model were shown to be both highly significant with $ p \leq 0.0001 $ across subjects (Cohen's $ d = 10.68 $ and $ d =  10.66 $ respectively). In addition to the MAE, we evaluated these models in \nameref{sec:supp_comparison_GNNs} using scale-free measures like R-squared ($ R^2 $) and the similarity of the predicted FC states. 
Despite their conceptual differences the results show that the RNN and WN based approach can both recover a comparable and consistent amount of temporal information from the fMRI data. In comparison to these, the TAtt architecture appears to be less suitable to accurately predict the BOLD signal with this limited amount of data. For this reason in the following we will focus on RNN and WN based approaches for identifying suitable models to model functional dynamics in brain networks.

In the next step we will study the impact of adding information on spatial relations between the different regions in the brain network. This will be implemented by invoking graph convolution operations in the predictive models, as outlined in detail in the \nameref{sec:spatial_dependencies} section. The definition of an adjacency matrix determines how information is propagated between the different nodes in our brain network, and in our evaluations we investigate three conceptually different possibilities. In the first approach we use the structural connectivity as derived from DTI as the substrate for information exchange between different ROIs. The SC based adjacency matrix $ \mb{A}_{SC} $ is illustrated in figure \ref{fig:comparison_temporal_spatial} (B). The information can propagate along direct connections in the graph $ (K = 1) $, but also higher orders ($ K = 2, 3, \ldots $) expressing the influence of indirect connections can considerably contribute to interactions between different areas in the brain \citep{Becker2018, Liang2017, Bettinardi2018}. A walk order of $ K = 0 $ denotes the case when including no spatial information exchange between network areas, exclusively incorporating temporal information for the predictions. Figure \ref{fig:comparison_temporal_spatial} (C) depicts the test MAE in dependence of the walk order $ K $ when using the SC derived from DTI as a basis for information propagation in space. The RNN based model in combination with graph convolution operations is referred to as DCRNN \citep{Li2018} and the MAE of its predictions, averaged across test samples, brain regions and predicted timepoints is depicted in blue. Figure \ref{fig:comparison_temporal_spatial} (C) shows that it has the lowest test MAE when incorporating walks on the structural graph up to a order of $ K = 2 $. The WN incorporating graph convolution operations is denoted as GWN \citep{Wu2019} and its average test MAE is shown in red in figure \ref{fig:comparison_temporal_spatial} (C). The influence of the walk order $ K $ on the GWN accuracy suggests that its performance can be successively improved by including first-order connections, followed by the second- and third-order connections. As an alternative thereto, the structural similarity between ROIs can be based on their CE similiarity $ \mb{A}_{CE} $, as illustrated in figure  \ref{fig:comparison_temporal_spatial} (B). The comparison between $ \mb{A}_{CE} $ and the structural connectivity matrix $ \mb{A}_{SC} $ highlights that in the adjacency relation defined by the structural embeddings, long range connections between brain regions are considerably more pronounced. Figure \ref{fig:comparison_temporal_spatial} (D) shows the test MAE of the models when incorporating  $ \mb{A}_{CE} $ in the graph convolution operations. In this case we can observe for both models a sharp drop in the error at walk order $ K = 1 $. This suggests that the node embeddings already inherently capture higher order relations between nodes in the brain network. Finally in figure \ref{fig:comparison_temporal_spatial} (E) the test MAE is shown when treating the connections between nodes as learnable weights. In this case we do not observe an improvement in the test error. This observation indicates that it is rather challenging to learn all $ N^2 $ connections between brain regions without any prior knowledge. In general both STGNN models could profit the most when using CEs to characterize the spatial layout for functional interactions between brain regions. For the DCRNN the test error was $ MAE = 0.1388 $ when incorporating no information from other brain regions in the network, and could be reduced to $ MAE = 0.1158 $ (for $ K = 1 $) when using CEs to model the information exchange within the brain network. To test whether incorporating information about structural connections significantly increases the prediction accuracy of our models, we at first recomputed the overall test MAE for each subject again. Then by using a paired t-test, we find that, for both STGNN models (DCRNN and GWN) and both adjacency types ($ \mb{A}_{SC} $ and $ \mb{A}_{CE} $), the impact of structural modeling is positive (Cohen's $ d > 1 $ for all comparisons) and significant ($ p \leq 0.0001 $ for all comparisons), compared to the case in which it is not considered. Although the performance differences between the GWN and DCRNN are quite small in general, the DCRNN slightly outperformed with a test error of $ MAE = 0.1158 $ the GWN with a test error of $ MAE = 0.1211 $ at $ K = 1 $ (significant with  $ p \leq 0.0001 $, Cohen's $ d = 0.49 $). In addition the distribution of the test error across subjects and ROIs, with and without the structural modeling in STGNNs is illustrated in \nameref{sec:supp_comparison_GNNs} in figure \ref{fig:supp_MAE_ROIs}. 
This demonstrates that around $ 17 \% $ more information on functional dynamics can be directly retrieved from nodes with similar context within the structural network. Using the SC to model transitions could only reduce the MAE of the DCRNN by $ 5 \% $ at $ K = 1 $. This observation supports the idea that structural node embeddings can strengthen the relationship between structural data derived from DTI with functional data observed in fMRI \citep{Rosenthal2018}. When applying a paired t-test, the improvement of the prediction accuracy when using the CE similarity in comparison to the SC became for both, the DCRNN and GWN model, significant with $ p \leq 0.0001 $ at $ K = 1 $ (Cohen's $ d = 1.45 $ for the DCRNN and $ d = 0.95 $ for the GWN). By inherently capturing higher order transitions in $ \mb{A}_{CE} $, only a low walk order $ K $ is required to capture information from structurally connected ROIs. In this manner, this technique can help to efficiently reduce the number of necessary parameters to account for spatial dependencies in STGNN models.



\subsection{Model accuracy and network scaling}\label{sec:network_scaling}

In this section we study the prediction accuracy of the above introduced STGNN based approaches and compare it to the VAR model, which is currently most often used for directed functional connectivity analysis \citep{Friston2013, Barnett2013}. In practicable applications the amount of available fMRI data may vary depending on the project size and on the recruited subject cohort. Also the size of the brain network of interest can range from a few specific areas in a single functional network to a large-scale whole brain analysis. For this purpose we consider different scenarios in our following evaluations, by analyzing the models accuracies in dependence of the brain network size and the fMRI dataset size. We consider one larger subject dataset consisting of resting-state fMRI sessions from $ 50 $ different subjects, one medium sized dataset of $ 25 $ subjects and one small dataset including data from $ 10 $ subjects. In addition we vary the size of the analyzed brain network. The first network consists of $ 22 $ ROIs per hemisphere involved in visual processing as defined by the Glasser parcellation \citep{Glasser2016} (a complete list of selected ROIs is provided in the \nameref{sec:supp_visual_ROIs}).
The second network includes the regions within one hemisphere, and for that purpose the $ 180 $ regions within the right hemisphere included in the Glasser atlas were selected \citep{Glasser2016}. Finally the whole brain network of in total $ 360 $ regions was incorporated. As discussed in the \nameref{sec:data_preparation} section, windowed input and output time sequence pairs were created from the data and the goal of the different models is accordingly to predict $ T_f = 60 $ TRs of neural activity from the past  $ T_p = 60 $ activity values. We fitted the VAR model using the ordinary least squares (OLS) method as implemented in the multivariate Granger causality (MVGC) toolbox \citep{Barnett2013}, and for each dataset we selected the VAR model with order $ p $ that achieved the best MAE on the test set, as outlined in more detail in the \nameref{sec:VAR} section. The hyperparmeters used for the STGNNs are described in the methods part in the \nameref{sec:model_training} section. Further for this comparison the CE similarity $ \mb{A}_{CE} $ with transition order of $ K = 1 $ was used to define the structural relations in the STGNN models, which has shown to improve the GNNs forecasting accuracy with low computational cost, as discussed in the section \nameref{sec:comparison_GNNs}. 

\begin{figure}[!htb]
\bc

\makebox[\textwidth][c]
{
  \begin{minipage}[b]{0.31\textwidth}
  \includegraphics[width=1\textwidth]{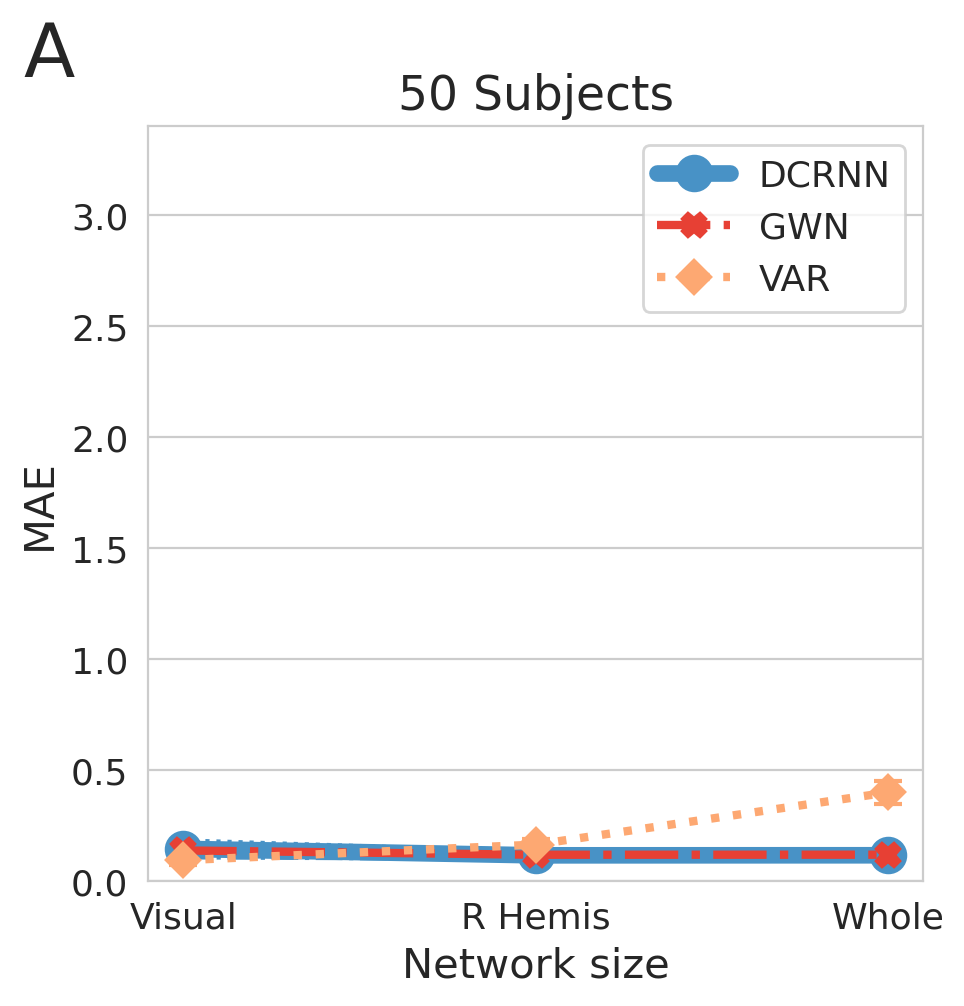}  
  \end{minipage}
    
  \begin{minipage}[b]{0.31\textwidth}
  \includegraphics[width=1\textwidth]{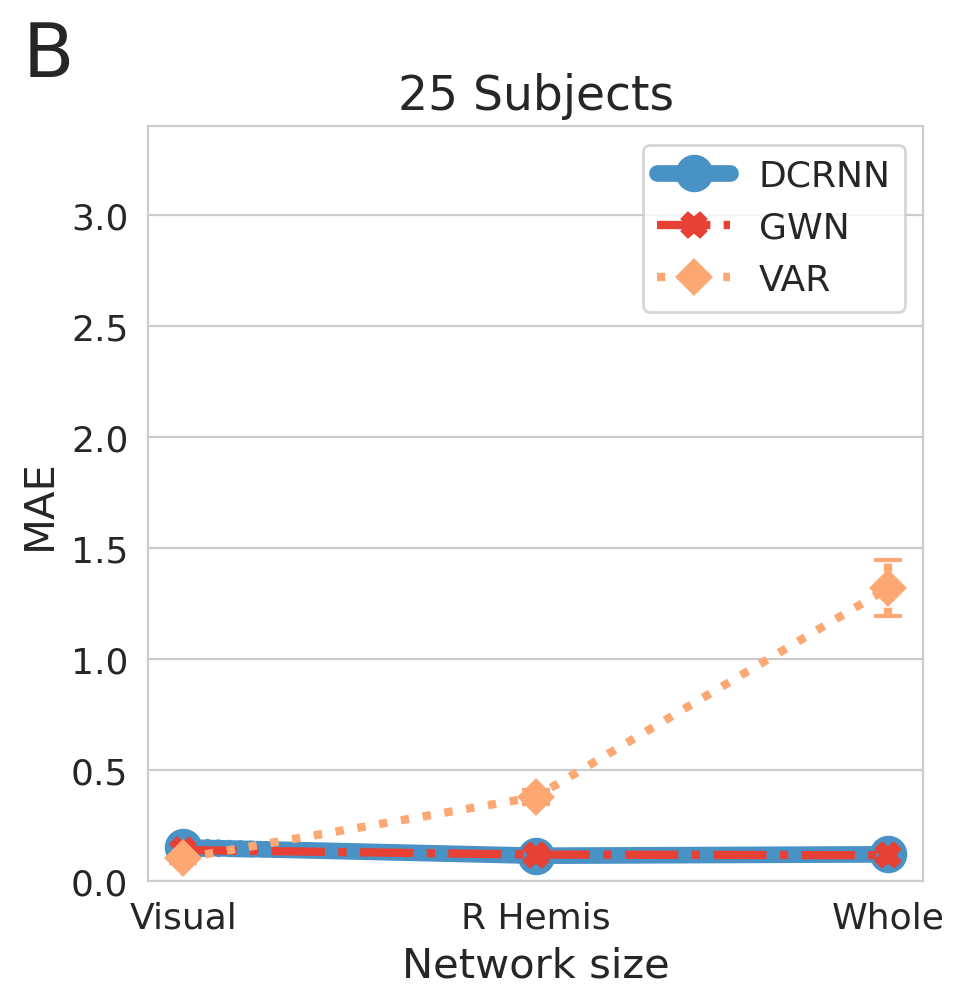}
  \end{minipage}  
  
  \begin{minipage}[b]{0.31\textwidth}
  \includegraphics[width=1\textwidth]{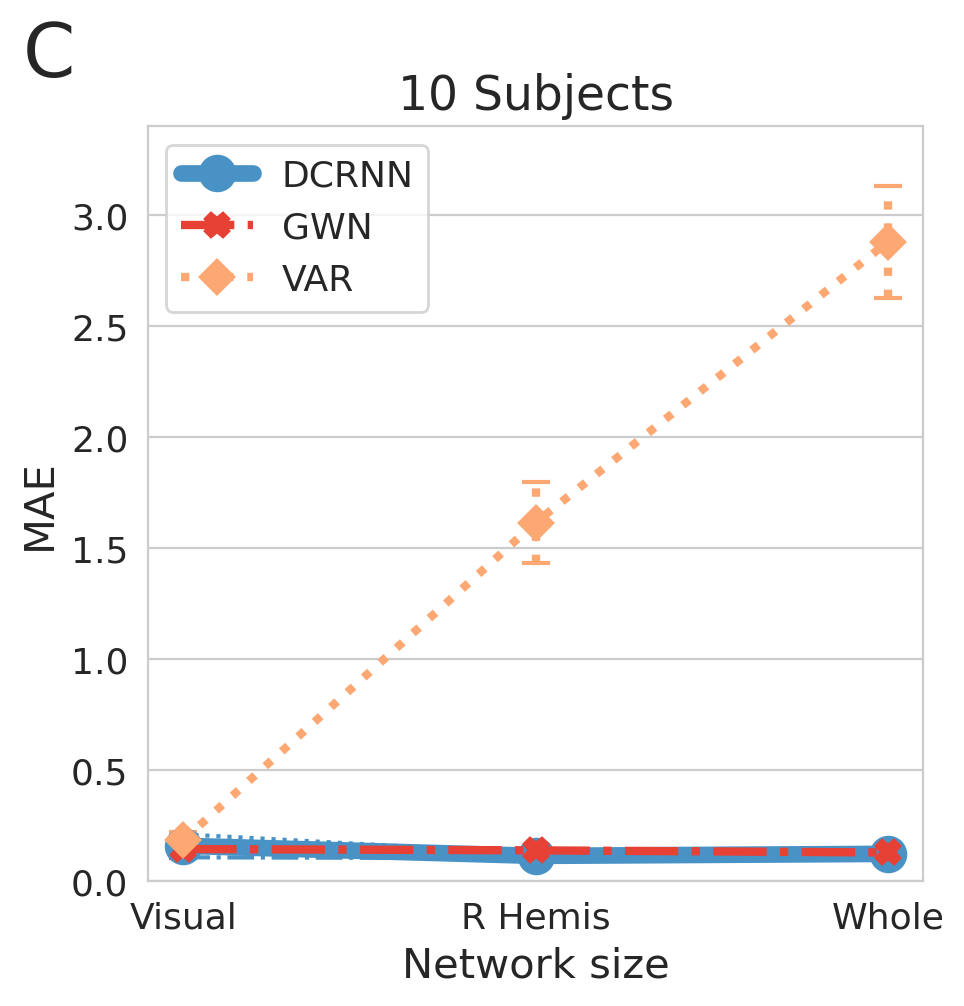}
  \end{minipage}  
}

\ec 
\caption{The figure shows a comparison of the model performances when varying the amount of data and the size of the network. The test MAE of the VAR is here depicted in orange, the MAE of the DCRNN in blue and the error of the GWN in red. The overall error was computed as an average across brain regions, timesteps and test samples. The error bars represent the standard deviation of the test error across subjects. In (A) the test MAE using a dataset of $ 50 $ subjects is shown for the visual network, the network within the right hemisphere and the whole brain network \citep{Glasser2016}. Figure (B) and (C) show the test performances in dependence of the network size using the $ 25 $ and $ 10 $ subject dataset, respectively.
}
\label{fig:comparison_models}
\end{figure}

Figure \ref{fig:comparison_models} shows the test accuracy of the VAR, DCRNN and GWN model in dependence on the dataset size and brain network size. It can be observed in figure \ref{fig:comparison_models} (A) that if a large dataset of $ 50 $ subjects is available, all models are able to accurately predict the BOLD signal with a low test MAE, and a notable increase in the test error only appears for the VAR model, when it is fitted to the whole brain network. Figure \ref{fig:comparison_models} (B) shows the test MAE when data from $ 25 $ subjects is incorporated. In this case the test error of the VAR model starts to increase noticeably when modeling activity distributions within one hemisphere and becomes quite large when including the whole brain network. In contrast to these, the prediction accuracies of the DCRNN and GWN models remain stable in all cases. Finally when only $ 10 $ subject datasets are available, the test MAE of VAR model is highly dependent on the analyzed network size, as illustrated in figure \ref{fig:comparison_models} (C). The DCRNN and GWN model can still achieve a high accuracy, also when a limited amount of data are available and the network size is relatively large. In addition, this comparison of the models is replicated in \nameref{sec:supp_network_scaling} using additional measures like $ R^2 $ and the similarity of predicted FC states. After applying a paired t-test, the differences between the DCRNN and GWN to the VAR were shown to be in all cases highly significant with $ p \leq 0.0001 $ (Cohen's d $ \gg 1 $), except when the VAR is only fitted to the single visual network, where it still could make reliable forecasts.

\begin{figure}[!htb]
\bc

\makebox[\textwidth][c]
{\includegraphics[width=0.65\textwidth]{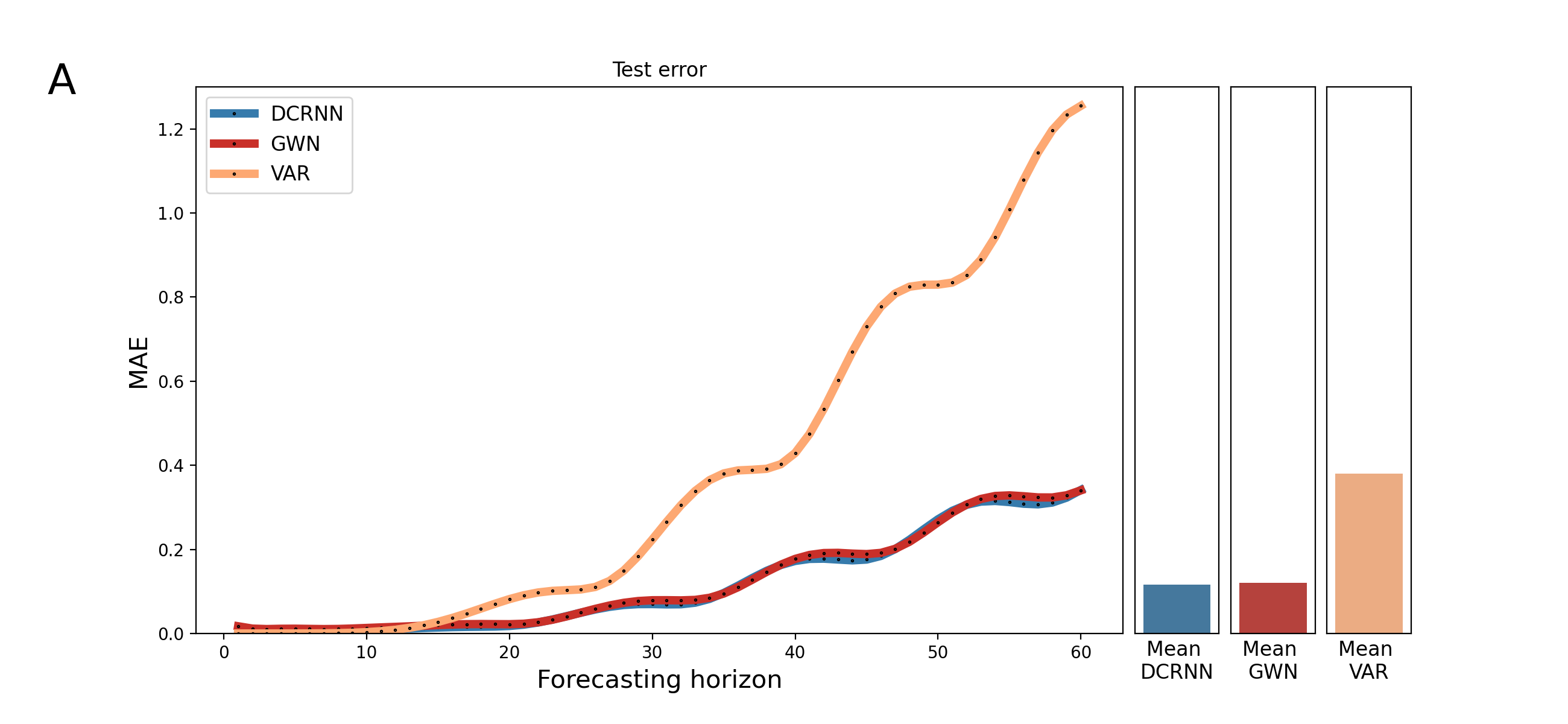}
}

\makebox[\textwidth][c]
{\includegraphics[width=0.86\textwidth]{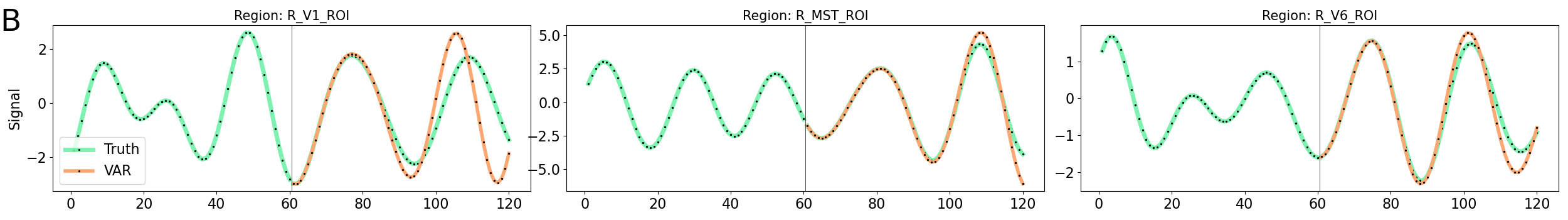}
}
\makebox[\textwidth][c]
{\includegraphics[width=0.86\textwidth]{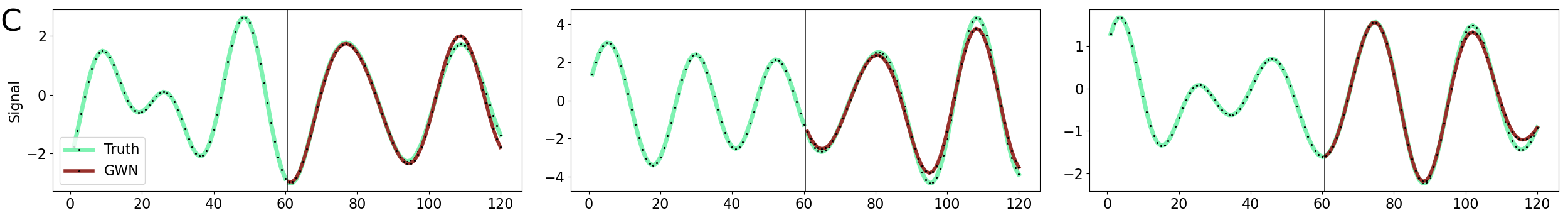}
}
\makebox[\textwidth][c]
{\includegraphics[width=0.86\textwidth]{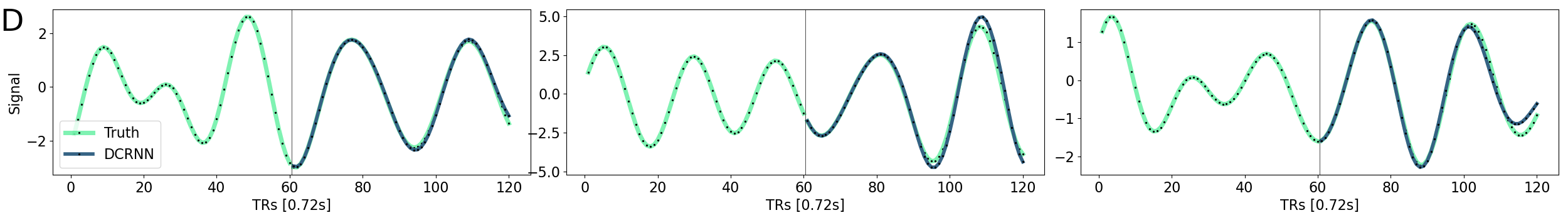}
}

\vspace{3mm}

\makebox[\textwidth][c]
{\includegraphics[width=0.99\textwidth]{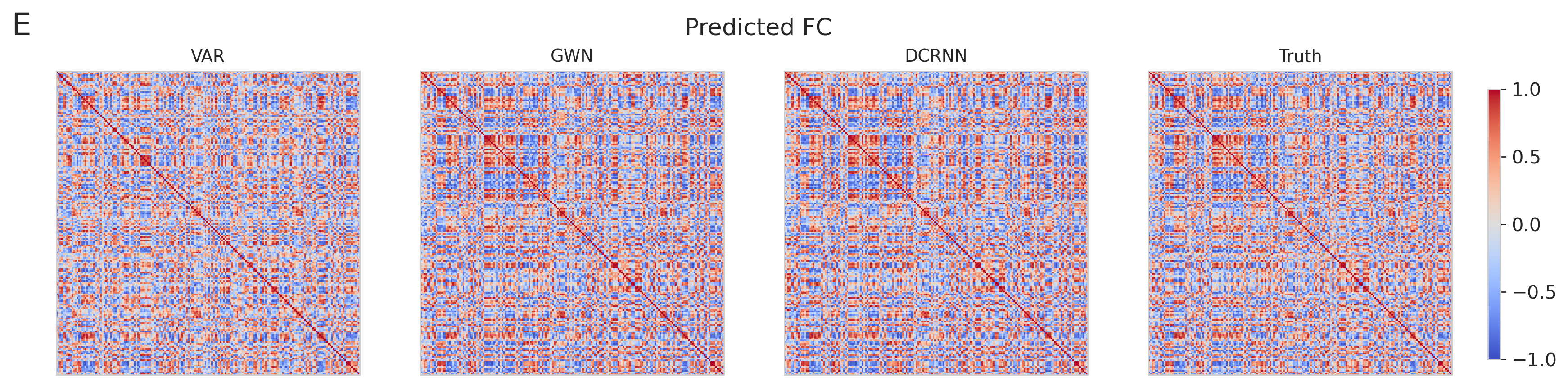}
}

\ec 
\caption{The prediction accuracy of the different models is presented in more detail for the $ 25 $ subject dataset and the brain network including the ROIs within the right hemisphere \citep{Glasser2016}. In (A) the test MAE in dependence of the forecasting horizon is shown, computed as an average across test samples and brain regions. Figure (B), (C) and (D) show examples of predictions generated by the VAR, GWN and DCRNN model respectively. The examples in this figures were chosen to be representative for the whole test set, by selecting only examples which errors maximally deviate by $ 0.02 $ from the corresponding average test MAE of the models.
Figure (E) depicts examples of predicted FC states of the different forecasting models, including the true FC state on the right side. These representative examples deviated maximally by $ 0.005 $ from their average correlation to the true FC state.}
\label{fig:comparison_detail}
\end{figure}

To illustrate the prediction accuracies of the different models in more detail, an example of the predictions using the dataset including $ 25 $ subjects, and modeling the activity within one hemisphere is shown in figure \ref{fig:comparison_detail}. Figure \ref{fig:comparison_detail} (A) shows the MAE of the models computed as an average across test samples and ROIs in dependence of the forecasting horizon. In this case within the first $ 15 $ predicted timesteps all three models can generate very accurate predictions, but after that period the error of the VAR model starts to accumulate, while the GNN based approaches remain considerably more stable and precise. The predicted BOLD signals of the different models in a few representative samples are shown in figures \ref{fig:comparison_detail} (B), (C) and (D). 
Additionally, the predicted FC states $ \mb{A}_{FC} \in \mbb{R}^{N \times N} $ were computed as the Pearson correlation between predicted BOLD signals of all $ N $ brain regions, and a comparison of representative predictions with the true FC state is illustrated in figure \ref{fig:comparison_detail} (E). The average correlation of the predicted FC state to the true FC state was for the VAR model $ r_{FC} =  0.635 $ on this dataset, while the GWN could achieve a correlation value of $ r_{FC} = 0.948 $, and the DCRNN a value of $ r_{FC} = 0.950 $. The overall FC similarity for all different datasets of all prediction models is given in \nameref{sec:supp_network_scaling} figure \ref{fig:supp_comparison_models_FC_simi}. Furthermore, in \nameref{sec:supp_liberal_bandpass} in figure \ref{fig:supp_liberal_bandpass} we performed the analysis on the same dataset using a more liberal frequency filtering within the $ 0.01 - 0.1 Hz $ frequency range. In this range the signal dynamic becomes more complex and we can observe an increase in the prediction error of the different models accordingly.

\begin{figure}[!htb]
\bc

\makebox[\textwidth][c]
{\includegraphics[width=0.95\textwidth]{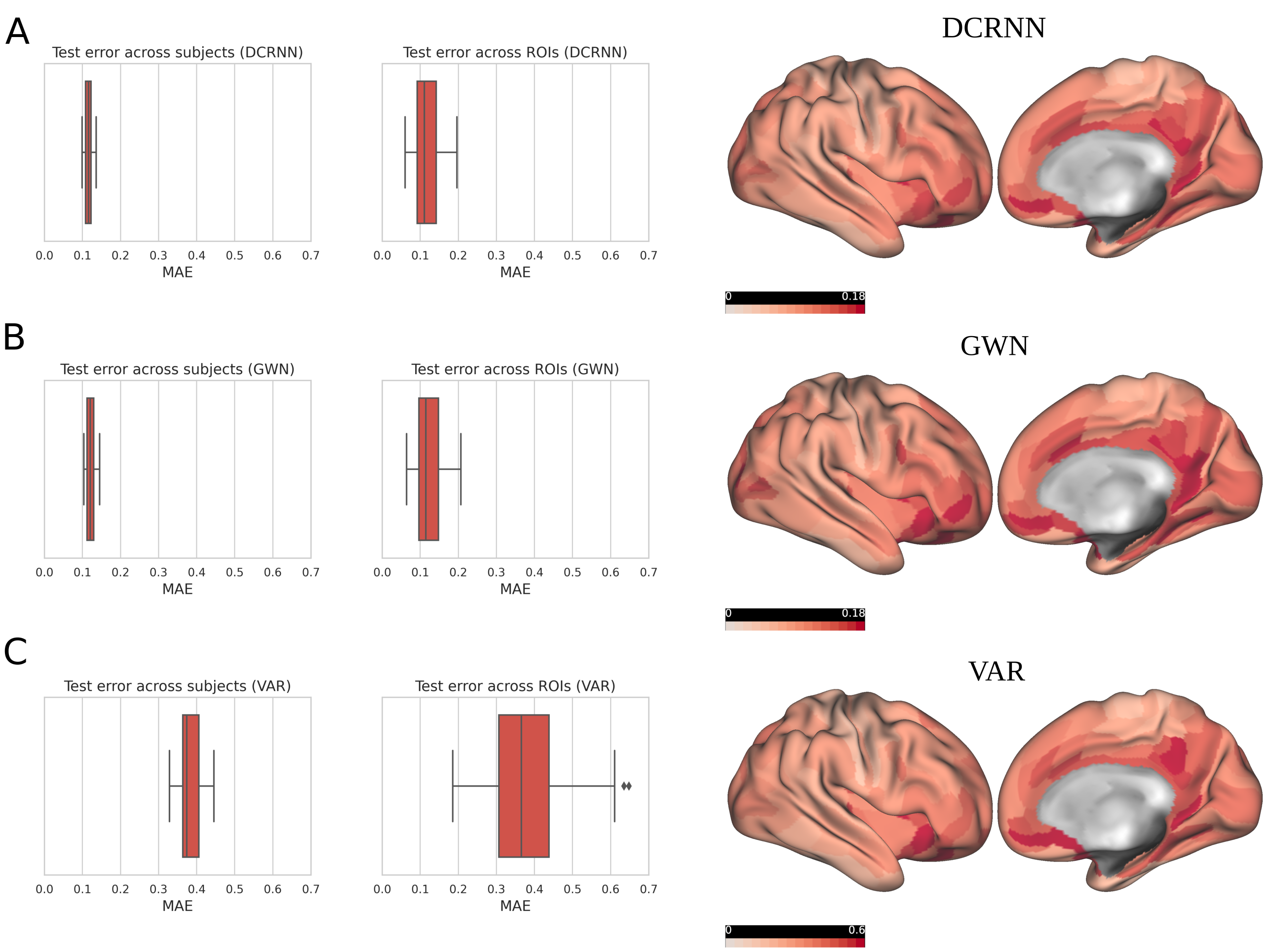}
}

\ec 
\caption{The distribution of the test error across subjects and brain regions is shown. In (A) the MAE across subjects and brain regions of the DCRNN is first visualized in a boxplot on the left side. Additionally on the right side of the figure, the MAE values are projected onto the cortical surface within the right hemisphere, where the colormap was linearly scaled between $ 0 $ and $ 0.18 $. In (B) the distribution of the test MAE of the GWN is shown and in (C) the MAE distribution of the VAR model. For the VAR the colormap was adjusted to account for larger error values by scaling it between $ 0 $ and $ 0.6 $.}
\label{fig:MAE_ROIs}
\end{figure}

In addition we evaluated in more detail how the prediction errors are distributed across different subjects and different ROIs. Figure \ref{fig:MAE_ROIs} shows the distribution of the test MAE of the DCRNN, GWN and VAR model across subjects, and in dependence of the brain region within the right hemisphere. For all three models we observe a consistently greater prediction error in the posterior cingulate cortex and medial orbitofrontal cortex, which could point towards a more complex BOLD dynamic in those regions. Alternatively, the prediction accuracy might be also affected by a lower signal-to-noise ratio observed in medial brain regions \citep{Olman2009}.

\subsection{Multi-modal directed connectivity}\label{sec:multimodal_connectivity}

In the \nameref{sec:comparison_GNNs} section different approaches have been investigated to model functional interactions between segregated regions in the brain network. The results showed that incorporating information on the spatial relation between ROIs in the form of the structural connectivity or connectome embedding similarity could considerably improve the prediction accuracy of the GNNs models. This points out that the GNNs are able to learn relevant and functional informative transitions of neural activity on their structural spatial layout. 
Based on the idea of Granger causality \citep{Granger1969} that the observation of one event $ A $ carries information about the occurrence of a future event $ B $, this might represent initial evidence for a potentially causal relation between $ A $ and $ B $. Due to the relatively low temporal sampling rate and physiological artifacts in fMRI \citep{Smith2011, Webb2013}, it still is a matter of discussion to what extent we can observe a \textit{causal} relationship between brain regions in this imaging modality \citep{Wang2014, Mill2017, Bielczyk2019}. But the observation that activity in one region carries additional information among activity in another region in the brain can also go beyond simple undirected FC or SC, and we therefore refer to this kind of relation as \textit{directed connectivity} in the following. Propagating the information between ROIs related to their SC or structural CE similarity has the potential to give us in this manner a multi-modal perspective of such a directed relationship between different brain areas.
For this purpose we choose a perturbation base approach to reconstruct the amount of information individual ROI carries about other ROIs \citep{Zeiler2013}. By learning a function $ h(\cdot) $, the GNN models try to infer from an input sequence of neural activity states $ [\mb{x}^{(1)}, \ldots, \mb{x}^{(T_p)}] $ a sequence of future activity states $ [\mb{\hat{x}}^{(T_p+1)}, \ldots, \mb{\hat{x}}^{(T_p+T_f)}] $, whereby $ \mb{x}^{(t)} \in \mbb{R}^{N} $ denotes the activity at timestep $ t $ in all regions $ n = 1, \ldots, N $. To induce an artificial perturbation into the system of neural dynamics, we remove all activity in a certain ROI $ n' $ by setting its activity values to the sample mean $ x_{n'} = 0 $. By using the perturbed timeseries as an input for our trained model $ h(\cdot) $ the model generates then a prediction $ [\mb{\hat{x}}'^{(T_p+1)}, \ldots, \mb{\hat{x}}'^{(T_p+T_f)}] $. To reconstruct the directed influence of ROI $ n' $  on ROI $ n $ we evaluate the overall difference between the original prediction and the prediction with perturbation in the input:

\beq
	I_n(n') = \frac{1}{S} \sum_{s=0}^{S} \frac{1}{T_f} \sum_{t=0}^{T_f} | \mb{\hat{x}}_n^{(t)}(s) - {\mb{\hat{x}}_n}'^{(t)}(s) | \label{eqn:directed_influence}
\eeq 
where $ I_n(n') $ denotes the impact of ROI $ n' $ on $ n $. Further $\mb{\hat{x}}_n^{(t)}(s)$ and ${\mb{\hat{x}}_n}'^{(t)}(s)$ denote the predictions in ROI $ n $ with and without the perturbation in $n'$ of one test sample $ s $ at time step $ t $. Note that this artificial perturbation approach has only the goal of making spatial relations between ROIs in STGNN models explainable, and should not be equated with the effect of an experimental perturbation applied to the human brain, like for example induced by transcranial magnetic stimulation (TMS).

\begin{figure}[!htb]
\bc

\includegraphics[width=0.65\textwidth]{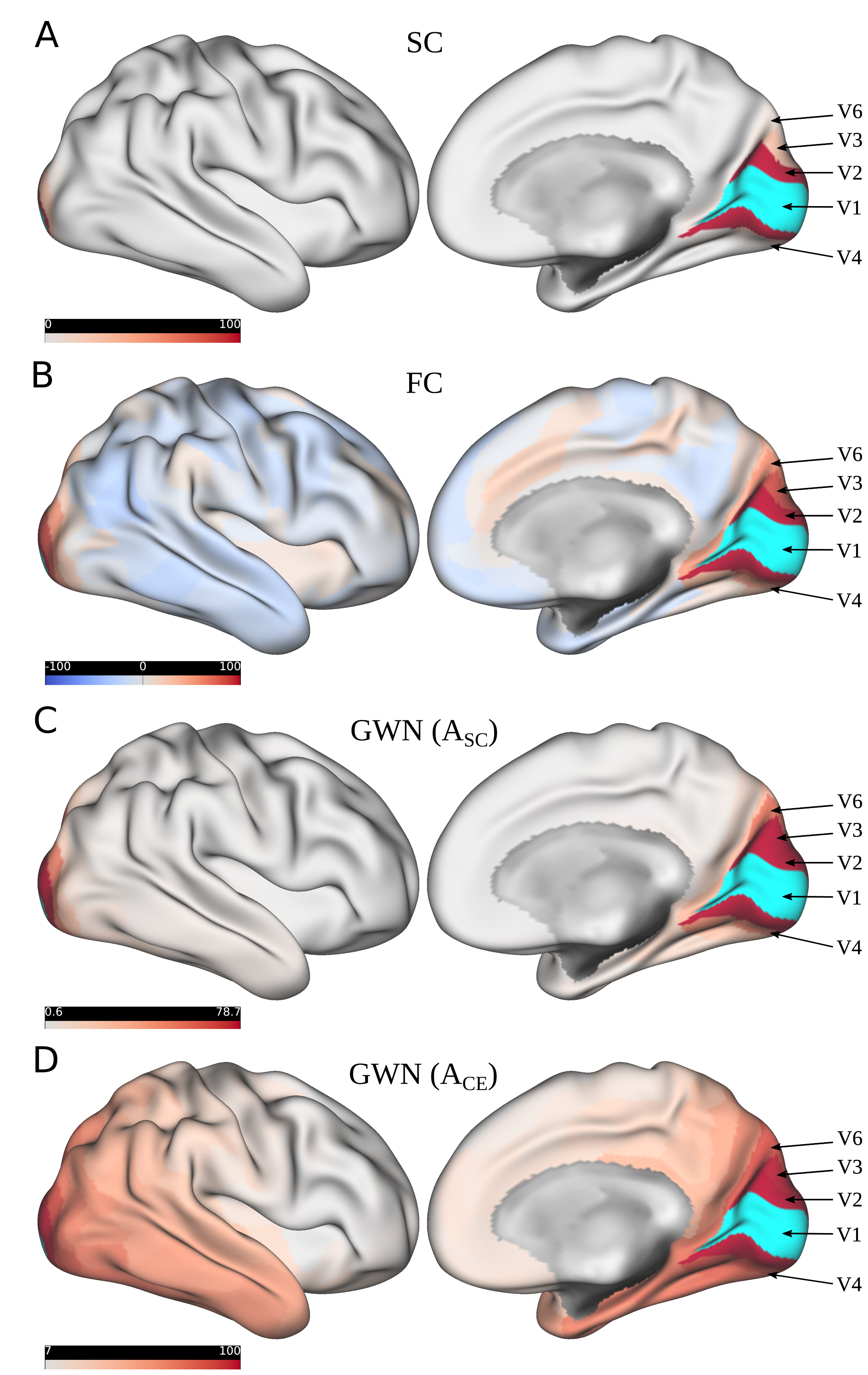}

\ec 
\caption{Different types of connectivity are illustrated between V1 and all other regions within the right brain hemisphere. In (A) the structural connectivity is shown, whereby the target region V1 is marked in light blue and the connectivity strength is encoded in red. In (B) the correlation based functional connectivity is illustrated, which was computed as an average across subjects. Further (C) shows the measures of influence $ \mb{I}(n') $, derived from the GWN model using the SC for information propagation and figure (D) depicts the influence when incorporating CEs for the information exchange between ROIs. The values of the connectivity measure were linearly mapped between $ 0 $ and $ 100 $ (and between $ -100 $ and $ 100 $ for FC). The default scaling of the color values provided by the \textit{connectome workbench} (version 1.4.2) was used, adjusting the colormap between the $ 2th $ and $ 98th $ percentile of the values respectively.}
\label{fig:comparison_connectivity}
\end{figure}

In the following we compare this proposed measure of directed influence $ \mb{I}(n') $ to the classical undirected types of brain connectivity. First we compare it to structural connectivity as derived from DTI, characterized by the number of fiber tracks connecting two brain regions. Then we incorporate functional connectivity, defined as the Pearson correlation of functional activity timecourses between two areas. We employ the above introduced GWN model to obtain a multi-modal measure of directed connectivity $ \mb{I}(n') $, first using the SC as substrate for information propagation, captured in $ \mb{A}_{SC} $, and then also employing the similarity of CEs, represented by $ \mb{A}_{CE} $. In the following example we study the connectivity of V1 within the right hemisphere by incorporating data of $ 25 $ subjects. For this purpose a perturbation was induced into the target region V1, and the impact of this perturbation on all other $ 179 $ regions within the right hemisphere was then characterized by computing the measure of directed influence $ \mb{I}(n') $, as defined in equation \ref{eqn:directed_influence}. These values for directed connectivity strength can then be visualized by projecting them onto the $ 179 $ other areas of the cortical surface. For the following comparison, all connectivity values were rescaled by normalizing them between $ 0 $ and $ 100 $. At first in figure \ref{fig:comparison_connectivity} (A) the structural connectivity between V1 and all other $ 179 $ regions within the right hemisphere is depicted. The target region V1 is marked here in light blue, and the strength of connectivity to all other regions is encoded in red. Figure \ref{fig:comparison_connectivity} (A) shows that we can mainly observe a pronounced structural connectivity between V1 and V2 and some structural connections leading to V3. Figure \ref{fig:comparison_connectivity} (B) shows the undirected functional connectivity in resting-state. In this type of connectivity we can observe predominantly correlations to the functional activity in V2 and V3, but also a considerable connection strength to V3, V4 and V6. In figure \ref{fig:comparison_connectivity} (C) the directed connectivity strength $ \mb{I}(n') $ is depicted, when using the SC as spatial backbone for the information exchange between brain regions in the GNN. In comparison to the SC, in this variant of brain connectivity we can observe in addition to V2 also a more pronounced relationship to areas V3 and V4, and to some anatomically more distant areas like V6 and the ventromedial visual area VMV1. This multi-modal type of connectivity also reflects the role of indirect structural connections by modeling higher order transitions on the structural scaffold captured by the STGNN model. As an alternative to the SC, in figure \ref{fig:comparison_connectivity} (D) the directed connectivity patterns when using CE similarity as the spatial layout in the GWN are displayed. Here we can see an even stronger integrity of V1 within the visual network, which is in agreement with the observation that CEs capture higher order topological information of anatomical connectivity \citep{Rosenthal2018}. Figure \ref{fig:supp_connectivity_DCRNN} in \nameref{sec:supp_connectivity} shows additionally the spatial relations learned by the DCRNN model. Here we can observe a pronounced similarity to the directed connectivity pattern learned by the GWN architecture, showing additionally strong relations to areas like V3 and V4. Based on this observation, such a GNN based connectivity approach can serve as a link between structural and functional connectivity, and as such they can provide a multi-modal perspective on directed influences between individual areas in brain networks. So far, we have only studied the effect of a single artificial perturbation in V1 directed to all other regions within the right hemisphere. This approach can further be extended to sample a full connectivity network, by systematically inducing perturbations into all regions of interest for the analysis, and then systematically observing the effect on the other network regions. In figure \ref{fig:supp_connectivity_visual} in \nameref{sec:supp_connectivity} we studied the effects of perturbations induced in some different additional areas of the visual network based on this approach.
In this manner, this technique can allow us to reconstruct the directed spatial relations between brain areas captured in STGNN models, and could be applied in practical applications by for example comparing these connectivity patterns between different conditions in task-based fMRI, or studying the difference between healthy and diseased brain states.

A study of \cite{Shinn2021} demonstrated that FC topology in resting-state fMRI is shaped by and can be predicted from spatial and temporal autocorrelations, as typically observed in fMRI data. To also investigate to what extend STGNN based connectivity patterns are related to such correlations, we computed the temporal autocorrelation as the Pearson correlation between the BOLD signal values and its lagged values with different lag orders $ j $, depicted in figure \ref{fig:supp_temp_autocorr} in \nameref{sec:supp_connectivity}. We could observe a relatively high temporal correlation of $ 0.89 $ around lag order $ j = 14 $, and of $ 0.62 $ around lag order $ j = 27 $, what shows that within these first $ 30 $ TRs of the signal, such temporal autocorrelations can still be detected. In section \nameref{sec:network_scaling} we could show that STGNNs were also able to make reliable long-term predictions of the BOLD signal up to a horizon of $ 60 $ TRs, what demonstrates that STGNNs capture properties of neural activity dynamics that go clearly beyond the range that is shaped by temporal autocorrelations.  
In addition, spatial autocorrelations can also play a distinctive role in shaping FC network properties \citep{Shinn2021}. The analysis in section \nameref{sec:comparison_GNNs} showed that by modeling the information exchange between structurally connected brain regions in the spatial domain, the prediction accuracy of the STGNNs could be significantly improved, in comparison to the null-models, which incorporated no spatial dynamics. The observation that spatially connected regions contain some \textit{additional} relevant information on functional dynamics points out that spatial interactions captured in STGNNs go beyond simple correlation based spatial network relations. A more detailed comparison of the individual differences and similarities between correlation based FC and the STGNN based connectivity pattern is additionally provided in a barplot in \nameref{sec:supp_connectivity}  (figure \ref{fig:supp_connectivity_comparison}).



\section{Conclusion}

In this study we have compared different STGNN architectures for learning the spatio-temporal dynamics in brain networks. First in the section \nameref{sec:comparison_GNNs} we studied different mechanisms for learning the temporal dynamics in the BOLD signal. We could show that a RNN based model and a WN based model exhibit very similar capabilities in learning the temporal characteristics in neural activity timeseries. Despite their conceptual differences in their architectures, they demonstrated almost the exact same prediction accuracy, which indicates that they are both very consistent in capturing the temporal information in the data. As an alternative, we also studied TAtt mechanisms to learn temporal characteristics of neural signals. The TAtt model showed to be less suitable to model the dynamics in the BOLD signal with a limited amount of fMRI data. Despite incorporating techniques into the TAtt model that in general stabilize the learning, like residual connections and batch normalization \citep{He2016, Ioffe2015}, its prediction error was considerably higher in comparison to the RNN and WN based approach. This indicates that the geometric assumptions which are realized by the temporally structured inference in the RNNs and WNs based on either recurrent computations or causal convolutions can contribute to the learning of the temporal characteristics of the BOLD signal. We then studied the impact of adding spatial dependencies to the temporal models, realized by invoking graph convolution operations. We have compared different spatial layouts for information propagation between ROIs, and therefore included either the structural connectivity ($ \mb{A}_{SC} $), the CE similarity ($ \mb{A}_{CE} $), or a self-adaptive adjacency matrix ($ \mb{A}_{Adap} $) into the STGNN models. While the model performance of the GWN and DCRNN steadily improved with higher walk orders $ K $ on the anatomical substrate, we could observe a more pronounced improvement already when using CEs with a walk order of only $ K = 1 $. This embedding strategy turns out to be therewith also interesting in applications of STGNNs, because it helps to effectively incorporate indirect structural connections with low computational cost. In addition, the observed characteristics of CEs in our application support the ideas of \cite{Rosenthal2018}, which showed in their study that embeddings of the structural network can naturally capture higher order topological relations between ROIs within the structural layout. In our context of modeling spatio-temporal dynamics this method also proved to strengthen the relationship between brain structure and functional dynamics. Learning all $ N^2 $ connections of the underlying structural graph during the model training has been shown to be challenging for the STGNN models, in case no prior knowledge is provided to them in form of the anatomical brain connectivity. While such highly parameterized artificial neural network models can be in theory quite flexible in learning complex relations \citep{Hornik1989, Bruel2020}, often the decisive limitation is the successful optimization of the parameters during model training \citep{Dauphin2014}. In the discussed applications of STGNNs in fMRI, where the amount of training data is most often quite limited, prior knowledge in form of the anatomical graph structure has been shown to considerably support the learning of spatial relations between brain areas captured in STGNN models.

So far methods based on biophysical modeling \citep{Honey2009, Deco2012, Deco2013, Messe2015, Messe2015b, Messe2014}, graph theory \citep{Liang2017, Abdelnour2018, Becker2018, Lim2019} or machine learning \citep{Deligianni2016, Amico2018, Rosenthal2018, Sarwar2021} have contributed already numerous valuable insights into the structure-function relation in brain networks, and could highlight the role and importance of the structural connectome in shaping functional connectivity patterns. While the majority of approaches studying the structure-function relationship infer brain dynamics by fitting the models to empirically observed FC patterns, STGNNs provide us with a possibility to directly predict the observed neural activity states. Similar to some other recently proposed predictive models \citep{Singh2020, Suarez2021}, this principle can allow us to investigate additional interesting aspects of dynamic brain functions. As discussed in section \nameref{sec:comparison_GNNs} this could enable us for example to study directly the amount of information on the activity of one ROI that is contained in the activity of other structurally connected ROIs. For a comparison with other currently used approaches investigating SC-FC mappings, the predicted BOLD signal states of STGNN can be used then again to reconstruct predictions for FC states, as shown in figure \ref{fig:comparison_detail} (E). The relatively high accuracy in predicting empirical FC states 
already point out the potential of STGNNs in this field. Moreover, in comparison to other currently popular approaches used in this area \citep{Messe2015b}, by learning localized graph filters in STGNNs, their forecasting accuracy is also robust with regards to the brain network size. While such highly parameterized artificial neural network models appear to be promising for achieving high prediction accuracies of FC states \citep{Sarwar2021}, they cannot provide us with the same mechanistic insights into physiological processes as biophysically inspired models. Still they can be used to supplement current biophysically inspired models, for studying different aspects of the structure-function relationship from a novel data-driven perspective. A more comprehensive comparison of these different new approaches, evaluating in detail their interrelations like in the study of \cite{Messe2015b}, could be  thereby interesting for future studies in this area.

In the \nameref{sec:network_scaling} section we have compared the STGNN models to a VAR model, which is currently most often used in Granger causality analysis for inference of directed relationships between brain regions \citep{Barnett2013}. We evaluated the accuracy of the different approaches on a variety of brain network sizes and data set sizes to account for different possible scenarios in their application in fMRI studies. The results showed that if a sufficiently large cohort of $ 50 $ subjects is available, also a VAR model is able to make very reliable long-term predictions, and only for a large network consisting of $ N = 360 $ there is a notable increase in the prediction error. But the dependency of the accuracy on the network size $ N $ becomes more apparent when data from only $ 25 $ subjects are used to fit the VAR model, and when only $ 10 $ subjects are available, the error grows strongly with $ N $. This demonstrates that a VAR is a very reliable and fast model for fMRI studies with a sufficiently large test subject size and for connectivity studies including a limited amount of pre-defined regions. However it can be desirable in some cases to include a larger amount of brain areas into the connectivity analysis, in order to avoid omitting relevant areas in the network of interest. Also in MRI studies it can be very costly and time-consuming to collect a large amount of data, which is, for example, especially challenging in studies on rare neurological disorders. A classical VAR model fits a parameter for every possible connection between the $ N $ regions in a network, so that the number of parameters in a VAR based approach grow strongly with an order of $ N^2 $. In contrast thereto, STGNNs utilize prior information in the form of the anatomical connectivity, and then model the functional information exchange based on this underlying structural substrate. By incorporating graph convolution operations in STGNNs, the amount of parameters only linearly scale with walk order $ K $, which can even be chosen to be $ K = 1 $, if higher order structural relations are already expressed in an adjacency matrix derived from connectome embeddings ($ \mb{A}_{CE} $). This property allowed STGNNs to make very robust inferences also on large networks and when only limited data are available, thereby providing a flexible method for various connectivity analysis scenarios.

Finally in the \nameref{sec:multimodal_connectivity} section we studied the individual spatial interactions within the brain network which were learned by the STGNN models. By integrating information on the anatomical connectivity into the GNN based models, we could derive a multi-modal connectivity measure for directed relationships between brain regions. When comparing this measure of influence to the original structural connectivity, we can observe that STGNNs have learned to include transitions along higher order structural connections in the network. The models could infer links between $ V1 $ and $ V2 $, but additionally strong connections to $ V3 $ and $ V4 $. Especially when incorporating the CE based similarity $ \mb{A}_{CE} $ to define spatial node relations in the STGNN models, we can observe a high integration of $ V1 $ within the visual system. However due to the relatively low temporal sampling rate in fMRI \citep{Friston2013}, and the indirect measurement of neural signals based on their hemodynamic response \citep{Webb2013}, one should also be aware of these limitations in the inference of directed and potentially causal connections in fMRI studies \citep{Smith2011}. Our lag-based predictive approach based on STGNN models might therefore also be affected by the same limitations as classical Granger causality in fMRI. On the other hand, a combined fMRI-MEG study by \cite{Mill2017} and different computational simulations of fMRI data \citep{Wang2014, Seth2012, Wen2013, Duggento2018} could establish evidence that Granger causality is still able to identify meaningful directed relationships between brain areas in fMRI, despite the indirect measurements based on the hemodynamic response. As an alternative, deconvolution based approaches can have the potential to infer from the measured BOLD signals the underlying neural timeseries \citep{Bush2015, Mill2017} for assessing \textit{effective} brain connectivity, rather than only estimating \textit{directed} functional connectivity. But the estimation of the underlying hemodynamic response from the data might come with the cost of introducing additional assumptions and uncertainties \citep{Roebroeck2009, Bielczyk2019}. A more detailed discussion on considerations concerning Granger causality, and in general causal inference in fMRI, is provided in the comprehensive review of \cite{Bielczyk2019}, as well as in the perspective on FC and its variants by \cite{Reid2019}. 
Despite these current limitations in fMRI, a multi-modal GNN based approach can allow us to join structural and functional imaging data in a new manner, and reveals thereby potential for supplementing current analysis methods in brain connectivity research by studying such directed relations under a novel perspective \citep{Reid2019}.

In conclusion, in our study we found that the DCRNN and GWN architecture are both suitable for the task of functional dynamics inference. Using CEs to characterize the structural similarities between brain regions could further improve their prediction accuracy. Their robust scaling properties and the possibility to combine the information in structural and functional MRI data reveal the potential of STGNNs in the field of brain connectivity analysis.
Besides their applications in fMRI, other functional neuroimaging techniques like electroencephalography (EEG) or magnetoencephalography (MEG) might be interesting for analyzing temporal dynamics with STGNNs in the high frequency range. While in this presented approach, we only incorporated a single temporal feature (the BOLD signal) into the STGNNs, in general such a flexible data-driven approach could be expanded to account for different types of data and annotations. For example the activity measured in a combined EEG-fMRI experiment \citep{Abreu2018, Mele2019} could be also simultaneously integrated in STGNNs as different temporal features, or adding the temporal response of a subject could be helpful to better predict activity patterns in task-based fMRI. Also alternative structural imaging techniques like neurite orientation dispersion and density imaging (NODDI) \citep{ZhangH2012} might capture additional aspects of the brain structure, which could be included as structural information in STGNN based models. In clinical applications multi-modal STGNNs could be interesting for studying how the relationship between structure and function is affected in the diseased brain \citep{Panda2021}, or which impact a structural lesion might have on the functional organization of the brain network \citep{Alstott2009}. Still the research on GNNs is a relatively new field in machine learning and recent developments in this field can make interesting contributions to our understanding of information processing in brain networks \citep{deHaan2020, Schnake2021}.



\section{Materials and methods}\label{sec:materials_methods}

\subsection{Dataset}\label{sec:dataset}

The MRI data set used in our study is provided by the HCP data repository \citep{Hodge2015, VanEssen2013}. As part of the HCP protocol, the study participants gave written informed consent to the HCP consortium. The MRI scanning protocols were approved by the Institutional Review Board at Washington University in St. Louis. We incorporated data of the \textit{S1200 release}, which provides data from resting-state fMRI sessions, each with a duration of $14.4$ minutes, whereby $1200$ volumes were sampled per session. The data was acquired with customized \textit{Siemens Connectome Skyra} magnetic resonance imaging scanners with a field strength of $B_0 = 3 T$, using multi-band (factor 8) acceleration \citep{Moeller2010, Feinberg2010, Setsompop2012, Xu2012}. A gradient-echo echo-planar imaging (EPI) sequences with a repetition time $TR = 720\ ms$ and an echo time $TE = 31.1\ ms$ was used. The field of view of the fMRI sequence was $FOV = 208\ mm \times 180\ mm$ and in total $N_s = 72$ slices with a slice thickness of $d_s = 2\ mm$ were collected, containing voxels with an isotropic size of $2\ mm$. The preprocessing of the HCP fMRI data includes corrections of gradient-nonlinearity-induced distortions, registration to a single-band reference image to account for subject motion, and registration to the structural T1w image  \citep{Glasser2013, Jenkinson2002, Jenkinson2012, Fischl2012}. Further ICA-FIX was applied to automatically classify and remove artifactual components in the resting-state fMRI data \citep{Smith2013, Salimi2014, Griffanti2014}. Finally the volumetric fMRI images are mapped into the CIFTI grayordinate space and Gaussian surface smoothing with a FWHM of  $2\ mm$ is performed. A detailed description of the standard minimal preprocessing pipelines of the HCP can be found in \cite{Glasser2013}. In a next step to define our brain network, the multi-modal parcellation proposed by \cite{Glasser2016} was applied, which divides the cortical surface into $ 180 $ segregated areas per hemisphere. The BOLD signal within each area was averaged, to obtain the temporal activity evolution for each node in our brain network. For this study we considered it useful to apply global signal regression in our preprocessing \citep{Power2016}. Firstly, in a systematic comparison of different preprocessing methods to address motion artifacts, \cite{Ciric2017} could show that an ICA-based de-nosing in combination with global signal regression is among the most effective methods to reduce movement artifacts. This result is in line with the study of \cite{Burgess2016}, investigating the effect of ICA-FIX in combination with global/grayordinate signal regression on resting-state fMRI data provided by the HCP. Furthermore in our study of functional interactions between specific brain regions, the objective was to extract the additional information, which certain regions contain about the activity in other regions. Therefore local interactions rather than global modulations in the signal were of main interest for our analysis \citep{Power2016}. The time courses were then bandpass filtered in the $0.04-0.07 Hz$ frequency range. In a summary of several different studies that account for different artifacts in the BOLD signal related to MRI scanner drift in the frequency range below $ 0.015 Hz $ \citep{Smith1999}, respiratory and cardiac frequencies around $ 0.3 Hz $ and $ 1-2 Hz $ respectively  \citep{Biswal1996}, and fluctuations in arterial carbon dioxide level around $ 0.0-0.05 Hz $ \citep{Wise2004}, the study of \cite{Glerean2012} identified the $0.04-0.07 Hz$ frequency band to be most reliable and relevant for gray matter activity in resting-state fMRI \citep{Achard2006, Zou2008, Bruckner2009}. 
To additionally ensure that the low frequency signals are not mainly related to respiratory artifacts, we studied the respiratory signals recorded with a \textit{Siemens} respiratory belt during resting-state fMRI, as provided by the HCP \citep{Smith2013}. The average respiratory frequency spectrum is depicted in \nameref{sec:supp_respiratory_freq} in figure \ref{fig:supp_respiratory_freq}, and we can observe that respiratory fluctuations are mainly present in the higher frequency range around $ 0.28 Hz $ in this resting-state fMRI dataset.
In addition, the different models were tested on data incorporating a more liberal frequency filtering within the $0.01-0.1 Hz$ range, as presented in \nameref{sec:supp_liberal_bandpass}. 

In the S1200 release, diffusion MRI data was collected in $ 6 $ runs, whereby approximately $ 90 $ directions were sampled during each run, using three shells of $b=1000, 2000,$ and $3000 \ s/mm^2$, with additionally 6 $b=0$ images \citep{Sotiropoulos2013b}. A spin-echo EPI sequence was incorporated with repetition time $TR = 5520\ ms$, echo time $ TE = 89.5\ ms$, using a multi-band factor of $ 3 $. In total $N_s = 111$ slices were collected, with field of view $FOV =  210\ mm \times 180\ mm$ and an isotropic voxel size of $1.25\ mm$. The minimal preprocessing pipeline of the HCP  includes intensity normalization across runs, EPI distortion correction using the FSL5 "topup" tool, correction of eddy-current induced field inhomogeneities and head motion artifacts using the FSL5 "eddy" tool, and finally includes gradient non-linearity corrections and registration to the structural T1w image \citep{Glasser2013, Sotiropoulos2013, Andersson2003, Andersson2015, Andersson2015b}. More details on the minimal preprocessing of the HCP diffusion MRI are described in \cite{Glasser2013}. To reconstruct the anatomical connection strengths between regions within the multi-modal parecellation \citep{Glasser2016}, the \textit{MRtrix3} software package was incorporated \citep{Tournier2019}. Multi-shell multi-tissue constrained spherical deconvolution \citep{Jeurissen2014} was applied to obtain response functions for fiber orientation distribution estimation \citep{Tournier2004, Tournier2007}. Then $ 10 $ million streamlines were created using anatomical constrained tractography \citep{SmithR2012}. Finally spherical-deconvolution informed filtering was used \citep{SmithR2013}, reducing the number of streamlines to 1 million. 
The strength of SC was defined as the number of streamlines connecting two brain regions, normalized by the region volumes. The group structural connectivity matrix $ \mb{A}_{SC} $ was obtained as the average SC across the first 10 subjects, because the variance in the SC strength was relatively low across subjects \citep{Zimmermann2019}, while probabilistic tractography methods are computationally demanding. For the HCP dataset, including only young healthy subjects, the similarity of the SC across subjects was quite high, and the Pearson correlation coefficient between SC values of the $ 10 $ subjects was on average $0.91$. But when comparing very different subject cohorts, like healthy and diseased subjects, the anatomical connectivity can differ considerably between those cohorts, and the SC matrix should then be computed for every studied group individually.


\subsection{Graph neural networks}

Different brain areas communicate via bioelectrical signals transmitted along neuronal axons and collected by neuronal dendrites. Spatio-temporal GNNs provide a novel possibility to incorporate such a structural scaffold into a graph-based prediction model \citep{Wein2021}. Due to cognitive information processing in the brain, the spatial interactions of the activity distribution changes dynamically. Spatio-temporal GNNs thus encompasses both the information about the layout of the physical scaffold encoded by the graph structure and the dynamical information about temporal activity correlations. Recently we used a DCRNN architecture to model the spatio-temporal brain dynamics in resting-state fMRI \citep{Wein2021}. In this study, spatial dependencies of brain activities were modeled via diffusion convolution operations based on the anatomical connectivity and the temporal dynamics of the graph signal were captured in an RNN based model architecture \citep{Li2018}. In our current study we evaluate some alternative spatial and temporal approaches to model dynamics in brain networks. In addition to RNNs, a CNN based architecture for temporal modeling has been introduced by \cite{Wu2019}. These authors built upon the WaveNets \citep{Oord2016} and stack dilated causal convolution layers to capture long-range temporal dependencies. Dilated convolutions support exponentially growing receptive fields in deeper layers of the network and allow us to handle long-range temporal sequences efficiently \citep{Oord2016}. In addition to the temporal processing based on RNNs and CNNs, we also follow ideas expressed in attention networks and incorporated a relevance score that was computed in temporal attention layers \citep{Vaswani2017, Zheng2020}. 

Based on these temporal approaches, we further study different concepts for representing the spatial dependency between brain regions. First we integrated the SC reconstructed from DTI to represent the anatomical substrate for information propagation in graph convolution operations. Then we additionally incorporated CEs of the structural graph to inherently capture higher order relations between ROIs. Finally we used no predefined spatial layout and treated the spatial connection strengths between ROIs as free parameters. These different spatio-temporal GNN architectures have to the best of our knowledge not been applied yet to analyze the dynamics of brain networks, and in our study we investigate their effectiveness in spatio-temporal modeling of functional MRI.


\subsection{Preliminaries}

Let us represent the brain network as a graph. Every specific brain area or region of interest (ROI), then forms a node in the graph. Let these $N$ ROIs form an graph $\mc{G} = (\mc{V}, \mc{E}, \mb{A}_w)$ encompassing $N$ vertices, i.e. the meta-voxels or ROIs, and a set $\mc{E}$ of edges connecting the vertices $v_n,v_{n'}$. The graph structure can then be captured in a weighted adjacency matrix $\mb{A}_w \in \mbb{R}^{N \times N}$, whose entries $w_{nn'}$ provide the connection strengths between vertices $v_n$ and $v_{n'}$, and implicitly define the spatial structure of the graph. As introduced above, in our study we compared three different variants to define the spatial relationship between ROIs. Once we incorporated the SC derived from DTI data as an adjacency matrix $ \mb{A}_{SC} $, we next employed CE to additionally capture higher order topological features in SC represented by  $ \mb{A}_{CE} $, and finally we treated the spatial relations as adaptive learnable parameters $ \mb{A}_{Adap} $ in the GNN models. The dynamics of the graph signal is then represented by the time-varying neural activity obtained from functional imaging data. Let us first assume that each node of the graph is associated with a single feature represented by the BOLD activity. By considering voxel time series of brain activity maps, then all data can be collected into a data matrix $\mb{X} = (\mb{x}^{(1)}, \ldots, \mb{x}^{(T)}) \in \mbb{R}^{N \times T}$ with $ \mb{x}^{(t)} \in \mbb{R}^N$. Given $N$ ROIs, taken from a brain atlas and each represented by a meta-voxel, and considering $T$ time points for each meta-voxel time series, which represents the activation time course of one of the ROIs, then we have, for the BOLD feature represented at node $n$ a related graph signal matrix or BOLD feature matrix:

\beq 
\mc{X}_{::m} \equiv \mb{X}^{(m)} = (\mb{x}^{(m)}_{1} \ldots \: \mb{x}^{(m)}_{T}) 
                 = \left( \begin{array}{ccc}
                 x^{(m)}_{11} & \cdots & x^{(m)}_{1T} \\
                 \vdots & x^{(m)}_{nt} & \vdots \\
                 x^{(m)}_{N1} & \cdots & x^{(m)}_{NT}
                \end{array}
               \right) \in \mbb{R}^{N \times T}
\eeq
Note that the columns $\mb{x}^{(m)}_{t} \in \mbb{R}^N$ of the data matrix describe the activation of all ROIs at any given time point $1 \le t \le T$, while its rows $\tilde{\mb{x}}^{(m)}_{n}$ represent the meta-voxel time course of every single ROI $1 \le n \le N$. More generally, if nodes not only represent a single feature $m$, like the input BOLD signal, but an $M$-dim  feature vector $\mc{X}_{nt:} \in \mbb{R}^M$,  then we obtain a feature tensor $\mc{X} \in \mbb{R}^{N \times T \times M}$, whose frontal, lateral (vertical) and horizontal slices, respectively, read $\mc{X}_{::m} \in \mbb{R}^{N \times T}$, $\mc{X}_{:t:} \in \mbb{R}^{N \times M}$ and $\mc{X}_{n::} \in \mbb{R}^{T \times M}$. 

In addition to above frontal slices $\mc{X}_{::m} \equiv \mb{X}^{(m)}$ of the data tensor $\mc{X}$, we thus have the lateral tensor slices:

\beq 
\mc{X}_{:t:} \equiv \mb{X}^{(t)} = \left( \mb{x}^{(t)}_1 \ldots \: \mb{x}^{(t)}_M \right)  = \left( \begin{array}{ccc}
                 x^{(t)}_{11} & \cdots & x^{(t)}_{1M} \\
                 \vdots & x^{(t)}_{nm} & \vdots \\
                 x^{(t)}_{N1} & \cdots & x^{(t)}_{NM}
                \end{array}
               \right)  \in \mbb{R}^{N \times M}
\eeq 
and the horizontal tensor slices:

\beq 
\mc{X}_{n::} \equiv \mb{X}^{(n)} =  \left( \mb{x}^{(n)}_1 \ldots \: \mb{x}^{(n)}_M \right)  = \left( \begin{array}{ccc}
                 x^{(n)}_{11} & \cdots & x^{(n)}_{1M} \\
                 \vdots & x^{(n)}_{mt} & \vdots \\
                 x^{(n)}_{T1} & \cdots & x^{(n)}_{TM}
                \end{array}
               \right)  \in \mbb{R}^{T \times M}
\eeq 
Note that the column fibers of the data tensor $\mc{X}_{:tm}$, denoted as $\mb{x}^{(m)}_{t}$ represent, at every time point $t$, the distribution of the activity of feature $m$ across all nodes $n$ of the graph. Correspondingly, the row fibers of the tensor $\mc{X}_{n:m}$, denoted as $\mb{x}^{(m)}_{n}$, represent the time course of every feature $m$ at node $n$. Finally the tube fibers of the tensor $\mc{X}_{nt:}$, denoted as $\mb{x}^{(n)}_t$, represent the distributions of features at every node $n$ and time point $t$. This notation will in the following provide the framework to introduce the different techniques to model either dependencies between nodes $ n $, time $ t $ or features $ m $.


\subsection{Spatial dependencies.}\label{sec:spatial_dependencies}

\paragraph{Diffusion convolution:} In the following we provide a short introduction on a variant of graph convolution denoted as \textit{diffusion convolution} in the context STGNNs \citep{Li2018, Wu2020}. The information flow in the underlying graph $\mc{G} = (\mc{V}, \mc{E}, \mb{A}_w)$  is considered as a stochastic random walk process modeled by a state transition matrix $\mb{T} = \mb{D}^{-1} \mb{A}_w = \left( \hat{\mb{w}}_1 \ldots \hat{\mb{w}}_N \right)$ where $\mb{A}_w$ represents a weighted adjacency matrix. The diagonal node degree matrix is given by: 

\beq 
\mb{D} = diag(\mb{A}_w \mb{1}) 
\eeq 

\nit where $\hat{\mb{w}}_n = \left( \hat{w}_{1n} \ldots \hat{w}_{Nn}  \right)^T  \in \mbb{R}^N \; \forall \; n =1, \ldots ,N$ with $\hat{w}_{nn'} = w_{nn'} / \sum_{n'} w_{nn'}$ denoted normalized edge strengths. State transitions were modeled as a diffusion process on an unstructured graph. The former was represented by a random walk Laplacian:

\beq 
\mb{L}_{rw} = \mb{I} - \mb{T} = \mb{U} \hat{\bs{\Lambda}} \mb{U}^T = \mb{U} \left( \mb{I} - \bs{\Lambda} \right) \mb{U}^T = \mb{I} - \mb{U} \bs{\Lambda} \mb{U}^T.
\eeq 

\nit where the transition operator $\mb{T}$ was replaced by its eigen-decomposition with $\mb{U}$ the matrix of eigenvectors and $\bs{\Lambda}$ the diagonal matrix of eigenvalues. Hence, the set of eigenvectors provided an orthogonal basis system for the spatial representation of the brain graph. With the help of these eigenvectors $\mb{u}_n$ the spatial structure of the graph could be implemented. A spectral representation in combination with the convolution theorem then provided a definition of the graph convolution operator $\mc{G}_C$ \citep{Shuman2013}, which served to compute the spatial convolution of the input signal and a spatial filter kernel to yield the output of the $\ell$-th convolution layer as:

\beqa 
\mb{y}^{(q)}_t 
&=& \mb{U} \bs{\Theta}^{(q)}_{\omega}\mb{x}^{(m)}_{\omega} = \mb{U} \bs{\Theta}^{(q)}_{\omega} \mb{U}^T \mb{x}^{(m)}_t \nonumber \\
&\approx& \sum_{k=0}^{K} \theta^{(q)}_k(\omega) \mb{T}^k \mb{x}^{(m)}_t 
\eeqa 

\nit Here the approximation resulted from a power series expansion of the convolution kernel with respect to the eigenvalue matrix $\bs{\Lambda}$ of the transition operator $\mb{T}$ \citep{Wein2021, Defferrard2016}. Finally considering a CNN architecture and applying the graph convolution operator $\mc{G}_C$, the filtered input signal $\mb{y}^{(q)}_t$ was transformed with an activation function $\sigma(\cdot)$ to yield the output $\mb{h}^{(q)}_{t}$ of each of the $q \in \{1,\ldots,Q\}$ graph convolution layers as follows:

\beq 
\mb{h}^{(q)}_{t} = \sigma \left( \mb{y}^{(q)}_{t} \right) = \sigma \left( \sum_{k=0}^{K} \theta^{(q)}_{k}(\omega) \mb{T}^k \mb{x}^{(m)}_t \right)\label{eqn:diffusion_conv}
\eeq 

\nit Hereby $\mb{x}^{(m)}_t \in \mbb{R}^{N}$ denotes the $m$-th input feature component at time $t$, $\mb{h}^{(q)}_{t} \in \mbb{R}^{N}$ the corresponding output component of the $q$-th convolution channel, $\bs{\Theta}^{(q)}_{k} \in \mbb{R}^{N}$ parameterizes the $q$-th convolutional kernel of order $k$ and $\sigma(\cdot) $ denotes any suitable activation function. Note that for deeper convolution layers $\ell > 1, \ell = 1, \ldots, L$, the input to the convolution $\ell$ layer is given by the output component of the convolution layer $\ell-1$ instead of the input signal. In summary, these graph convolution layers can learn to represent graph structured data and could be trained with gradient descent based optimization techniques.

\paragraph{Structural connectivity:} One possibility to define the spatial layout of the brain network characterized by the weighted adjacency matrix $ \mb{A}_w $ is to directly incorporated the structural connection strength as reconstructed from DTI data. The weights $ w_{nn'} $ in our adjacency matrix would accordingly reflect the number of fibers connecting two brain regions $ n $ and $ n' $, derived from probabilistic fiber tracking \citep{Tournier2004}. This type of structural adjacency relation is denoted as $ \mb{A}_{SC}  \in \mbb{R}^{N \times N}$. The acquisition parameters of the DTI data and the structural connectome generation are outlined in detail in the \nameref{sec:dataset} section.

\paragraph{Connectome embeddings:} As an alternative to the original SC, connectome embeddings (CEs) can generate node embeddings that capture also higher order topological features of the structural layout \citep{Rosenthal2018}. The idea of such a graph embedding is to represent each node in the graph by a $ M $-dimensional feature vector. This technique is originally inspired by the word2vec algorithm introduced by \cite{Mikolov2013} who proposed a technique to learn vector-valued representations for words in a text which preserve linguistic regularities in their embedding space. Similarly the node2vec algorithm can be used to embed vertices of a graph into a subspace where similar embeddings capture the $ k $-step ($ k = 1, 2, \ldots , K $) relation between the vertices and their $k$-step neighbors \citep{Rosenthal2018, Grover2016}. We used this technique to embed each brain region $ n $ in the SC graph into a $ 64 $-dimensional vector representation. We therefore employed the \textit{gensim} python package \citep{Rehurek2010} using the skip-gram model to learn the node representations \citep{Mikolov2013}. Briefly, in this context the idea of the skip-gram model is to predict from a target node in a network its neighboring nodes, whereby a sequence of neighboring nodes is created by performing a biased random walk on the structural graph \citep{Grover2016}. To generate the node sequences in total $ 100 $ random walks were performed for each node with walk a length of $ 80 $ nodes. The return parameter of the random walk was set to $ p = 2 $ and the in-out parameter to $ q = 1 $. The similarity between the $ N $ brain regions in their embedding space was computed using the Pearson correlation coefficient, yielding a connectivity matrix denoted with $ \mb{A}_{CE} \in \mbb{R}^{N \times N} $. As illustrated in figure \ref{fig:comparison_temporal_spatial} (B), the embeddings could yield meaningful representations that revealed long-range connections between regions which were not present in the original SC \citep{Rosenthal2018}. 

\paragraph{Adaptive adjacency matrix:} So far the spatial layout of the brain graph has been represented with the help of the orthogonal eigenbasis system $\mb{U}$ of the transition operator proportional to the random walk Laplacian. This presupposed a thorough knowledge about the spatial structure of the underlying brain network that entered the related adjacency matrix. Remember that the weights of the adjacency matrix were deduced from DTI measurements based on SC or their CE similarity. However there may exist hidden activity correlations that are not represented in the original adjacency matrix used to construct the random walk Laplacian. Hence, one may wish to introduce an additional self-adaptive, normalized adjacency matrix $\mb{A}_{Adap} \in \mbb{R}^{N \times N}$ \citep{Wu2019}. The latter has been constructed as a matrix of trainable weights $\mb{V}_{Adap} \in \mbb{R}^{N \times N}$, which were at first initialized as zero and then again optimized via gradient descent \citep{Kingma2014}. Inspired by the study of \cite{Wu2019}, a normalized self-adaptive adjacency matrix was computed as:

\beq 
\mb{A}_{Adap} =  \frac{\sigma\left(\mb{V}_{Adap}\right)}{N}
\eeq
The transformation function $ \sigma(\cdot) \equiv \tanh(\cdot)$ confined the adaptive weights to the range $ [-1, 1] $, which then were normalized by the number of nodes $ N $ in the network. This self-adaptive adjacency matrix can help to uncover any hidden, still unknown dependencies between ROIs of a given graph structure. Thus it may extend any graph diffusion convolution layer to yield its output activity as:

\beq 
\mb{h}^{(q)}_{t} 
= \sigma \left( \mb{y}^{(q)}_{t} \right) 
= \sum_{k=0}^{K} \left[ \theta^{(q)}_k \mb{T}^k + \theta_k^{(q),Adap} \left( \mb{A}_{Adap} \right)^k \right] \mb{x}^{(m)}_t
\eeq 
Note that the normalized self-adaptive adjacency matrix $\mb{A}_{Adap}$ may be considered as an additional transition operator here. In an attempt to decouple the temporal processing from any underlying spatial layout of the graph connectivity, the first term within parentheses may be skipped and the self-adaptive adjacency matrix may possibly identify the underlying graph structure from the data alone. This may be applicable to situations, where no predefined graph structure is known or involved. The output of the $q$-th convolution channel can in this case be obtained with:

\beq 
\mb{h}^{(q)}_{t} 
= \sigma \left( \mb{y}^{(q)}_{t} \right) 
= \sigma \left( \sum_{k=0}^{K}  \theta_k^{(q), Adap} \left( \mb{A}_{Adap} \right)^k  \mb{x}^{(m)}_t \right)
\eeq


\subsection{Temporal dependencies.}

\begin{figure}[!htb]
\bc
\makebox[\textwidth][c]
{	
\includegraphics[width=1.0\textwidth]{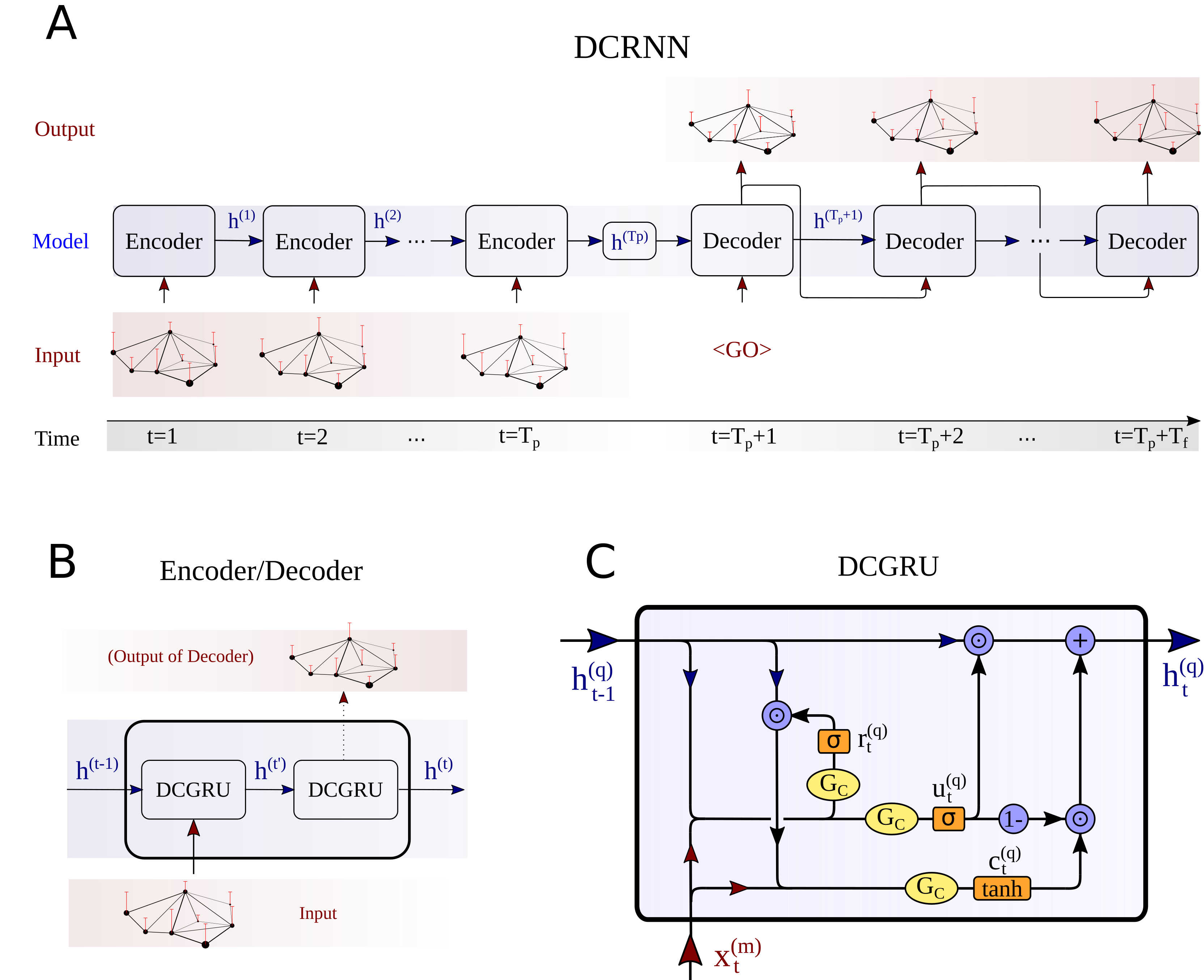}
}

\ec 
\caption{The overview of the complete DCRNN model is provided in (A). The RNN architecture consists of an encoder and decoder, which recursively process the graph structured signals. The encoder receives a sequence of inputs $[\mb{x}^{(1)},\ldots,\mb{x}^{(T_p)}]$, and iteratively updates the hidden state $\mb{h}^{(t)}$. The final state of the encoder $\mb{h}^{(T_p)}$ is passed to the decoder branch, which then recursively predicts the output sequence of future signals $[\mb{x}^{(T_p+1)}, \ldots ,\mb{x}^{(T_p + T_f)}]$. The encoder, as well as the decoder (B) consists of multiple diffusion convolution gated recurrent unit cells (DCGRU). The first DCGRU cell receives the input graph signal, and then passes its hidden state to the subsequent cell. During decoding, the final cell of the decoder then generates the predictions for the signal. For testing and validation, the decoder uses its own prediction as input for generating the subsequent prediction. The first input of the decoding branch ($<GO>$ symbol) is simply a vector of zeros.
The processing steps in an individual DCGRU cell are shown in (C). The input $\mb{x}^{(m)}_{t}$, as well as the previous hidden state $\mb{h}^{(q)}_{t-1}$ are concatenated and passed to the reset gate $\mb{r}^{(q)}_t$, as well as to the update gate $\mb{u}^{(q)}_t$. The reset gate $\mb{r}^{(q)}_t$ determines the proportion of $\mb{h}^{(q)}_{t-1}$, which enters $\mb{c}^{(q)}_t$, together with input $\mb{x}^{(m)}_{t}$. Then the hidden state $\mb{h}^{(q)}_{t-1}$ is updated by $\mb{c}^{(q)}_t$, whereby the amount of new information is controlled by $\mb{u}^{(q)}_t$.}
\label{fig:DCRNN}
\end{figure}

\paragraph{Recurrent neural networks:} In the DCRNN model, the temporal variations of the signal $\mb{x}^{(m)}_t  \in \mbb{R}^{N}$ in $N$ brain regions at $T_p$ past time points were explored with sequence-to-sequence learning in RNNs \citep{Sutskever2014}, where an encoder network compresses the information into a compact new representation. The latter is fed into a decoding network, which generates predictions of the graph signal at $T_f$ future time points representing the intended prediction horizon, as illustrated in figure \ref{fig:DCRNN} (A).

Given that the graph convolution operation to effect the spatial layout of graph structure at any time point $t$, temporal dynamics on the graph can be modeled in the DCRNN via GRUs \citep{Chung2014}. The idea is to replace convolution operations in the spatial domain by corresponding matrix multiplications in the conjugate spatial-frequency domain employing the diffusion convolution operator. This leads to the diffusion convolution gated recurrent unit (DCGRU) \citep{Li2018}:

\beqa
\mb{r}^{(q)}_t &=& \sigma \left( \mc{G}_C \left( \bs{\Theta}^{(q)}_r, \left[ \mb{x}^{(m)}_{t} \parallel \mb{h}^{(q)}_{t-1}\right] \right) + \mb{b}_r \right) \nonumber \\
\mb{u}^{(q)}_t &=& \sigma \left( \mc{G}_C \left( \bs{\Theta}^{(q)}_u, \left[ \mb{x}^{(m)}_{t} \parallel \mb{h}^{(q)}_{t-1} \right] \right) + \mb{b}_u \right) \nonumber \\
\mb{c}^{(q)}_t &=& \tanh \left( \mc{G}_C \left( \bs{\Theta}^{(q)}_c, \left[ \mb{x}^{(m)}_{t} \parallel (\mb{r}^{(q)}_t \odot \mb{h}^{(q)}_{t-1}) \right] \right) + \mb{b}_c \right) \nonumber \\
\mb{h}^{(q)}_t &=& \mb{u}^{(q)}_t \odot \mb{h}^{(q)}_{t-1} + (1 - \mb{u}^{(q)}_t) \odot \mb{c}^{(q)}_t
\eeqa 
where $\mb{x}^{(m)}_{t}, \mb{h}^{(q)}_{t}$ denote the $m$-th input and $q$-th output graph signal feature component of the GRU, respectively, at time $t$ and $[\mb{x}^{(m)}_{t} \parallel \mb{h}^{(q)}_{t-1}]$ denotes their concatenation. Also $\mb{r}^{(q)}_t, \mb{u}^{(q)}_t$ represent reset and update gates at time $t$, and $\mb{b}_r, \mb{b}_u, \mb{b}_c$, respectively, denote bias terms. Furthermore, $\bs{\Theta}^{(q)}_r, \bs{\Theta}^{(q)}_u, \bs{\Theta}^{(q)}_c$ denote the parameter sets of the corresponding filters. An illustration of the complete sequence-to-sequence architecture incorporating DCGRU cells is provided in figure \ref{fig:DCRNN}.

\clearpage

\begin{figure}[!htb]
\bc
\makebox[\textwidth][c]
{	
\includegraphics[width=1.0\textwidth]{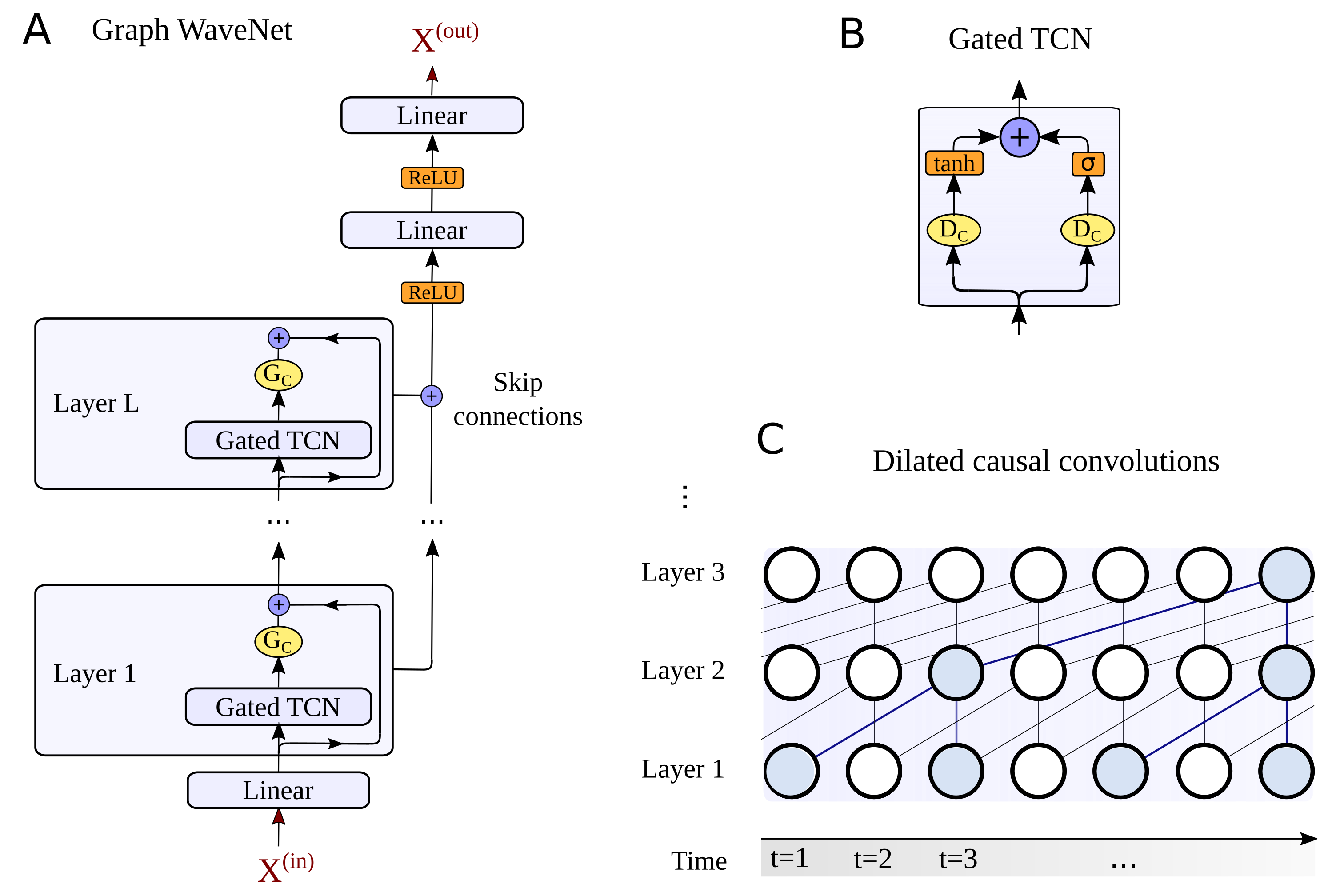}
}

\ec 
\caption{An overview of the complete GWN model is provided in (A). The GWN model consists of $ L $ layers. For the temporal modeling the GWN applies first the gated TCN mechanism and then for the spatial aspects utilizes graph convolution operations ($ \mc{G}_C $) in each layer. Each layer additionally incorporates residual connections to stabilize the gradient during learning \citep{He2016}. The information in each layer is combined by using skip connections, and the final predictions are generated by passing the output of the skip connections through two fully connected layers. The gated temporal convolution network (TCN) mechanism (B) applies a dilated causal convolution ($ \mc{D}_C $) in combination with a $ \tanh (\cdot) $ and a $\sigma(\cdot)$ activation function to control the information flow. In (C) the dilated causal convolutions are illustrated. In each layer a temporal convolution is applied whereby the dilation factor can be increased in subsequent layers. These dilations lead to exponentially growing receptive fields for neurons in higher layers. The receptive field of a neuron in layer is highlighted in blue.}
\label{fig:GWN}
\end{figure}

\paragraph{WaveNets:} Rather than incorporating diffusion convolution layers into RNNs, dilated causal convolution (DCC) layers \citep{Oord2016} have been instead employed in the GWN architecture \citep{Wu2019}. The full GWN model is illustrated in figure \ref{fig:GWN}. The DCC was defined through a dilated causal convolution operator $\mc{D}_C$:

\beq 
\mc{D}_C \left(\bs{\Theta}^{(q)}_t, \mb{x}^{(m)}_{t} \right) = \sum_{r=0}^{R-1} \bs{\Theta}^{(q)}_r \mb{x}^{(m)}_{t-d \cdot r}
\eeq 
whereby $d$ denoted the dilation factor and $\bs{\Theta}^{(q)}_t$ represented the filter kernel. DCC could be implemented by sliding over the input time series $\mb{x}^{(m)}_{t}$ while skipping input values while, from layer to layer, increasing step size $d \cdot r$. This procedure leads to an exponential growth of the receptive field with increasing layer depth as is schematically illustrated in figure \ref{fig:GWN} (C). The information flow was controlled by a gated temporal convolution network (TCN) as shown in figure \ref{fig:GWN} (B), which is obtained as:

\beq 
\mb{h}^{(q)}_t = \tanh \left( \mc{D}_C \left(\bs{\Theta}^{(q)}_1, \mb{x}^{(m)}_t + \mb{b}_1 \right) \right) \odot \sigma \left( \mc{D}_C \left( \bs{\Theta}^{(q)}_2, \mb{x}^{(m)}_t + \mb{b}_2 \right) \right)
\eeq  
Here $\tanh(\cdot)$ denotes the output activation function, and $\bs{\Theta}^{(q)}_1, \bs{\Theta}^{(q)}_1$, and $ \mb{b}_1, \mb{b}_2 $ represent the convolution filers and bias terms respectively. Further $\mc{D}_C$ represents the causal convolution operator, $\odot$ the Hadamard product and $\sigma(\cdot)$ denotes the logistic function, which controls the information passed to the next layer. To achieve large receptive fields, the layers in a WN architecture are organized in blocks, whereby in each block the dilation factor $ d $ is doubled with $ d = 1, 2, 4, \ldots $ up to a certain limit and then repeated in the same manner in the next block \citep{Oord2016}. After each such dilated convolution layer a diffusion convolution layer $ \mathcal{G}_C $ (equation \ref{eqn:diffusion_conv}) is subsequently applied to account for the spatial dependencies, as illustrated in figure \ref{fig:GWN} (A).

\clearpage

\begin{figure}[!htb]
\bc
\makebox[\textwidth][c]
{	
\includegraphics[width=0.25\textwidth]{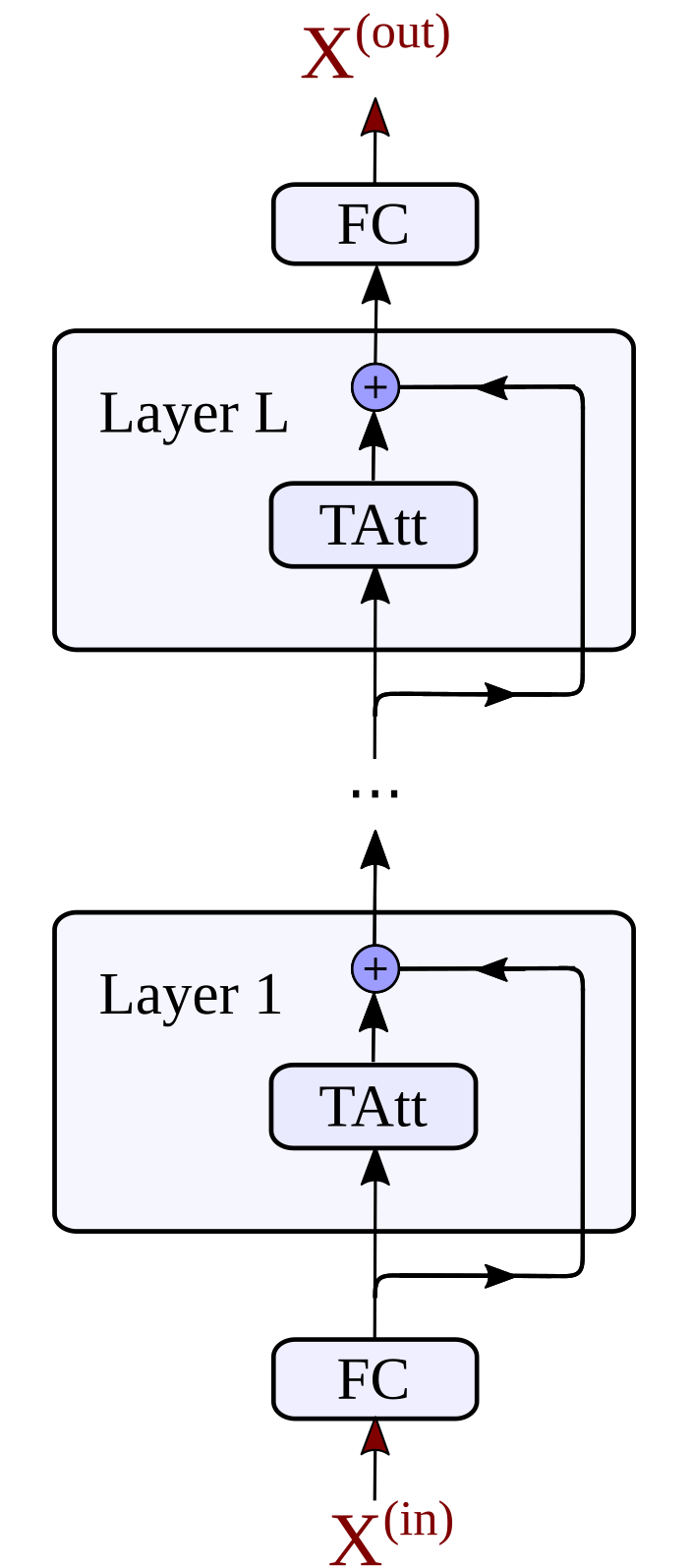}
}

\ec 
\caption{An overview over the temporal relevance or attention model. The single feature input $ X^{(in)} $, representing the BOLD signal, is first projected by a fully connected layer onto $ M $ output features. The temporal relevance scores are computed in each of the $ L $ attention layers, and to further account for vanishing gradients, additionally residual connects are incorporated \citep{He2016}. The output of the final layer $ L $ is then projected back onto a single feature, representing the predicted neural signal.}
\label{fig:tATTs}
\end{figure}

\paragraph{Temporal relevance:} Yet another approach to solve spatio-temporal time series prediction problems considers attention mechanisms in spatial and temporal domains to capture dynamic correlations \citep{Vaswani2017, Zheng2020}. In this study we therefore additionally explore non-linear temporal correlations via a temporal relevance mechanism for modeling temporal fluctuations in the BOLD signal. Let the temporal state of the brain network be represented by the multivariate signal tensor $ \mc{X} \in \mbb{R}^{N \times T \times M} $ such that the temporal states of any node $n$ be collected in the signal matrix $\mc{X}_{n::} \equiv \mb{X}^{(n)} \in \mbb{R}^{T \times M} $. The activity at any node $n$ and at any time $t$ was then represented by the tube fibers $\tilde{\mb{x}}^{(n)}_t \in \mbb{R}^M$, where $M$ denoted the number of features characterizing the node activity. Temporal correlations between different node states could be estimated by filtering the multivariate signals in a cascade of temporal relevance blocks, as illustrated in figure \ref{fig:tATTs}. The queries and keys are computed from the input in the $\ell$-th block at time point $t$ with a simple non-linear transformation $ g_r(\tilde{\mb{x}}^{(n)}_t) = \mathrm{ReLU}(\mb{W}_r \tilde{\mb{x}}^{(n)}_t + \mb{b}_r) $ with parameters $ \mb{W}_r  \in \mbb{R}^{D \times M} $ and $ \mb{b}_r \in \mbb{R}^{D} $. For any node $n$ and any time point $t_i$ the relevance of its states $\tilde{\mb{x}}^{(n)}_{t_i} $ at time points $ t_j < t_i $ with respect to the considered state $ \tilde{\mb{x}}^{(n)}_{t_j} $ could then be assessed by computing the inner product between the queries and keys:

\beq
\delta^{(n)}_{t_i,t_j} = \frac{\left( g_r(\tilde{\mb{x}}^{(n)}_{t_i}) \right) \cdot \left( g_r(\tilde{\mb{x}}^{(n)}_{t_j}) \right)^T}{\sqrt{D}}.
\eeq 
A normalized temporal relevance score $\hat{\delta}^{(n)}_{(t_i,t_j)}$ could then be computed according to:

\beqa 
\hat{\delta}_{t_i,t_j}^{(n)} = \frac{\exp\left(\delta^{(n)}_{t_i,t_j}\right)}{\sum_{t_j < t_i} \exp\left(\delta_{t_i,t_j}^{(n)}\right)}
\eeqa 
Finally, $t_j < t_i, j \in \{ 1, \dots , T_p \}$ denoted a set of time steps before time point $t_i$. After computing the temporal relevance score $\hat{\delta}_{t_i,t_j}^{(n)}$, the hidden state of node $n$ at time $t_i$ could be derived as:

\beq 
\tilde{\mb{h}}_{t_i} =  g_{r}\left( \sum_{t_j < t_i}  \hat{\delta}_{t_i,t_j}^{(n)} \cdot g_{r} ( \tilde{\mb{x}}^{(n)}_{t_j} )  \right)
\eeq 

\nit Whereby $g_{r}(\cdot)$ denotes a non-linear projection again. Note that all parameters $\mb{W}_r$ and $\mb{b}_r$ to be learned were shared across all nodes and time steps. In total $ L $ layers of temporal attention mechanisms were stacked to generate a final prediction for the BOLD signal. After each layer batch normalization was applied and additionally residual connections were incorporated to stabilize the gradient \citep{He2016}.

\clearpage


\subsection{Model training}\label{sec:model_training}

In this section we outline the training procedures which were used for the different neural network models to learn the temporal and spatial dynamics in the BOLD signal. Before training, the fMRI data of each session was linearly scaled between $ 0 $ and $ 1 $, to get gradients of a small magnitude during the backpropagation learning, what facilitates the fine-tuning of the learning rate. For all models, the mean absolute error (MAE) was used as an objective function to quantify the overall difference between the true BOLD signal $\mb{x}^{(t)}$ and predicted signal $\hat{\mb{x}}^{(t)}$ in all $ N $ brain regions:

\beq
	\mb{MAE}(\mb{x} , \hat{\mb{x}}) = \frac{1}{N} \sum_{n=1}^{N} \frac{1}{T_f} \sum_{t=1}^{T_f} |x_n^{(t)} - \hat{x}_n^{(t)}|\label{eqn:mae}
\eeq
\paragraph{DCRNN} The DCRNN model, based on a RNN architecture, was trained with backpropagation through time \citep{Werbos1990}, with the objective to maximize the likelihood of generating the target timeseries. To additionally account for a mismatch between training and testing distributions of stimuli, a scheduled sampling strategy was used \citep{Bengio2015}. The probability of using a true label as a decoder input decayed according to:

\beq
	\epsilon(i) = \frac{\tau}{\tau +  \textrm{exp}(i/\tau)} \in (0,1)
\eeq  

\nit with $\tau > 0$ the decay parameter and $i \in \mbb{N}$ counting the iterations. During supervised learning, instances to be predicted were, of course, known.
For this optimization problem, the Adam algorithm \citep{Kingma2014} was employed, and the model was trained for $70$ epochs on mini-batches of $ 16 $ training samples. To further improve convergence, an annealing learning rate was used, initialized as $\eta = 0.1$, and decreased by a factor of $0.1$ at epochs $20$, $40$ and $60$, or if the validation error did not improve for more than $10$ epochs. Before lowering the learning rate, the weights with lowest validation error were restored, in order to avoid getting stuck in local optima. For the training data set including only $ 10 $ subjects used in the \nameref{sec:network_scaling} section, the number of training epochs was increased to $ 140 $ and the learning rate decay applied at epochs $ 40 $, $ 80 $ and $ 120 $. The influence of the DCRNN model hyperparameters are discussed in \nameref{sec:supp_hyperparameters} (figure \ref{fig:hyperparameters_DCRNN}) and were chosen to yield a reasonable trade-off between accuracy and computational requirements. The encoder and decoder of the sequence-to-sequence architecture consist to two diffusion convolution GRU layers each, and the hidden state size was set to $ 64 $. The computations were performed on a \textit{Nvidia RTX 2080 Ti} GPU, running on a desktop PC with an \textit{Intel(R) Core(TM) i7-9800X} CPU under \textit{Linux Ubuntu 20.04}. With this setup one epoch on the dataset including $ 25 $ subjects and predicting the activity within one hemisphere including $ 180 $ ROIs took approximately $ 3.4 $ minutes. To have error values of a magnitudes that are easier to interpret, for the evaluations in the \nameref{sec:results} section the whole dataset was rescaled to zero mean and unit variance after the training of all STGNN models was finished. 

\paragraph{GWN} The GWN model was also trained incorporating the Adam optimizer \citep{Kingma2014} to minimize the MAE defined in equation \ref{eqn:mae}. For GWN model it was sufficient to train it $ 30 $ epochs with a batch size of $ 8 $, thereby initializing the learning rate with $ \eta = 0.0001 $ and decreasing it by a factor of $0.1$ at epochs $10$ and $20$. For the $ 10 $ subject dataset the number of epochs was also increased to $ 60 $ and the learning rate decay adapted to epochs $ 20 $ and $ 40 $ correspondingly. The influence of the hyperparameters of the GWN is evaluated in \nameref{sec:supp_hyperparameters} (figure \ref{fig:hyperparameters_GWN}). A good trade-off between model accuracy and complexity could be found using $ 32 $ neurons. The number of layers per block were defined as $ 2 $ with a total number of $ 12 $ blocks. With this setup one epoch on the $ 25 $ subjects' dataset including $ 180 $ ROIs took around $ 12.2 $ minutes.

\paragraph{TAtt} The TAtt model was trained using the Adam optimizer \citep{Kingma2014} for in total $ 40 $ epochs, minimizing the MAE defined in equation \ref{eqn:mae}, using a batch size of $ 16 $. The learning rate was initialized with $ \eta = 0.1 $ and decreased by a factor of $ 0.1 $ at epochs $ 10 $, $ 20 $ and $ 30 $. The influence of the hyperparmeters is evaluated in \nameref{sec:supp_hyperparameters} (figure \ref{fig:hyperparameters_TAtt}). The number of neurons in the temporal attentions were set to $ 32 $ thereby using $ 4 $ attention heads in the $ 4 $ TAtt layers. With this setup of hyperparameters one epoch of the TAtt model took around $ 7.7 $ minutes.



\subsection{Vector auto regressive model}\label{sec:VAR}

Granger causality \citep{Granger1969} is currently most often based on linear vector autoregressive (VAR) models for stochastic time series data. Therefore we compare our GNN based approach with a VAR model, as implemented in the multivariate Granger causality (MVGC) toolbox \citep{Barnett2013}. An autoregressive process (AR) is based on the idea that a signal $x^{(t)}$ can be described as a linear superposition of the first $T_p$ of its lagged values \citep{Luetkepohl2005}:

\beq
	x^{(t)} = \beta + \alpha_1 x^{(t-1)} + \alpha_2 x^{(t-2)} + \dots + \alpha_p x^{(t-T_p)} + u^{(t)} 
\eeq
with coefficients or weights $ \alpha_1, \dots, \alpha_p $, an intercept $\beta$ and the error term $u^{(t)}$. This univariate formulation can be then extended to a multivariate VAR model including $N$ time series $ \mb{x}^{(t)} = [x_1^{(t)}, \dots , x_N^{(t)}] $ \citep{Luetkepohl2005}: 

\beq
	\mb{x}^{(t)} = \mb{b} + \mb{A}_1 \mb{x}^{(t-1)} + \mb{A}_2 \mb{x}^{(t-2)} + \dots + \mb{A}_p \mb{x}^{(t-T_p)} + \mb{u}^{(t)} 
\eeq
whereby the coefficients are now collected in matrices $ \mb{A} \in \mbb{R}^{N \times N} $, and intercepts and errors are characterized by vectors $\mb{b} \in \mbb{R}^{N} $ and $\mb{u}^{(t)}  \in \mbb{R}^{N}$ respectively. In our study the multivariate time series $\mb{x}^{(t)}$ reflect the BOLD signal strength in the $N$ brain regions, sampled at different timesteps $t$.  

To estimate parameters of the VAR model we used the ordinary least squares (OLS) fit provided in the MVGC toolbox \citep{Barnett2013}. As outlined in the \nameref{sec:data_preparation} section we used the first $80 \%$ of the data from each fMRI session to fit the model. Then for the comparison to the GNN approaches in the \nameref{sec:network_scaling} section we tested the model order $ p $ in steps of $ 5 $ with $ p = 5, 10, \ldots, T_p $ and chose the model with highest accuracy individually for each dataset. To check for stationarity of the signals, an augmented Dickey-Fuller test for unit roots was applied to the BOLD timecourses \citep{Hamilton1994, Mackinnon1994}, using a p-value of $p<0.01$. For the $ 25 $ subject dataset, around $10.0\%$ of the BOLD time courses do not fulfill the stationarity criteria of the augmented Dickey-Fuller test ($p>0.01$) when using such a high lag order of $ T_p=60 $. But as the objective criterion of the evaluation in the \nameref{sec:network_scaling} section was to assess the capabilities of the models to predict empirically observed neural activity patterns, we chose the VAR model with best prediction accuracy for comparisons with the GNNs.


\clearpage

\section*{Data availability}
 
A demo version for MRI data preparation and training the DCRNN model is provided under: 

\url{https://github.com/simonvino/DCRNN_brain_connectivity}. 

\nit In addition, a demo version for the GWN model is provided under: 

\url{https://github.com/simonvino/GraphWaveNet_brain_connectivity}.

\nit Preprocessed HCP data is publicly available under: 

\url{https://db.humanconnectome.org}.


\section*{Acknowledgements}

The work was supported by the DFG projects GR988/25-1 and ISNT89/393-1 granted to M.W.G.. Data were provided by the Human Connectome Project, WU-Minn Consortium (Principal Investigators: David Van Essen and Kamil Ugurbil; 1U54MH091657) funded by the 16 NIH Institutes and Centers that support the NIH Blueprint for Neuroscience Research; and by the McDonnell Center for Systems Neuroscience at Washington University.



\bibliography{GNN_comparison}

\begin{thebibliography}{129}
\providecommand{\natexlab}[1]{#1}
\providecommand{\url}[1]{\texttt{#1}}
\expandafter\ifx\csname urlstyle\endcsname\relax
  \providecommand{\doi}[1]{doi: #1}\else
  \providecommand{\doi}{doi: \begingroup \urlstyle{rm}\Url}\fi

\bibitem[Abdelnour et~al.(2018)Abdelnour, Dayan, Devinsky, Thesen, and
  Raj]{Abdelnour2018}
F.~Abdelnour, M.~Dayan, O.~Devinsky, T.~Thesen, and A.~Raj.
\newblock Functional brain connectivity is predictable from anatomic network's
  laplacian eigen-structure.
\newblock \emph{NeuroImage}, 172:\penalty0 728--739, 2018.

\bibitem[Abreu et~al.(2018)Abreu, Leal, and Figueiredo]{Abreu2018}
R.~Abreu, A.~Leal, and P.~Figueiredo.
\newblock Eeg-informed fmri: A review of data analysis methods.
\newblock \emph{Frontiers in Human Neuroscience}, 12:\penalty0 29, 2018.

\bibitem[Achard et~al.(2006)Achard, Salvador, Whitcher, Suckling, and
  Bullmore]{Achard2006}
S.~Achard, R.~Salvador, B.~Whitcher, J.~Suckling, and E.~Bullmore.
\newblock A resilient, low-frequency, small-world human brain functional
  network with highly connected association cortical hubs.
\newblock \emph{J Neurosci}, 3:\penalty0 e17, 2006.

\bibitem[Alstott et~al.(2009)Alstott, Breakspear, Hagmann, Cammoun, and
  Sporns]{Alstott2009}
J.~Alstott, M.~Breakspear, P.~Hagmann, L.~Cammoun, and O.~Sporns.
\newblock Modeling the impact of lesions in the human brain.
\newblock \emph{PLoS computational biology}, 5:\penalty0 e1000408, 2009.

\bibitem[Amico and Go{\~n}i(2018)]{Amico2018}
E.~Amico and J.~Go{\~n}i.
\newblock Mapping hybrid functional-structural connectivity traits in the human
  connectome.
\newblock \emph{Network Neuroscience}, 2:\penalty0 306--322, 2018.

\bibitem[Andersson and Sotiropoulos(2015{\natexlab{a}})]{Andersson2015}
J.~Andersson and S.~Sotiropoulos.
\newblock Non-parametric representation and prediction of single- and
  multi-shell diffusion-weighted {MRI} data using gaussian processes.
\newblock \emph{NeuroImage}, 122:\penalty0 166--76, 2015{\natexlab{a}}.
\newblock \doi{10.1016/j.neuroimage.2015.07.067}.

\bibitem[Andersson and Sotiropoulos(2015{\natexlab{b}})]{Andersson2015b}
J.~Andersson and S.~Sotiropoulos.
\newblock An integrated approach to correction for off-resonance effects and
  subject movement in diffusion {MR} imaging.
\newblock \emph{NeuroImage}, 125:\penalty0 1063--1078, 2015{\natexlab{b}}.
\newblock \doi{10.1016/j.neuroimage.2015.10.019}.

\bibitem[Andersson et~al.(2003)Andersson, Skare, and Ashburner]{Andersson2003}
J.~Andersson, S.~Skare, and J.~Ashburner.
\newblock How to correct susceptibility distortions in spin-echo echo-planar
  images: Application to diffusion tensor imaging.
\newblock \emph{NeuroImage}, 20:\penalty0 870--88, 2003.
\newblock \doi{10.1016/S1053-8119(03)00336-7}.

\bibitem[Arslan et~al.(2018)Arslan, Ktena, Glocker, and Rueckert]{Arslan2018}
S.~Arslan, S.~I. Ktena, B.~Glocker, and D.~Rueckert.
\newblock Graph saliency maps through spectral convolutional networks:
  Application to sex classification with brain connectivity.
\newblock In \emph{GRAIL/Beyond-MIC@MICCAI}, 2018.

\bibitem[Barnett and Seth(2013)]{Barnett2013}
L.~Barnett and A.~Seth.
\newblock The {MVGC} multivariate granger causality toolbox: A new approach to
  granger-causal inference.
\newblock \emph{Journal of neuroscience methods}, 223:\penalty0 50--68, 2013.
\newblock \doi{10.1016/j.jneumeth.2013.10.018}.

\bibitem[Becker et~al.(2018)Becker, Pequito, Pappas, Miller, T.~Grafton,
  S.~Bassett, and Preciado]{Becker2018}
C.~Becker, S.~Pequito, G.~Pappas, M.~Miller, S.~T.~Grafton, D.~S.~Bassett, and
  V.~Preciado.
\newblock Spectral mapping of brain functional connectivity from diffusion
  imaging.
\newblock \emph{Scientific Reports}, 8, 12 2018.
\newblock \doi{10.1038/s41598-017-18769-x}.

\bibitem[Bengio et~al.(2015)Bengio, Vinyals, Jaitly, and Shazeer]{Bengio2015}
S.~Bengio, O.~Vinyals, N.~Jaitly, and N.~Shazeer.
\newblock Scheduled sampling for sequence prediction with recurrent neural
  networks.
\newblock In \emph{NIPS}, pages 1171--1179, 2015.

\bibitem[Bettinardi et~al.(2018)Bettinardi, Deco, Karlaftis, Hartevelt,
  Fernandes, Kourtzi, Kringelbach, and Zamora-L\'opez]{Bettinardi2018}
R.~G. Bettinardi, G.~Deco, V.~M. Karlaftis, T.~J.~V. Hartevelt, H.~M.
  Fernandes, Z.~Kourtzi, M.~L. Kringelbach, and G.~Zamora-L\'opez.
\newblock How structure sculpts function: unveiling the contribution of
  anatomical connectivity to the brain’s spontaneous correlation structure.
\newblock \emph{Chaos: An Interdisciplinary Journal of Nonlinear Science}, 27,
  2018.

\bibitem[Bielczyk et~al.(2019)Bielczyk, Uithol, van Mourik, Anderson, Glennon,
  and Buitelaar]{Bielczyk2019}
N.~Bielczyk, S.~Uithol, T.~van Mourik, P.~Anderson, J.~Glennon, and
  J.~Buitelaar.
\newblock Disentangling causal webs in the brain using functional magnetic
  resonance imaging: A review of current approaches.
\newblock \emph{Network Neuroscience}, 3, 02 2019.

\bibitem[Biswal et~al.(1996)Biswal, Deyoe, and Hyde]{Biswal1996}
B.~Biswal, E.~A. Deyoe, and J.~S. Hyde.
\newblock Reduction of physiological fluctuations in fmri using digital
  filters.
\newblock \emph{Magnetic Resonance in Medicine}, 35\penalty0 (1):\penalty0
  107--113, 1996.
\newblock \doi{https://doi.org/10.1002/mrm.1910350114}.

\bibitem[Biswal et~al.(1995)Biswal, Yetkin, Haughton, and Hyde]{Biswal1995}
B.~B. Biswal, F.~Z. Yetkin, V.~Haughton, and J.~S. Hyde.
\newblock Functional connectivity in the motor cortex of resting human brain
  using echo-planar {MRI}.
\newblock \emph{Magnetic resonance in medicine}, 34 4:\penalty0 537--41, 1995.

\bibitem[Bronstein et~al.(2017)Bronstein, Bruna, LeCun, Szlam, and
  Vandergheynst]{Bronstein2017}
M.~M. Bronstein, J.~Bruna, Y.~LeCun, A.~D. Szlam, and P.~Vandergheynst.
\newblock Geometric deep learning: Going beyond euclidean data.
\newblock \emph{IEEE Signal Processing Magazine}, 34:\penalty0 18--42, 2017.

\bibitem[Br\"{u}el~Gabrielsson(2020)]{Bruel2020}
R.~Br\"{u}el~Gabrielsson.
\newblock Universal function approximation on graphs.
\newblock In H.~Larochelle, M.~Ranzato, R.~Hadsell, M.~Balcan, and H.~Lin,
  editors, \emph{Advances in Neural Information Processing Systems}, volume~33,
  pages 19762--19772. Curran Associates, Inc., 2020.
\newblock URL
  \url{https://proceedings.neurips.cc/paper/2020/file/e4acb4c86de9d2d9a41364f93951028d-Paper.pdf}.

\bibitem[Buckner et~al.(2009)Buckner, Sepulcre, Talukdar, Krienen, Liu, Hedden,
  Andrews-Hanna, Sperling, and Johnson]{Bruckner2009}
R.~Buckner, J.~Sepulcre, T.~Talukdar, F.~Krienen, H.~Liu, T.~Hedden,
  J.~Andrews-Hanna, R.~Sperling, and K.~Johnson.
\newblock Cortical hubs revealed by intrinsic functional connectivity: Mapping,
  assessment of stability, and relation to alzheimer's disease.
\newblock \emph{The Journal of neuroscience : the official journal of the
  Society for Neuroscience}, 29:\penalty0 1860--73, 2009.

\bibitem[Bullmore and Bassett(2011)]{Bullmore2011}
E.~T. Bullmore and D.~S. Bassett.
\newblock Brain graphs: graphical models of the human brain connectome.
\newblock \emph{Annu. Rev. Clin. Psychol.}, 7:\penalty0 113--140, 2011.

\bibitem[Burgess et~al.(2016)Burgess, Kandala, Nolan, Laumann, Power, Adeyemo,
  Harms, Petersen, and Barch]{Burgess2016}
G.~Burgess, S.~Kandala, D.~Nolan, T.~Laumann, J.~Power, B.~Adeyemo, M.~Harms,
  S.~Petersen, and D.~Barch.
\newblock Evaluation of denoising strategies to address motion-correlated
  artifact in resting state {fMRI} data from the human connectome project.
\newblock \emph{Brain Connectivity}, 6, 2016.

\bibitem[Bush et~al.(2015)Bush, Cisler, Bian, Hazaroglu, Hazaroglu, and
  Kilts]{Bush2015}
K.~Bush, J.~Cisler, J.~Bian, G.~Hazaroglu, O.~Hazaroglu, and C.~Kilts.
\newblock Improving the precision of fmri bold signal deconvolution with
  implications for connectivity analysis.
\newblock \emph{Magnetic resonance imaging}, 33, 07 2015.
\newblock \doi{10.1016/j.mri.2015.07.007}.

\bibitem[Chen and Wang(2018)]{ChenX2018}
X.~Chen and Y.~Wang.
\newblock Predicting {R}esting-state {F}unctional {C}onnectivity with
  {E}fficient {S}tructural {C}onnectivity.
\newblock \emph{EEE/CAA Journal of Automatica Sinica}, 5\penalty0 (6):\penalty0
  1079--1088, 2018.

\bibitem[Chung et~al.(2014)Chung, Gulcehre, Cho, and Bengio]{Chung2014}
J.~Chung, C.~Gulcehre, K.~Cho, and Y.~Bengio.
\newblock Empirical evaluation of gated recurrent neural networks on sequence
  modeling, 2014.

\bibitem[Ciric et~al.(2017)Ciric, Wolf, Power, Roalf, Baum, Ruparel, Shinohara,
  Elliott, Eickhoff, Davatzikos, Gur, Gur, Bassett, and
  Satterthwaite]{Ciric2017}
R.~Ciric, D.~H. Wolf, J.~D. Power, D.~R. Roalf, G.~L. Baum, K.~Ruparel, R.~T.
  Shinohara, M.~A. Elliott, S.~B. Eickhoff, C.~Davatzikos, R.~C. Gur, R.~E.
  Gur, D.~S. Bassett, and T.~D. Satterthwaite.
\newblock Benchmarking of participant-level confound regression strategies for
  the control of motion artifact in studies of functional connectivity.
\newblock \emph{NeuroImage}, 154:\penalty0 174--187, 2017.
\newblock \doi{https://doi.org/10.1016/j.neuroimage.2017.03.020}.

\bibitem[Cole et~al.(2016)Cole, Ito, Bassett, and Schultz]{Cole2016}
M.~Cole, T.~Ito, D.~Bassett, and D.~Schultz.
\newblock Activity flow over resting-state networks shapes cognitive task
  activations.
\newblock \emph{Nature neuroscience}, 19, 10 2016.
\newblock \doi{10.1038/nn.4406}.

\bibitem[Dauphin et~al.(2014)Dauphin, Pascanu, Gulcehre, Cho, Ganguli, and
  Bengio]{Dauphin2014}
Y.~Dauphin, R.~Pascanu, C.~Gulcehre, K.~Cho, S.~Ganguli, and Y.~Bengio.
\newblock Identifying and attacking the saddle point problem in
  high-dimensional non-convex optimization.
\newblock \emph{NIPS}, 27, 06 2014.

\bibitem[de~Haan et~al.(2020)de~Haan, Cohen, and Welling]{deHaan2020}
P.~de~Haan, T.~S. Cohen, and M.~Welling.
\newblock Natural graph networks.
\newblock In H.~Larochelle, M.~Ranzato, R.~Hadsell, M.~F. Balcan, and H.~Lin,
  editors, \emph{Advances in Neural Information Processing Systems}, volume~33,
  pages 3636--3646. Curran Associates, Inc., 2020.
\newblock URL
  \url{https://proceedings.neurips.cc/paper/2020/file/2517756c5a9be6ac007fe9bb7fb92611-Paper.pdf}.

\bibitem[Deco et~al.(2012)Deco, Senden, and Jirsa]{Deco2012}
G.~Deco, M.~Senden, and V.~Jirsa.
\newblock How anatomy shapes dynamics: a semi-analytical study of the brain at
  rest by a simple spin model.
\newblock \emph{Frontiers in computational neuroscience}, 6:\penalty0 68, 2012.

\bibitem[Deco et~al.(2013)Deco, Ponce-Alvarez, Mantini, Romani, Hagmann, and
  Corbetta]{Deco2013}
G.~Deco, A.~Ponce-Alvarez, D.~Mantini, G.-L. Romani, P.~Hagmann, and
  M.~Corbetta.
\newblock Resting-state functional connectivity emerges from structurally and
  dynamically shaped slow linear fluctuations.
\newblock \emph{The Journal of neuroscience : the official journal of the
  Society for Neuroscience}, 33:\penalty0 11239--11252, 07 2013.
\newblock \doi{10.1523/JNEUROSCI.1091-13.2013}.

\bibitem[Deco et~al.(2017)Deco, Kringelbach, Jirsa, and Ritter]{Deco2017}
G.~Deco, M.~L. Kringelbach, V.~K. Jirsa, and P.~Ritter.
\newblock The dynamics of resting fluctuations in the brain: metastability and
  its dynamical cortical core.
\newblock \emph{Scientific Reports}, 7, 2017.

\bibitem[Defferrard et~al.(2016)Defferrard, Bresson, and
  Vandergheynst]{Defferrard2016}
M.~Defferrard, X.~Bresson, and P.~Vandergheynst.
\newblock Convolutional neural networks on graphs with fast localized spectral
  filtering.
\newblock In \emph{NIPS}, pages 3837--3845, 2016.

\bibitem[Deligianni et~al.(2016)Deligianni, Carmichael, H~Zhang, Clark, and
  Clayden]{Deligianni2016}
F.~Deligianni, D.~Carmichael, G.~H~Zhang, C.~Clark, and J.~Clayden.
\newblock Noddi and tensor-based microstructural indices as predictors of
  functional connectivity.
\newblock \emph{PloS one}, 11:\penalty0 e0153404, 04 2016.

\bibitem[Demirtaş et~al.(2019)Demirtaş, Burt, Helmer, Ji, Adkinson, Glasser,
  Van~Essen, Sotiropoulos, Anticevic, and Murray]{Demirtas2019}
M.~Demirtaş, J.~Burt, M.~Helmer, J.~L. Ji, B.~Adkinson, M.~Glasser,
  D.~Van~Essen, S.~Sotiropoulos, A.~Anticevic, and J.~Murray.
\newblock Hierarchical heterogeneity across human cortex shapes large-scale
  neural dynamics.
\newblock \emph{Neuron}, 101, 02 2019.
\newblock \doi{10.1016/j.neuron.2019.01.017}.

\bibitem[Duggento et~al.(2018)Duggento, Passamonti, Valenza, Barbieri,
  Guerrisi, and Toschi]{Duggento2018}
A.~Duggento, L.~Passamonti, G.~Valenza, R.~Barbieri, M.~Guerrisi, and
  N.~Toschi.
\newblock Multivariate granger causality unveils directed parietal to
  prefrontal cortex connectivity during task-free mri.
\newblock \emph{Scientific Reports}, 8, 04 2018.

\bibitem[Feinberg et~al.(2010)Feinberg, Moeller, M~Smith, Auerbach, Ramanna,
  Günther, F~Glasser, Miller, Ugurbil, and Yacoub]{Feinberg2010}
D.~Feinberg, S.~Moeller, S.~M~Smith, E.~Auerbach, S.~Ramanna, M.~Günther,
  M.~F~Glasser, K.~Miller, K.~Ugurbil, and E.~Yacoub.
\newblock Multiplexed echo planar imaging for sub-second whole brain {FMRI} and
  fast diffusion imaging.
\newblock \emph{PloS one}, 5:\penalty0 e15710, 2010.

\bibitem[Fischl(2012)]{Fischl2012}
B.~Fischl.
\newblock Freesurfer.
\newblock \emph{NeuroImage}, 62\penalty0 (2):\penalty0 774 -- 781, 2012.
\newblock \doi{10.1016/j.neuroimage.2012.01.021}.

\bibitem[Friston et~al.(2013)Friston, Moran, and Seth]{Friston2013}
K.~Friston, R.~Moran, and A.~K. Seth.
\newblock Analysing connectivity with granger causality and dynamic causal
  modelling.
\newblock \emph{Curr Opin Neurobiol}, 23:\penalty0 172--178, 2013.

\bibitem[Frässle et~al.(2018)Frässle, Lomakina-Rumyantseva, Kasper, Manjaly,
  Leff, Pruessmann, Buhmann, and Stephan]{Frassle2018}
S.~Frässle, E.~Lomakina-Rumyantseva, L.~Kasper, Z.~Manjaly, A.~Leff,
  K.~Pruessmann, J.~Buhmann, and K.~Stephan.
\newblock A generative model of whole-brain effective connectivity.
\newblock \emph{NeuroImage}, 179, 05 2018.
\newblock \doi{10.1016/j.neuroimage.2018.05.058}.

\bibitem[Glasser et~al.(2013)Glasser, Sotiropoulos, Wilson, Coalson, Fischl,
  Andersson, Xu, Jbabdi, Webster, Polimeni, DC, and Jenkinson]{Glasser2013}
M.~Glasser, S.~Sotiropoulos, J.~Wilson, T.~Coalson, B.~Fischl, J.~Andersson,
  J.~Xu, S.~Jbabdi, M.~Webster, J.~Polimeni, V.~DC, and M.~Jenkinson.
\newblock The minimal preprocessing pipelines for the human connectome project.
\newblock \emph{NeuroImage}, 80, 2013.

\bibitem[Glasser et~al.(2016)Glasser, Coalson, Robinson, Hacker, Harwell,
  Yacoub, Ugurbil, Andersson, Beckmann, Jenkinson, Smith, and
  Van~Essen]{Glasser2016}
M.~Glasser, T.~Coalson, E.~Robinson, C.~Hacker, J.~Harwell, E.~Yacoub,
  K.~Ugurbil, J.~Andersson, C.~Beckmann, M.~Jenkinson, S.~Smith, and
  D.~Van~Essen.
\newblock A multi-modal parcellation of human cerebral cortex.
\newblock \emph{Nature}, 536, 2016.

\bibitem[Glerean et~al.(2012)Glerean, Salmi, Lahnakoski, Jääskeläinen, and
  Sams]{Glerean2012}
E.~Glerean, J.~Salmi, J.~Lahnakoski, I.~Jääskeläinen, and M.~Sams.
\newblock Functional magnetic resonance imaging phase synchronization as a
  measure of dynamic functional connectivity.
\newblock \emph{Brain connectivity}, 2:\penalty0 91--101, 2012.

\bibitem[Granger(1969)]{Granger1969}
C.~W.~J. Granger.
\newblock Investigating causal relations by econometric models and
  cross-spectral methods.
\newblock \emph{Econometrica}, 37:\penalty0 424--438, 1969.

\bibitem[Griffanti et~al.(2014)Griffanti, Salimi-Khorshidi, Beckmann, Auerbach,
  Douaud, Sexton, Zsoldos, Ebmeier, Filippini, Mackay, Moeller, Xu, Yacoub,
  Baselli, Ugurbil, Miller, and Smith]{Griffanti2014}
L.~Griffanti, G.~Salimi-Khorshidi, C.~F. Beckmann, E.~J. Auerbach, G.~Douaud,
  C.~E. Sexton, E.~Zsoldos, K.~P. Ebmeier, N.~Filippini, C.~E. Mackay,
  S.~Moeller, J.~Xu, E.~Yacoub, G.~Baselli, K.~Ugurbil, K.~L. Miller, and S.~M.
  Smith.
\newblock {ICA-based artefact removal and accelerated fMRI acquisition for
  improved resting state network imaging}.
\newblock \emph{NeuroImage}, 95:\penalty0 232 -- 247, 2014.
\newblock \doi{10.1016/j.neuroimage.2014.03.034}.

\bibitem[Grover and Leskovec(2016)]{Grover2016}
A.~Grover and J.~Leskovec.
\newblock node2vec: Scalable feature learning for networks.
\newblock volume 2016, pages 855--864, 07 2016.
\newblock \doi{10.1145/2939672.2939754}.

\bibitem[Hamilton(1994)]{Hamilton1994}
J.~Hamilton.
\newblock \emph{Time Series Analysis}.
\newblock Princeton University Press, Princeton, NJ., 1994.

\bibitem[He et~al.(2016)He, Zhang, Ren, and Sun]{He2016}
K.~He, X.~Zhang, S.~Ren, and J.~Sun.
\newblock Deep residual learning for image recognition.
\newblock \emph{2016 IEEE Conference on Computer Vision and Pattern Recognition
  (CVPR)}, pages 770--778, 2016.

\bibitem[Hodge et~al.(2015)Hodge, Horton, Brown, Herrick, Olsen, Hileman,
  McKay, Archie, Cler, Harms, Burgess, Glasser, Elam, Curtiss, Barch,
  Oostenveld, Larson-Prior, Ugurbil, Van~Essen, and Marcus]{Hodge2015}
M.~Hodge, W.~Horton, T.~Brown, R.~Herrick, T.~Olsen, M.~Hileman, M.~McKay,
  K.~Archie, E.~Cler, M.~Harms, G.~Burgess, M.~Glasser, J.~Elam, S.~Curtiss,
  D.~Barch, R.~Oostenveld, L.~Larson-Prior, K.~Ugurbil, D.~Van~Essen, and
  D.~Marcus.
\newblock {ConnectomeDB} – sharing human brain connectivity data.
\newblock \emph{NeuroImage}, 124, 2015.
\newblock \doi{10.1016/j.neuroimage.2015.04.046}.

\bibitem[Honey et~al.(2009)Honey, Sporns, Cammoun, Gigandet, Thiran, Meuli, and
  Hagmann]{Honey2009}
C.~J. Honey, O.~Sporns, L.~Cammoun, X.~Gigandet, J.~P. Thiran, R.~Meuli, and
  P.~Hagmann.
\newblock Predicting human resting-state functional connectivity from
  structural connectivity.
\newblock \emph{Proceedings of the National Academy of Sciences of the United
  States of America}, 106 6:\penalty0 2035--40, 2009.

\bibitem[Hornik et~al.(1989)Hornik, Stinchcombe, and White]{Hornik1989}
K.~Hornik, M.~Stinchcombe, and H.~White.
\newblock Multilayer feedforward networks are universal approximators.
\newblock \emph{Neural Networks}, 2\penalty0 (5):\penalty0 359--366, 1989.
\newblock ISSN 0893-6080.
\newblock \doi{https://doi.org/10.1016/0893-6080(89)90020-8}.

\bibitem[Ioffe and Szegedy(2015)]{Ioffe2015}
S.~Ioffe and C.~Szegedy.
\newblock Batch normalization: Accelerating deep network training by reducing
  internal covariate shift.
\newblock In F.~Bach and D.~Blei, editors, \emph{Proceedings of the 32nd
  International Conference on Machine Learning}, volume~37 of \emph{Proceedings
  of Machine Learning Research}, pages 448--456, Lille, France, 07--09 Jul
  2015. PMLR.

\bibitem[Ito et~al.(2020)Ito, Hearne, Mill, Cocuzza, and Cole]{Ito2020}
T.~Ito, L.~Hearne, R.~Mill, C.~Cocuzza, and M.~W. Cole.
\newblock Discovering the computational relevance of brain network
  organization.
\newblock \emph{Trends in Cognitive Sciences}, 24\penalty0 (1):\penalty0
  25--38, 2020.
\newblock ISSN 1364-6613.
\newblock \doi{https://doi.org/10.1016/j.tics.2019.10.005}.

\bibitem[Jenkinson et~al.(2002)Jenkinson, Bannister, Brady, and
  Smith]{Jenkinson2002}
M.~Jenkinson, P.~Bannister, M.~Brady, and S.~Smith.
\newblock Improved optimization for the robust and accurate linear registration
  and motion correction of brain images.
\newblock \emph{NeuroImage}, 17:\penalty0 825 -- 841, 2002.
\newblock \doi{10.1006/nimg.2002.1132}.

\bibitem[Jenkinson et~al.(2012)Jenkinson, Beckmann, Behrens, Woolrich, and
  Smith]{Jenkinson2012}
M.~Jenkinson, C.~F. Beckmann, T.~E. Behrens, M.~W. Woolrich, and S.~M. Smith.
\newblock {FSL}.
\newblock \emph{NeuroImage}, 62\penalty0 (2):\penalty0 782 -- 790, 2012.
\newblock \doi{10.1016/j.neuroimage.2011.09.015}.

\bibitem[Jeurissen et~al.(2014)Jeurissen, Tournier, Dhollander, Connelly, and
  Sijbers]{Jeurissen2014}
B.~Jeurissen, J.-D. Tournier, T.~Dhollander, A.~Connelly, and J.~Sijbers.
\newblock Multi-tissue constrained spherical deconvolution for improved
  analysis of multi-shell diffusion {MRI} data.
\newblock \emph{NeuroImage}, 103:\penalty0 411--426, 2014.

\bibitem[Kim and Ye(2020)]{Kim2020}
B.-H. Kim and J.~C. Ye.
\newblock Understanding graph isomorphism network for rs-{fMRI} functional
  connectivity analysis.
\newblock \emph{Frontiers in Neuroscience}, 14:\penalty0 630, 2020.
\newblock \doi{10.3389/fnins.2020.00630}.

\bibitem[Kingma and Ba(2014)]{Kingma2014}
D.~Kingma and J.~Ba.
\newblock Adam: A method for stochastic optimization.
\newblock 2014.

\bibitem[Ktena et~al.(2018)Ktena, Parisot, Ferrante, Rajchl, Lee, Glocker, and
  Rueckert]{Ktena2018}
S.~I. Ktena, S.~Parisot, E.~Ferrante, M.~Rajchl, M.~C.~H. Lee, B.~Glocker, and
  D.~Rueckert.
\newblock Metric learning with spectral graph convolutions on brain
  connectivity networks.
\newblock \emph{NeuroImage}, 169:\penalty0 431--442, 2018.

\bibitem[Lang et~al.(2012)Lang, Tomé, Keck, Gorriz, and Puntonet]{Lang2012}
E.~Lang, A.~Tomé, I.~Keck, J.~Gorriz, and C.~Puntonet.
\newblock Brain connectivity analysis: A short survey.
\newblock \emph{Computational intelligence and neuroscience}, 2012.
\newblock \doi{10.1155/2012/412512}.

\bibitem[Li et~al.(2019)Li, Dvornek, Zhou, Zhuang, Ventola, and
  Duncan]{LiX2019}
X.~Li, N.~Dvornek, Y.~Zhou, J.~Zhuang, P.~Ventola, and J.~Duncan.
\newblock Graph neural network for interpreting task-f{MRI} biomarkers.
\newblock pages 485--493, 2019.

\bibitem[Li et~al.(2018)Li, Yu, Shahabi, and Liu]{Li2018}
Y.~Li, R.~Yu, C.~Shahabi, and Y.~Liu.
\newblock Diffusion convolutional recurrent neural network: Data-driven traffic
  forecasting, 2018.

\bibitem[Liang and Wang(2017)]{Liang2017}
H.~Liang and H.~Wang.
\newblock Structure-{F}unction {N}etwork {M}apping and its {A}ssessment via
  {P}ersistent {H}omology.
\newblock \emph{PLoS Computational Biology}, 2017.

\bibitem[Liang et~al.(2019)Liang, Jiang, Cao, Kalantidis, Li, and
  Hauptmann]{Liang2019}
J.~Liang, L.~Jiang, L.~Cao, Y.~Kalantidis, L.-J. Li, and A.~G. Hauptmann.
\newblock Focal visual-text attention for memex question answering.
\newblock \emph{IEEE Transactions on Pattern Analysis and Machine
  Intelligence}, 41\penalty0 (8):\penalty0 1893--1908, 2019.
\newblock \doi{10.1109/TPAMI.2018.2890628}.

\bibitem[Lim et~al.(2019)Lim, Radicchi, P.~van~den Heuvel, and Sporns]{Lim2019}
S.~Lim, F.~Radicchi, M.~P.~van~den Heuvel, and O.~Sporns.
\newblock Discordant attributes of structural and functional brain connectivity
  in a two-layer multiplex network.
\newblock \emph{Scientific Reports}, 9, 12 2019.

\bibitem[Luetkepohl(2005)]{Luetkepohl2005}
H.~Luetkepohl.
\newblock \emph{The New Introduction to Multiple Time Series Analysis}.
\newblock Springer, 2005.
\newblock \doi{10.1007/978-3-540-27752-1}.

\bibitem[Mackinnon(1994)]{Mackinnon1994}
J.~Mackinnon.
\newblock Approximate asymptotic distribution functions for unit-root and
  cointegration tests.
\newblock \emph{Journal of Business and Economic Statistics}, 12:\penalty0
  167--76, 1994.
\newblock \doi{10.1080/07350015.1994.10510005}.

\bibitem[Mele et~al.(2019)Mele, Cavaliere, Alfano, Orsini, Salvatore, and
  Aiello]{Mele2019}
G.~Mele, C.~Cavaliere, V.~Alfano, M.~Orsini, M.~Salvatore, and M.~Aiello.
\newblock Simultaneous eeg-fmri for functional neurological assessment.
\newblock \emph{Frontiers in Neurology}, 10, 08 2019.
\newblock \doi{10.3389/fneur.2019.00848}.

\bibitem[Mess\'e et~al.(2014)Mess\'e, Rudrauf, Benali, and Marrelec]{Messe2014}
A.~Mess\'e, D.~Rudrauf, H.~Benali, and G.~Marrelec.
\newblock Relating {S}tructure and {F}unction in the {H}uman {B}rain:
  {R}elative {C}ontributions of {A}natomy, {S}tationary {D}ynamics, and
  {N}on-stationarities.
\newblock \emph{PLoS Computational Biology}, 10\penalty0 (3), 2014.

\bibitem[Mess\'e et~al.(2015)Mess\'e, Hütt, König, and Hilgetag]{Messe2015}
A.~Mess\'e, M.~T. Hütt, P.~König, and C.~C. Hilgetag.
\newblock A closer look at the apparent correlation of structural and
  functional connectivity in excitable neural networks.
\newblock \emph{Scientific Reports}, 5:\penalty0 7870, 2015.

\bibitem[Messé et~al.(2015)Messé, Rudrauf, Giron, and Marrelec]{Messe2015b}
A.~Messé, D.~Rudrauf, A.~Giron, and G.~Marrelec.
\newblock Predicting functional connectivity from structural connectivity via
  computational models using mri: An extensive comparison study.
\newblock \emph{NeuroImage}, 111, 05 2015.
\newblock \doi{10.1016/j.neuroimage.2015.02.001}.

\bibitem[Mikolov et~al.(2013)Mikolov, Sutskever, Chen, Corrado, and
  Dean]{Mikolov2013}
T.~Mikolov, I.~Sutskever, K.~Chen, G.~Corrado, and J.~Dean.
\newblock Distributed representations of words and phrases and their
  compositionality.
\newblock \emph{Advances in Neural Information Processing Systems}, 26, 10
  2013.

\bibitem[Mill et~al.(2017)Mill, Bagic, Bostan, Schneider, and Cole]{Mill2017}
R.~D. Mill, A.~Bagic, A.~Bostan, W.~Schneider, and M.~W. Cole.
\newblock Empirical validation of directed functional connectivity.
\newblock \emph{NeuroImage}, 146:\penalty0 275--287, 2017.

\bibitem[Moeller et~al.(2010)Moeller, Yacoub, Olman, Auerbach, Strupp, Harel,
  and Ugurbil]{Moeller2010}
S.~Moeller, E.~Yacoub, C.~A. Olman, E.~Auerbach, J.~Strupp, N.~Y. Harel, and
  K.~Ugurbil.
\newblock Multiband multislice ge-epi at 7 tesla, with 16-fold acceleration
  using partial parallel imaging with application to high spatial and temporal
  whole-brain {fMRI}.
\newblock \emph{Magnetic resonance in medicine}, 63 5:\penalty0 1144--53, 2010.

\bibitem[Olman et~al.(2009)Olman, Davachi, and Inati]{Olman2009}
C.~A. Olman, L.~Davachi, and S.~J. Inati.
\newblock Distortion and signal loss in medial temporal lobe.
\newblock \emph{PLoS ONE}, 4, 2009.

\bibitem[Panda et~al.(2021)Panda, Thibaut, Lopez-Gonzalez, Escrichs, Bahri,
  Hillebrand, Deco, Laureys, Gosseries, Annen, and Tewarie]{Panda2021}
R.~Panda, A.~Thibaut, A.~Lopez-Gonzalez, A.~Escrichs, M.~A. Bahri,
  A.~Hillebrand, G.~Deco, S.~Laureys, O.~Gosseries, J.~Annen, and P.~Tewarie.
\newblock Disruption in structural-functional network repertoire and
  time-resolved subcortical-frontoparietal connectivity in disorders of
  consciousness.
\newblock \emph{bioRxiv}, 2021.
\newblock \doi{10.1101/2021.12.10.472068}.

\bibitem[Power et~al.(2016)Power, Plitt, Laumann, and Martin]{Power2016}
J.~Power, M.~Plitt, T.~Laumann, and A.~Martin.
\newblock Sources and implications of whole-brain fmri signals in humans.
\newblock \emph{NeuroImage}, 146, 10 2016.
\newblock \doi{10.1016/j.neuroimage.2016.09.038}.

\bibitem[Prando et~al.(2020)Prando, Zorzi, Bertoldo, Corbetta, Zorzi, and
  Chiuso]{Prando2020}
G.~Prando, M.~Zorzi, A.~Bertoldo, M.~Corbetta, M.~Zorzi, and A.~Chiuso.
\newblock Sparse dcm for whole-brain effective connectivity from resting-state
  fmri data.
\newblock \emph{NeuroImage}, 208:\penalty0 116367, 2020.
\newblock ISSN 1053-8119.

\bibitem[Ramsey et~al.(2011)Ramsey, Hanson, and Glymour]{Ramsey2011}
J.~Ramsey, S.~Hanson, and C.~Glymour.
\newblock Multi-subject search correctly identifies causal connections and most
  causal directions in the dcm models of the smith et al. simulation study.
\newblock \emph{NeuroImage}, 58:\penalty0 838--48, 07 2011.
\newblock \doi{10.1016/j.neuroimage.2011.06.068}.

\bibitem[Reid et~al.(2019)Reid, Headley, Mill, sanchez romero, Uddin,
  Marinazzo, Lurie, Valdés-Sosa, Hanson, Biswal, Calhoun, Poldrack, and
  Cole]{Reid2019}
A.~Reid, D.~Headley, R.~Mill, R.~sanchez romero, L.~Uddin, D.~Marinazzo,
  D.~Lurie, P.~Valdés-Sosa, S.~Hanson, B.~Biswal, V.~Calhoun, R.~Poldrack, and
  M.~Cole.
\newblock Advancing functional connectivity research from association to
  causation.
\newblock \emph{Nature Neuroscience}, 22:\penalty0 1--10, 10 2019.
\newblock \doi{10.1038/s41593-019-0510-4}.

\bibitem[Roebroeck et~al.(2009)Roebroeck, Formisano, and Goebel]{Roebroeck2009}
A.~Roebroeck, E.~Formisano, and R.~Goebel.
\newblock The identification of interacting networks in the brain using fmri:
  Model selection, causality and deconvolution.
\newblock \emph{NeuroImage}, 58:\penalty0 296--302, 09 2009.
\newblock \doi{10.1016/j.neuroimage.2009.09.036}.

\bibitem[Rosenthal et~al.(2018)Rosenthal, Váša, Griffa, Hagmann, Amico,
  Goñi, Avidan, and Sporns]{Rosenthal2018}
G.~Rosenthal, F.~Váša, A.~Griffa, P.~Hagmann, E.~Amico, J.~Goñi, G.~Avidan,
  and O.~Sporns.
\newblock Mapping higher-order relations between brain structure and function
  with embedded vector representations of connectomes.
\newblock \emph{Nature Communications}, 9, 12 2018.

\bibitem[Rumelhart et~al.(1986)Rumelhart, Hinton, and Williams]{Rumelhart1986}
D.~Rumelhart, G.~E. Hinton, and R.~J. Williams.
\newblock Learning representations by back-propagating errors.
\newblock \emph{Nature}, 323:\penalty0 533--536, 1986.

\bibitem[Salimi-Khorshidi et~al.(2014)Salimi-Khorshidi, Douaud, Beckmann,
  Glasser, Griffanti, and Smith]{Salimi2014}
G.~Salimi-Khorshidi, G.~Douaud, C.~F. Beckmann, M.~F. Glasser, L.~Griffanti,
  and S.~M. Smith.
\newblock Automatic denoising of functional {MRI} data: Combining independent
  component analysis and hierarchical fusion of classifiers.
\newblock \emph{NeuroImage}, 90:\penalty0 449 -- 468, 2014.
\newblock \doi{10.1016/j.neuroimage.2013.11.046}.

\bibitem[Sarwar et~al.(2021)Sarwar, Tian, Yeo, Ramamohanarao, and
  Zalesky]{Sarwar2021}
T.~Sarwar, Y.~Tian, B.~Yeo, K.~Ramamohanarao, and A.~Zalesky.
\newblock Structure-function coupling in the human connectome: A machine
  learning approach.
\newblock \emph{NeuroImage}, 226:\penalty0 117609, 02 2021.
\newblock \doi{10.1016/j.neuroimage.2020.117609}.

\bibitem[Schnake et~al.(2021)Schnake, Eberle, Lederer, Nakajima, Schutt,
  Muller, and Montavon]{Schnake2021}
T.~Schnake, O.~Eberle, J.~Lederer, S.~Nakajima, K.~T. Schutt, K.-R. Muller, and
  G.~Montavon.
\newblock Higher-order explanations of graph neural networks via relevant
  walks.
\newblock \emph{IEEE transactions on pattern analysis and machine
  intelligence}, PP, 2021.

\bibitem[Seo et~al.(2018)Seo, Defferrard, Vandergheynst, and Bresson]{Seo2018}
Y.~Seo, M.~Defferrard, P.~Vandergheynst, and X.~Bresson.
\newblock Structured sequence modeling with graph convolutional recurrent
  networks.
\newblock In L.~Cheng, A.~C.~S. Leung, and S.~Ozawa, editors, \emph{Neural
  Information Processing}, pages 362--373. Springer International Publishing,
  2018.

\bibitem[Seth et~al.(2012)Seth, Chorley, and Barnett]{Seth2012}
A.~Seth, P.~Chorley, and L.~Barnett.
\newblock Granger causality analysis of fmri bold signals is invariant to
  hemodynamic convolution but not downsampling.
\newblock \emph{NeuroImage}, 65:\penalty0 540--55, 2012.
\newblock \doi{10.1016/j.neuroimage.2012.09.049}.

\bibitem[Setsompop et~al.(2012)Setsompop, Gagoski, Polimeni, Witzel, Wedeen,
  and Wald]{Setsompop2012}
K.~Setsompop, B.~Gagoski, J.~R. Polimeni, T.~Witzel, V.~J. Wedeen, and L.~L.
  Wald.
\newblock Blipped-controlled aliasing in parallel imaging for simultaneous
  multislice echo planar imaging with reduced g-factor penalty.
\newblock \emph{Magnetic resonance in medicine}, 67 5:\penalty0 1210--24, 2012.

\bibitem[Shinn et~al.(2021)Shinn, Hu, Turner, Noble, Achard, Anticevic,
  Scheinost, Constable, Lee, Bullmore, and Murray]{Shinn2021}
M.~Shinn, A.~Hu, L.~Turner, S.~Noble, S.~Achard, A.~Anticevic, D.~Scheinost,
  R.~T. Constable, D.~Lee, E.~T. Bullmore, and J.~D. Murray.
\newblock Spatial and temporal autocorrelation weave human brain networks.
\newblock \emph{bioRxiv}, 2021.
\newblock \doi{10.1101/2021.06.01.446561}.
\newblock URL
  \url{https://www.biorxiv.org/content/early/2021/06/01/2021.06.01.446561}.

\bibitem[Shuman et~al.(2013)Shuman, Narang, Frossard, Ortega, and
  Vandergheynst]{Shuman2013}
D.~I. Shuman, S.~K. Narang, P.~Frossard, A.~Ortega, and P.~Vandergheynst.
\newblock The emerging field of signal processing on graphs: Ex-tending
  high-dimensional data analysis to networks and other irregular domains.
\newblock \emph{IEEE Signal Processing Magazine}, 30\penalty0 (3):\penalty0
  83--98, 2013.

\bibitem[Singh et~al.(2020)Singh, Braver, Cole, and Ching]{Singh2020}
M.~F. Singh, T.~S. Braver, M.~W. Cole, and S.~Ching.
\newblock Estimation and validation of individualized dynamic brain models with
  resting state fmri.
\newblock \emph{NeuroImage}, 221:\penalty0 117046, 2020.
\newblock ISSN 1053-8119.
\newblock \doi{https://doi.org/10.1016/j.neuroimage.2020.117046}.

\bibitem[Smith et~al.(1999)Smith, Lewis, Ruttimann, Ye, Sinnwell, Yang, Duyn,
  and Frank]{Smith1999}
A.~M. Smith, B.~K. Lewis, U.~E. Ruttimann, F.~Q. Ye, T.~M. Sinnwell, Y.~Yang,
  J.~H. Duyn, and J.~A. Frank.
\newblock Investigation of low frequency drift in fmri signal.
\newblock \emph{NeuroImage}, 9:\penalty0 526--533, 1999.

\bibitem[Smith et~al.(2012)Smith, Tournier, Calamante, and
  Connelly]{SmithR2012}
R.~Smith, J.-D. Tournier, F.~Calamante, and A.~Connelly.
\newblock Anatomically-constrained tractography: Improved diffusion mri
  streamlines tractography through effective use of anatomical information.
\newblock \emph{NeuroImage}, 62:\penalty0 1924--38, 2012.
\newblock \doi{10.1016/j.neuroimage.2012.06.005}.

\bibitem[Smith et~al.(2013{\natexlab{a}})Smith, Tournier, Calamante, and
  Connelly]{SmithR2013}
R.~Smith, J.-D. Tournier, F.~Calamante, and A.~Connelly.
\newblock Sift: Spherical-deconvolution informed filtering of tractograms.
\newblock \emph{NeuroImage}, 67:\penalty0 298--312, 2013{\natexlab{a}}.
\newblock \doi{10.1016/j.neuroimage.2012.11.049}.

\bibitem[Smith et~al.(2011)Smith, Miller, Salimi-Khorshidi, Webster, Beckmann,
  Nichols, Ramsey, and Woolrich]{Smith2011}
S.~M. Smith, K.~L. Miller, G.~Salimi-Khorshidi, M.~Webster, C.~F. Beckmann,
  T.~E. Nichols, J.~D. Ramsey, and M.~W. Woolrich.
\newblock Network modelling methods for fmri.
\newblock \emph{NeuroImage}, 54\penalty0 (2):\penalty0 875--891, 2011.

\bibitem[Smith et~al.(2013{\natexlab{b}})Smith, Beckmann, Andersson, Auerbach,
  Bijsterbosch, Douaud, Duff, Feinberg, Griffanti, Harms, Kelly, Laumann,
  Miller, Moeller, Petersen, Power, Salimi-Khorshidi, Snyder, Vu, Woolrich, Xu,
  Yacoub, Uğurbil, Essen, and Glasser]{Smith2013}
S.~M. Smith, C.~F. Beckmann, J.~Andersson, E.~J. Auerbach, J.~Bijsterbosch,
  G.~Douaud, E.~Duff, D.~A. Feinberg, L.~Griffanti, M.~P. Harms, M.~Kelly,
  T.~Laumann, K.~L. Miller, S.~Moeller, S.~Petersen, J.~Power,
  G.~Salimi-Khorshidi, A.~Z. Snyder, A.~T. Vu, M.~W. Woolrich, J.~Xu,
  E.~Yacoub, K.~Uğurbil, D.~C.~V. Essen, and M.~F. Glasser.
\newblock Resting-state {fMRI} in the human connectome project.
\newblock \emph{NeuroImage}, 80:\penalty0 144 -- 168, 2013{\natexlab{b}}.
\newblock \doi{10.1016/j.neuroimage.2013.05.039}.

\bibitem[Sotiropoulos et~al.(2013{\natexlab{a}})Sotiropoulos, Jbabdi, Xu,
  Andersson, Moeller, Auerbach, Glasser, Hernandez~Fernandez, Sapiro,
  Jenkinson, Feinberg, Yacoub, Lenglet, DC, Ugurbil, and
  Behrens]{Sotiropoulos2013b}
S.~Sotiropoulos, S.~Jbabdi, J.~Xu, J.~Andersson, S.~Moeller, E.~Auerbach,
  M.~Glasser, M.~Hernandez~Fernandez, G.~Sapiro, M.~Jenkinson, D.~Feinberg,
  E.~Yacoub, C.~Lenglet, V.~DC, K.~Ugurbil, and T.~Behrens.
\newblock Advances in diffusion mri acquisition and processing in the human
  connectome project.
\newblock \emph{NeuroImage}, 80:\penalty0 125, 2013{\natexlab{a}}.
\newblock \doi{10.1016/j.neuroimage.2013.05.057}.

\bibitem[Sotiropoulos et~al.(2013{\natexlab{b}})Sotiropoulos, Moeller, Jbabdi,
  Xu, Andersson, Auerbach, Yacoub, Feinberg, Setsompop, Wald, Behrens, Ugurbil,
  and Lenglet]{Sotiropoulos2013}
S.~Sotiropoulos, S.~Moeller, S.~Jbabdi, J.~Xu, J.~Andersson, E.~Auerbach,
  E.~Yacoub, D.~Feinberg, K.~Setsompop, L.~Wald, T.~Behrens, K.~Ugurbil, and
  C.~Lenglet.
\newblock Effects of image reconstruction on fibre orientation mapping from
  multi-channel diffusion {MRI}: Reducing the noise floor using {SENSE}.
\newblock \emph{Magnetic resonance in medicine : official journal of the
  Society of Magnetic Resonance in Medicine / Society of Magnetic Resonance in
  Medicine}, \penalty0 (70), 2013{\natexlab{b}}.
\newblock \doi{10.1002/mrm.24623}.

\bibitem[Suarez et~al.(2020)Suarez, Markello, Betzel, and Misic]{Suarez2020}
L.~Suarez, R.~Markello, R.~Betzel, and B.~Misic.
\newblock Linking structure and function in macroscale brain networks.
\newblock \emph{Trends in Cognitive Sciences}, 24, 02 2020.
\newblock \doi{10.1016/j.tics.2020.01.008}.

\bibitem[Suarez et~al.(2021)Suarez, Richards, Lajoie, and Misic]{Suarez2021}
L.~Suarez, B.~Richards, G.~Lajoie, and B.~Misic.
\newblock Learning function from structure in neuromorphic networks.
\newblock \emph{Nature Machine Intelligence}, 3, 09 2021.
\newblock \doi{10.1038/s42256-021-00376-1}.

\bibitem[Sutskever et~al.(2014)Sutskever, Vinyals, and Le]{Sutskever2014}
I.~Sutskever, O.~Vinyals, and Q.~V. Le.
\newblock Sequence to sequence learning with neural networks.
\newblock \emph{CoRR}, abs/1409.3215, 2014.
\newblock URL \url{http://arxiv.org/abs/1409.3215}.

\bibitem[Thomas et~al.(2014)Thomas, Ye, Irfanoglu, Modi, Saleem, Leopold, and
  Pierpaoli]{Thomas2014}
C.~Thomas, F.~Q. Ye, M.~O. Irfanoglu, P.~D. Modi, K.~S. Saleem, D.~A. Leopold,
  and C.~Pierpaoli.
\newblock Anatomical accuracy of brain connections derived from diffusion {MRI}
  tractography is inherently limited.
\newblock \emph{Proceedings of the National Academy of Sciences of the United
  States of America}, 111 46:\penalty0 16574--9, 2014.

\bibitem[Tournier et~al.(2004)Tournier, Calamante, Gadian, and
  Connelly]{Tournier2004}
J.-D. Tournier, F.~Calamante, D.~Gadian, and A.~Connelly.
\newblock Direct estimation of the fiber orientation density function from
  diffusion-weighted mri data using spherical deconvolution.
\newblock \emph{NeuroImage}, 23:\penalty0 1176--85, 2004.
\newblock \doi{10.1016/j.neuroimage.2004.07.037}.

\bibitem[Tournier et~al.(2007)Tournier, Calamante, and Connelly]{Tournier2007}
J.-D. Tournier, F.~Calamante, and A.~Connelly.
\newblock Robust determination of the fibre orientation distribution in
  diffusion mri: Non-negativity constrained super-resolved spherical
  deconvolution.
\newblock \emph{NeuroImage}, 35:\penalty0 1459--72, 2007.
\newblock \doi{10.1016/j.neuroimage.2007.02.016}.

\bibitem[Tournier et~al.(2019)Tournier, Smith, Raffelt, Tabbara, Dhollander,
  Pietsch, Christiaens, Jeurissen, Yeh, and Connelly]{Tournier2019}
J.-D. Tournier, R.~Smith, D.~Raffelt, R.~Tabbara, T.~Dhollander, M.~Pietsch,
  D.~Christiaens, B.~Jeurissen, C.-H. Yeh, and A.~Connelly.
\newblock {MRtrix3}: A fast, flexible and open software framework for medical
  image processing and visualisation.
\newblock \emph{NeuroImage}, 202, 2019.
\newblock \doi{10.1101/551739}.

\bibitem[Uğurbil et~al.(2013)Uğurbil, Xu, Auerbach, Moeller, Vu,
  Duarte-Carvajalino, Lenglet, Wu, Schmitter, Van~de Moortele, Strupp, Sapiro,
  De~Martino, Wang, Harel, Garwood, Chen, Feinberg, Smith, and
  Yacoub]{Ugurbil2013}
K.~Uğurbil, J.~Xu, E.~Auerbach, S.~Moeller, A.~Vu, J.~Duarte-Carvajalino,
  C.~Lenglet, X.~Wu, S.~Schmitter, P.-F. Van~de Moortele, J.~Strupp, G.~Sapiro,
  F.~De~Martino, D.~Wang, N.~Harel, M.~Garwood, L.~Chen, D.~Feinberg, S.~Smith,
  and E.~Yacoub.
\newblock Pushing spatial and temporal resolution for functional and diffusion
  {MRI} in the human connectome project.
\newblock \emph{NeuroImage}, 80, 2013.

\bibitem[van~den Oord et~al.(2016)van~den Oord, Dieleman, Zen, Simonyan,
  Vinyals, Graves, Kalchbrenner, Senior, and Kavukcuoglu]{Oord2016}
A.~van~den Oord, S.~Dieleman, H.~Zen, K.~Simonyan, O.~Vinyals, A.~Graves,
  N.~Kalchbrenner, A.~Senior, and K.~Kavukcuoglu.
\newblock Wavenet: A generative model for raw audio.
\newblock In \emph{Arxiv}, 2016.
\newblock URL \url{https://arxiv.org/abs/1609.03499}.

\bibitem[Van~Essen et~al.(2013)Van~Essen, Smith, Barch, Behrens, Yacoub, and
  Ugurbil]{VanEssen2013}
D.~Van~Essen, S.~Smith, D.~Barch, T.~Behrens, E.~Yacoub, and K.~Ugurbil.
\newblock The wu-minn human connectome project: an overview.
\newblock \emph{NeuroImage}, 80, 2013.

\bibitem[Vaswani et~al.(2017)Vaswani, Shazeer, Parmar, Uszkoreit, Jones, Gomez,
  Kaiser, and Polosukhin]{Vaswani2017}
A.~Vaswani, N.~Shazeer, N.~Parmar, J.~Uszkoreit, L.~Jones, A.~N. Gomez, L.~u.
  Kaiser, and I.~Polosukhin.
\newblock Attention is all you need.
\newblock In I.~Guyon, U.~V. Luxburg, S.~Bengio, H.~Wallach, R.~Fergus,
  S.~Vishwanathan, and R.~Garnett, editors, \emph{Advances in Neural
  Information Processing Systems}, volume~30. Curran Associates, Inc., 2017.
\newblock URL
  \url{https://proceedings.neurips.cc/paper/2017/file/3f5ee243547dee91fbd053c1c4a845aa-Paper.pdf}.

\bibitem[Vézquez-Rodríguez et~al.(2020)Vézquez-Rodríguez, Liu, Hagmann, and
  Misic]{Rodriguez2020}
B.~Vézquez-Rodríguez, Z.-Q. Liu, P.~Hagmann, and B.~Misic.
\newblock Signal propagation via cortical hierarchies.
\newblock \emph{Network Neuroscience}, 4\penalty0 (4):\penalty0 1072--1090, 11
  2020.
\newblock ISSN 2472-1751.

\bibitem[Wang et~al.(2014)Wang, Bénar, Quilichini, Friston, Jirsa, and
  Bernard]{Wang2014}
H.~E. Wang, C.~G. Bénar, P.~P. Quilichini, K.~J. Friston, V.~K. Jirsa, and
  C.~Bernard.
\newblock A systematic framework for functional connectivity measures.
\newblock \emph{Frontiers in Neuroscience}, 8, 2014.

\bibitem[Wang et~al.(2019)Wang, Kong, Kong, Liégeois, Orban, Deco, van~den
  Heuvel, and Yeo]{Wang2019}
P.~Wang, R.~Kong, X.~Kong, R.~Liégeois, C.~Orban, G.~Deco, M.~P. van~den
  Heuvel, and B.~T. Yeo.
\newblock Inversion of a large-scale circuit model reveals a cortical hierarchy
  in the dynamic resting human brain.
\newblock \emph{Science Advances}, 5\penalty0 (1):\penalty0 eaat7854, 2019.
\newblock \doi{10.1126/sciadv.aat7854}.

\bibitem[Webb et~al.(2013)Webb, Ferguson, Nielsen, and Anderson]{Webb2013}
J.~T. Webb, M.~A. Ferguson, J.~A. Nielsen, and J.~S. Anderson.
\newblock Bold granger causality reflects vascular anatomy.
\newblock \emph{PLOS ONE}, 8\penalty0 (12):\penalty0 null, 12 2013.

\bibitem[Wein et~al.(2021{\natexlab{a}})Wein, Deco, Tomé, Goldhacker, Malloni,
  Greenlee, and Lang]{Wein2021b}
S.~Wein, G.~Deco, A.~Tomé, M.~Goldhacker, W.~Malloni, M.~Greenlee, and
  E.~Lang.
\newblock Brain connectivity studies on structure-function relationships: A
  short survey with an emphasis on machine learning.
\newblock \emph{Computational Intelligence and Neuroscience}, 2021:\penalty0
  1--31, 05 2021{\natexlab{a}}.
\newblock \doi{10.1155/2021/5573740}.

\bibitem[Wein et~al.(2021{\natexlab{b}})Wein, Malloni, Tomé, Frank, Henze,
  Wüst, Greenlee, and Lang]{Wein2021}
S.~Wein, W.~Malloni, A.~Tomé, S.~Frank, G.-I. Henze, S.~Wüst, M.~Greenlee,
  and E.~Lang.
\newblock A graph neural network framework for causal inference in brain
  networks.
\newblock \emph{Scientific Reports}, 11, 04 2021{\natexlab{b}}.
\newblock \doi{10.1038/s41598-021-87411-8}.

\bibitem[Wen et~al.(2013)Wen, Rangarajan, and Ding]{Wen2013}
X.~Wen, G.~Rangarajan, and M.~Ding.
\newblock Is granger causality a viable technique for analyzing fmri data?
\newblock \emph{PloS one}, 8:\penalty0 e67428, 07 2013.
\newblock \doi{10.1371/journal.pone.0067428}.

\bibitem[Werbos(1990)]{Werbos1990}
P.~Werbos.
\newblock Backpropagation through time: what it does and how to do it.
\newblock \emph{Proceedings of the IEEE}, 78\penalty0 (10):\penalty0
  1550--1560, 1990.

\bibitem[Wise et~al.(2004)Wise, Ide, Poulin, and Tracey]{Wise2004}
R.~G. Wise, K.~Ide, M.~J. Poulin, and I.~Tracey.
\newblock Resting fluctuations in arterial carbon dioxide induce significant
  low frequency variations in bold signal.
\newblock \emph{NeuroImage}, 21:\penalty0 1652--1664, 2004.

\bibitem[Wu et~al.(2019)Wu, Pan, Long, Jiang, and Zhang]{Wu2019}
Z.~Wu, S.~Pan, G.~Long, J.~Jiang, and C.~Zhang.
\newblock Graph wavenet for deep spatial-temporal graph modeling.
\newblock pages 1907--1913, 08 2019.
\newblock \doi{10.24963/ijcai.2019/264}.

\bibitem[{Wu} et~al.(2020){Wu}, {Pan}, {Chen}, {Long}, {Zhang}, and
  {Yu}]{Wu2020}
Z.~{Wu}, S.~{Pan}, F.~{Chen}, G.~{Long}, C.~{Zhang}, and P.~S. {Yu}.
\newblock A comprehensive survey on graph neural networks.
\newblock \emph{IEEE Transactions on Neural Networks and Learning Systems},
  pages 1--21, 2020.

\bibitem[Xu et~al.(2012)Xu, Moeller, Strupp, Auerbach, Chen, A.~Feinberg,
  Ugurbil, and Yacoub]{Xu2012}
J.~Xu, S.~Moeller, J.~Strupp, E.~Auerbach, L.~Chen, D.~A.~Feinberg, K.~Ugurbil,
  and E.~Yacoub.
\newblock Highly accelerated whole brain imaging using
  aligned-blipped-controlled-aliasing multiband {EPI}.
\newblock \emph{Proceedings of the 20th Annual Meeting of ISMRM}, page 2036,
  2012.

\bibitem[Xu et~al.(2015)Xu, Ba, Kiros, Cho, Courville, Salakhutdinov, Zemel,
  and Bengio]{Xu2015}
K.~Xu, J.~Ba, R.~Kiros, K.~Cho, A.~C. Courville, R.~Salakhutdinov, R.~S. Zemel,
  and Y.~Bengio.
\newblock Show, attend and tell: Neural image caption generation with visual
  attention.
\newblock In \emph{ICML}, 2015.

\bibitem[Yan et~al.(2021)Yan, Liu, Ai, Shi, Zhang, Pei, and Wu]{Yan2021}
T.~Yan, T.~Liu, J.~Ai, Z.~Shi, J.~Zhang, G.~Pei, and J.~Wu.
\newblock Task-induced activation transmitted by structural connectivity is
  associated with behavioral performance.
\newblock \emph{Brain Structure and Function}, 226, 06 2021.
\newblock \doi{10.1007/s00429-021-02249-0}.

\bibitem[Zeiler and Fergus(2013)]{Zeiler2013}
M.~Zeiler and R.~Fergus.
\newblock Visualizing and understanding convolutional neural networks.
\newblock \emph{ECCV 2014, Part I, LNCS 8689}, 8689, 2013.

\bibitem[Zhang et~al.(2012)Zhang, Schneider, Gandini Wheeler-Kingshott, and
  Alexander]{ZhangH2012}
H.~Zhang, T.~Schneider, C.~Gandini Wheeler-Kingshott, and D.~Alexander.
\newblock {NODDI}: Practical in vivo neurite orientation dispersion and density
  imaging of the human brain.
\newblock \emph{NeuroImage}, 61:\penalty0 1000--16, 2012.

\bibitem[Zheng et~al.(2020)Zheng, Fan, Wang, and Qi]{Zheng2020}
C.~Zheng, X.~Fan, C.~Wang, and J.~Qi.
\newblock Gman: A graph multi-attention network for traffic prediction.
\newblock \emph{Proceedings of the AAAI Conference on Artificial Intelligence},
  34:\penalty0 1234--1241, 04 2020.
\newblock \doi{10.1609/aaai.v34i01.5477}.

\bibitem[Zimmermann et~al.(2019)Zimmermann, Griffiths, Schirner, Ritter, and
  McIntosh]{Zimmermann2019}
J.~Zimmermann, J.~Griffiths, M.~Schirner, P.~Ritter, and A.~R. McIntosh.
\newblock Subject-specificity of the correlation between large-scale structural
  and functional connectivity.
\newblock \emph{Network Neuroscience}, pages 1--35, 2019.

\bibitem[Zou et~al.(2008)Zou, Zhu, Yang, Zuo, Long, Cao, Wang, and
  Zang]{Zou2008}
Q.~Zou, C.-Z. Zhu, Y.~Yang, X.-N. Zuo, X.~Long, Q.-J. Cao, Y.-F. Wang, and
  Y.-F. Zang.
\newblock An improved approach to detection of amplitude of low-frequency
  fluctuation (alff) for resting-state fmri: Fractional alff.
\newblock \emph{Journal of neuroscience methods}, 172:\penalty0 137--41, 08
  2008.

\bibitem[Řehůřek and Sojka(2010)]{Rehurek2010}
R.~Řehůřek and P.~Sojka.
\newblock Software framework for topic modelling with large corpora.
\newblock In \emph{Proceedings of the LREC 2010 Workshop on New Challenges for
  NLP Frameworks}, pages 45--50, 05 2010.
\newblock \doi{10.13140/2.1.2393.1847}.

\end{thebibliography}

\clearpage


\setcounter{figure}{0}
\renewcommand{\thefigure}{S\arabic{figure}}
\renewcommand{\theHfigure}{Supplement.\thefigure}

\section*{Supplementary Information}

\subsection*{Supplement I} \label{sec:supp_hyperparameters}

\begin{figure}[!htb]
\bc

\makebox[\textwidth][c]
{
  \begin{minipage}[b]{0.4\textwidth}
  \includegraphics[width=1\textwidth]{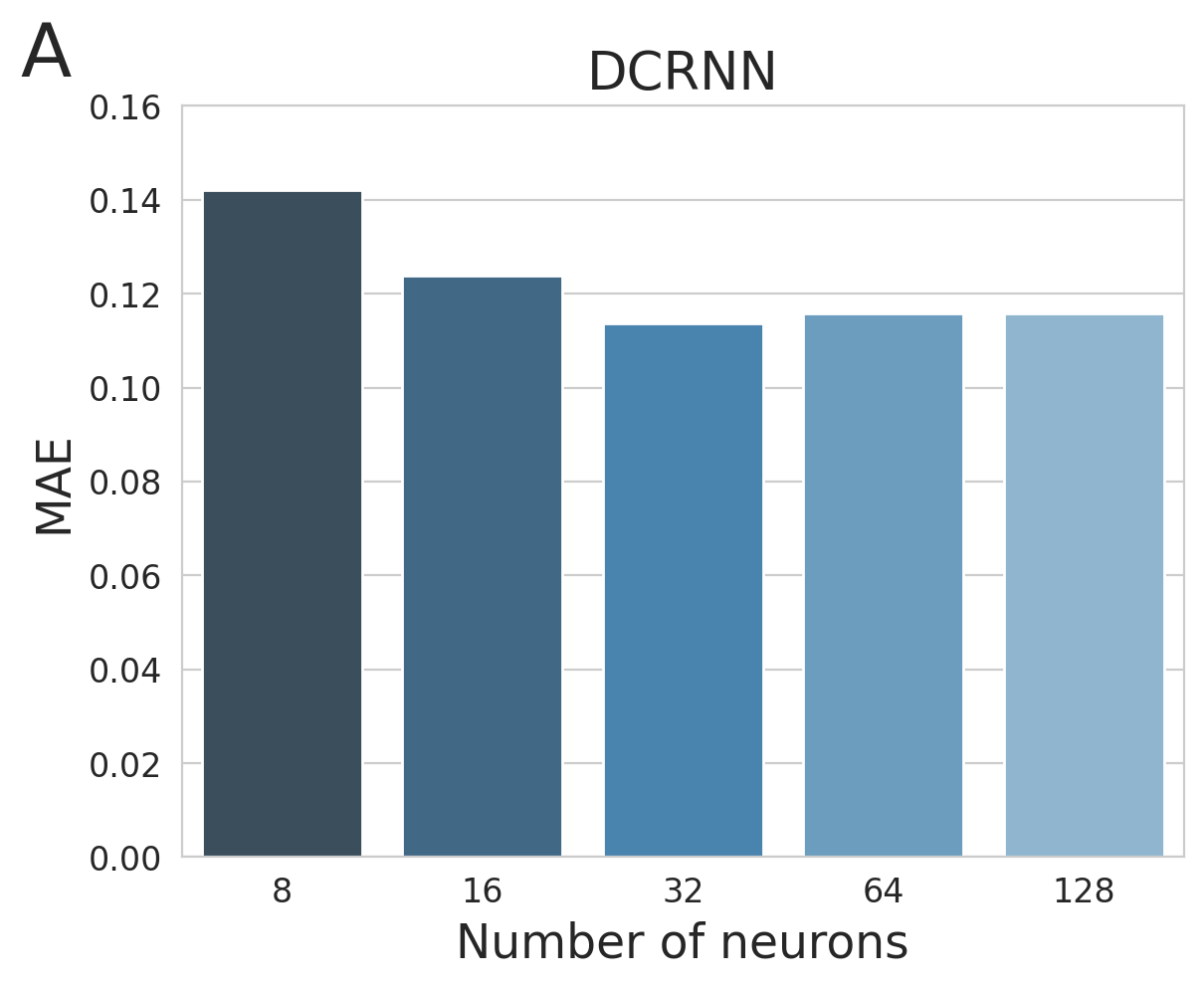}  
  \end{minipage}
    
  \begin{minipage}[b]{0.4\textwidth}
  \includegraphics[width=1\textwidth]{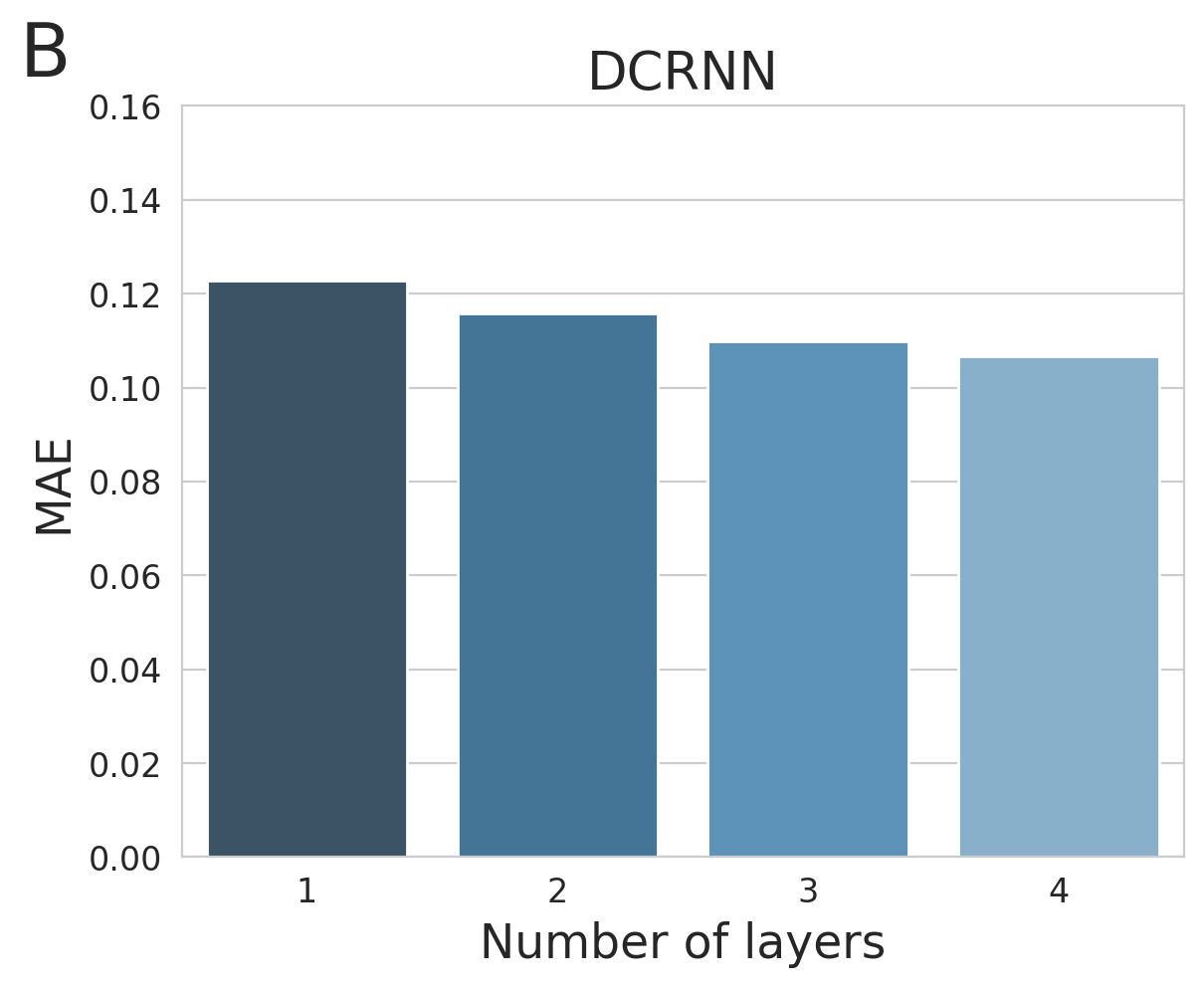}
  \end{minipage}   
}

\ec 
\caption{Here the influence of the hyperparameters on the prediction accuracy of the DCRNN is depicted. In (A) the test error is shown in dependence of the number of neurons in each layer, and in (B) the error in dependence of the number of DCGRU layers.}
\label{fig:hyperparameters_DCRNN}
\end{figure}

\begin{figure}[!htb]
\bc

\makebox[\textwidth][c]
{
  \begin{minipage}[b]{0.33\textwidth}
  \includegraphics[width=1\textwidth]{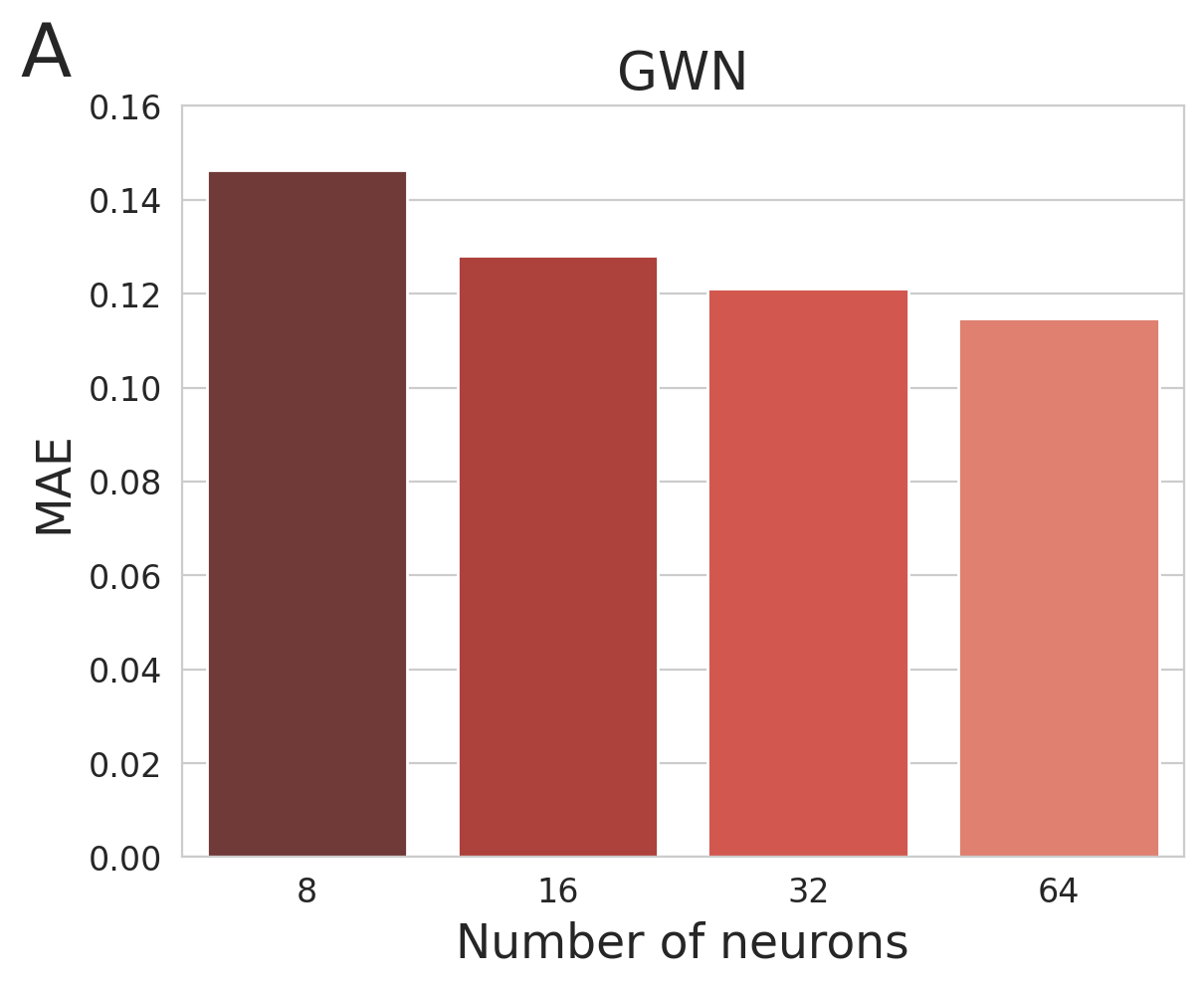}  
  \end{minipage}
    
  \begin{minipage}[b]{0.33\textwidth}
  \includegraphics[width=1\textwidth]{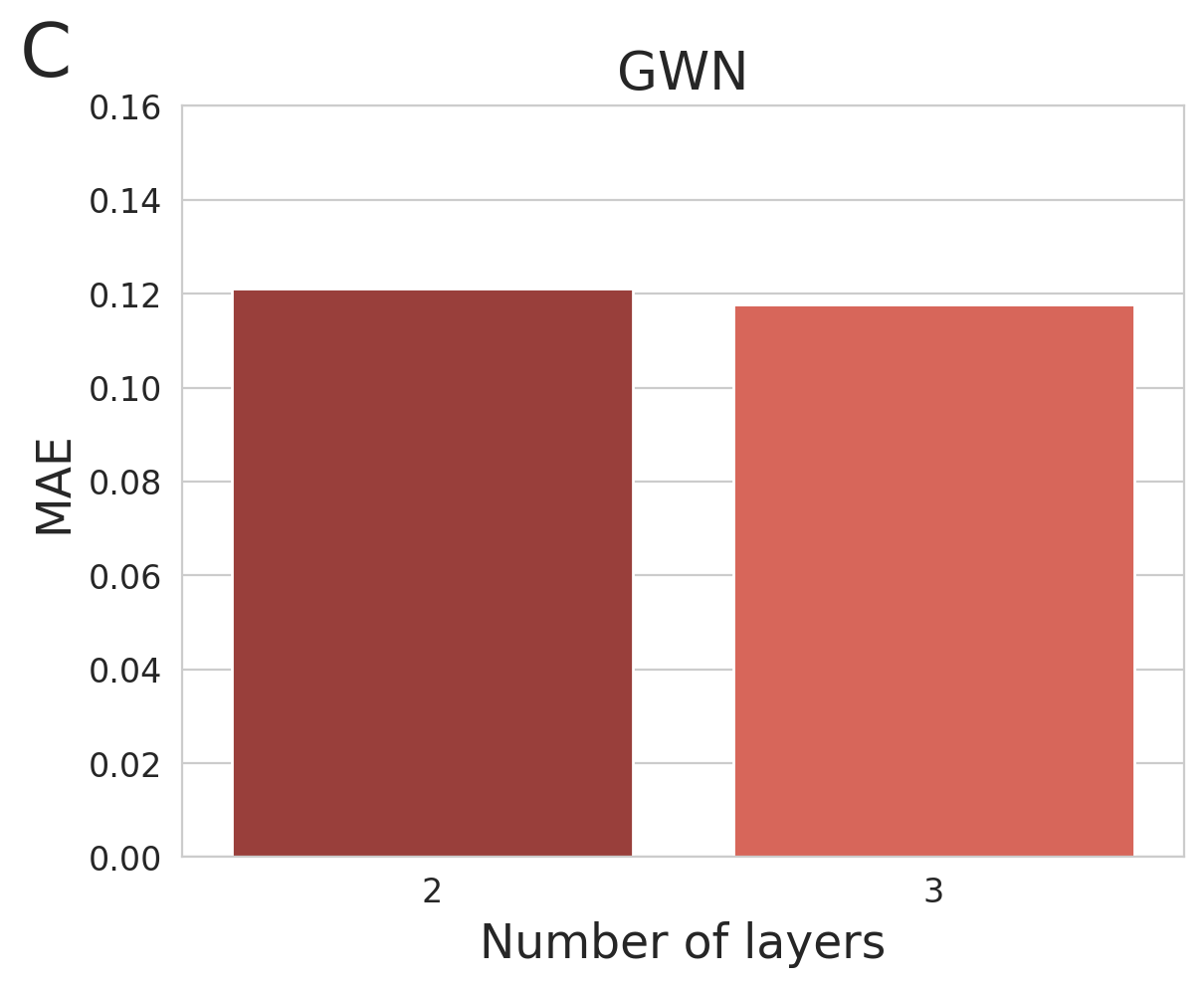}
  \end{minipage}   
    
  \begin{minipage}[b]{0.33\textwidth}
  \includegraphics[width=1\textwidth]{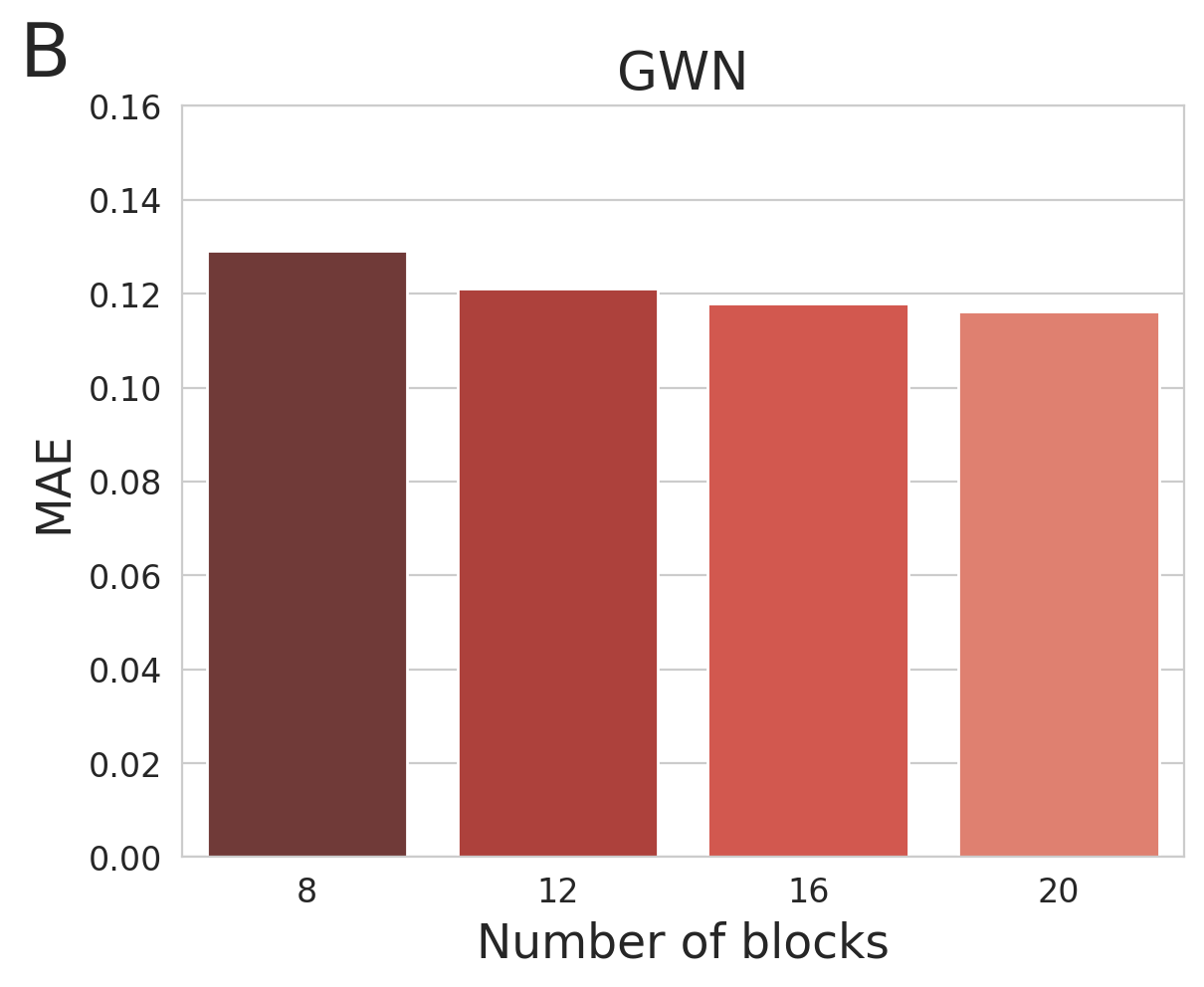}
  \end{minipage} 
  
}

\ec 
\caption{In this figure the influence of the GWN hyperparameters on the prediction accuracy is shown. In (A) the test error in dependence of number of neurons (or feature maps) is illustrated. Here (B) shows the influence of the number of DCC blocks used in the GWN architecture and (C) shows the impact of the number of layers per DCC block.}
\label{fig:hyperparameters_GWN}
\end{figure}

\begin{figure}[!htb]
\bc

\makebox[\textwidth][c]
{
  \begin{minipage}[b]{0.33\textwidth}
  \includegraphics[width=1\textwidth]{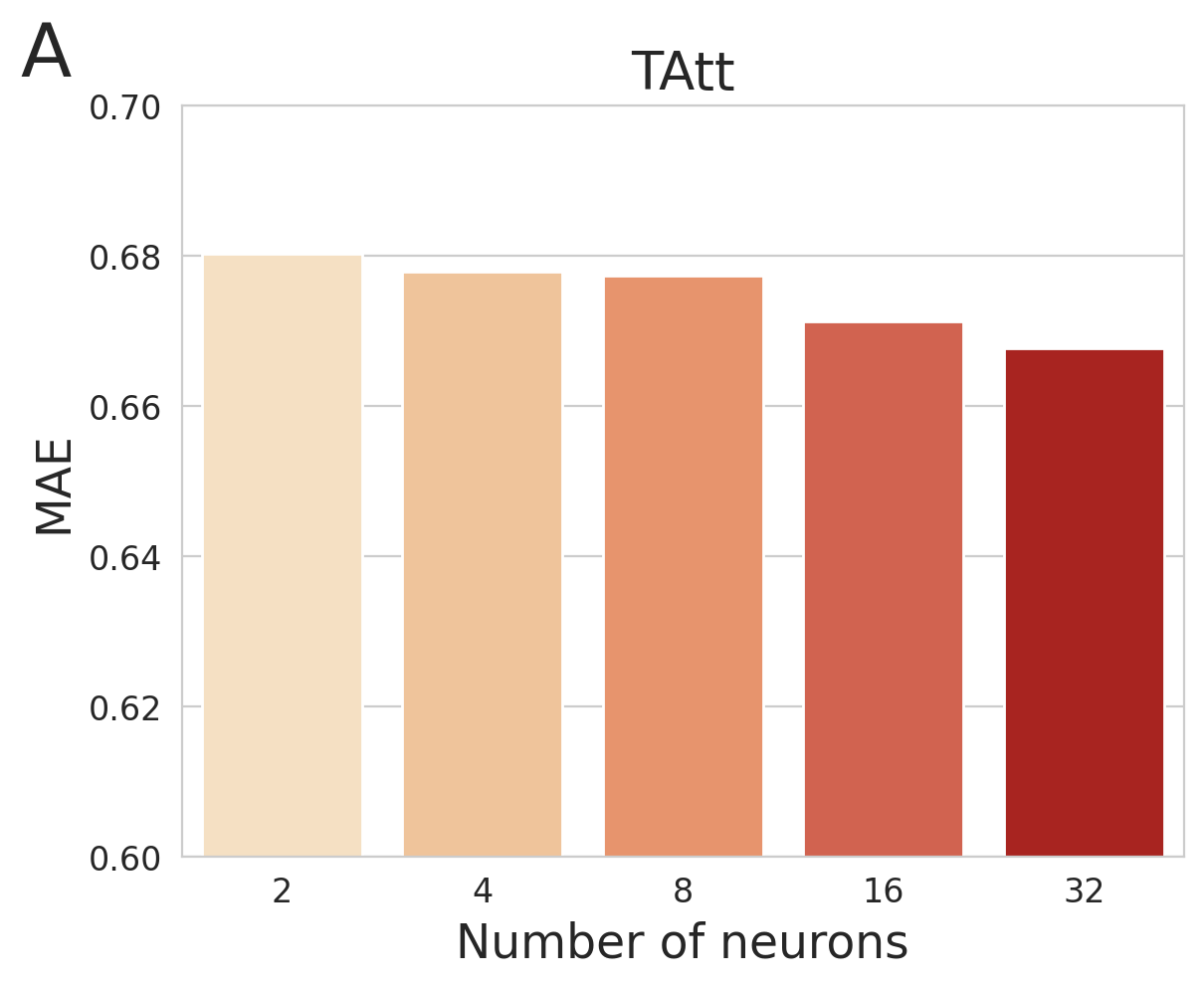}  
  \end{minipage}
    
  \begin{minipage}[b]{0.33\textwidth}
  \includegraphics[width=1\textwidth]{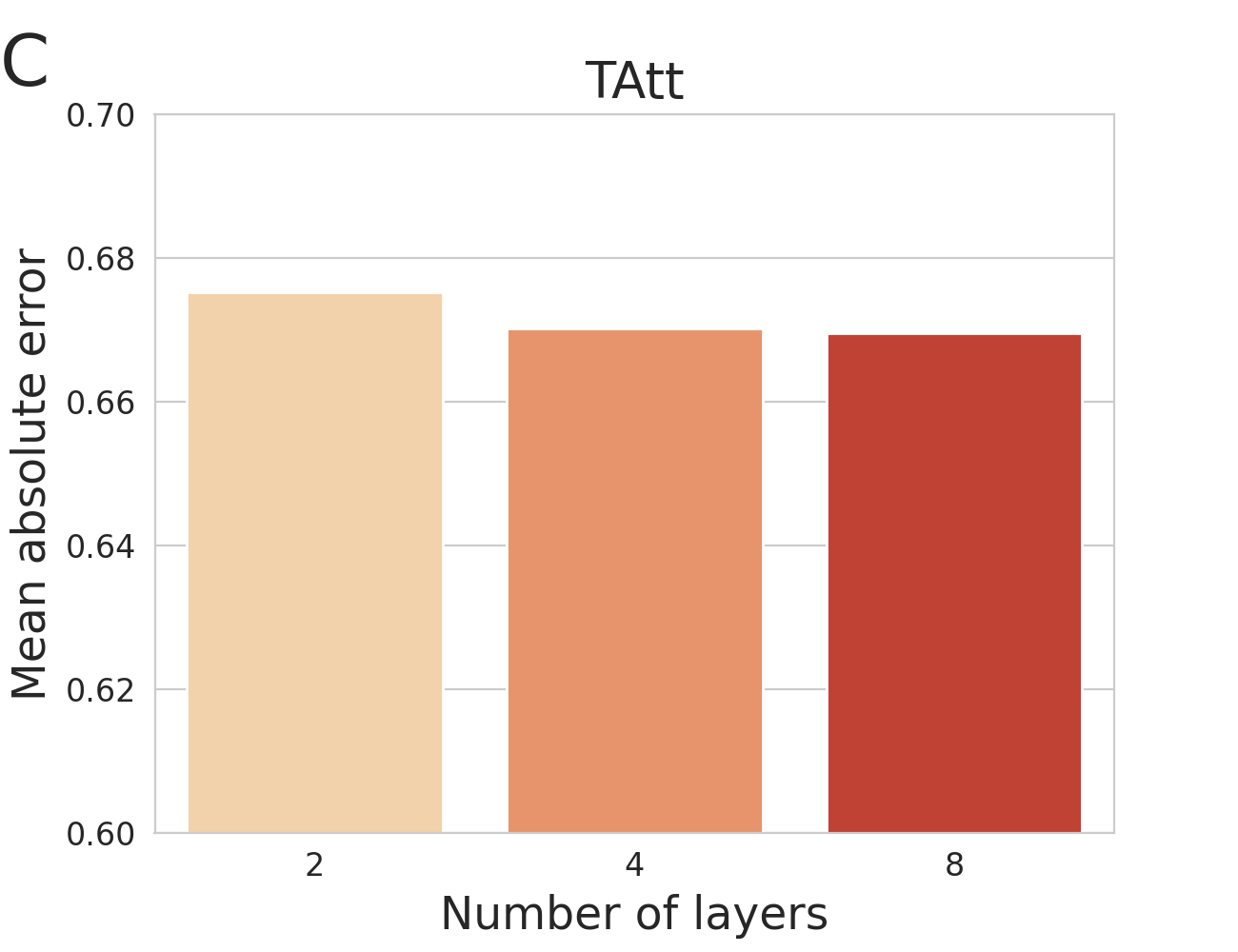}
  \end{minipage}  
  
    \begin{minipage}[b]{0.33\textwidth}
  \includegraphics[width=1\textwidth]{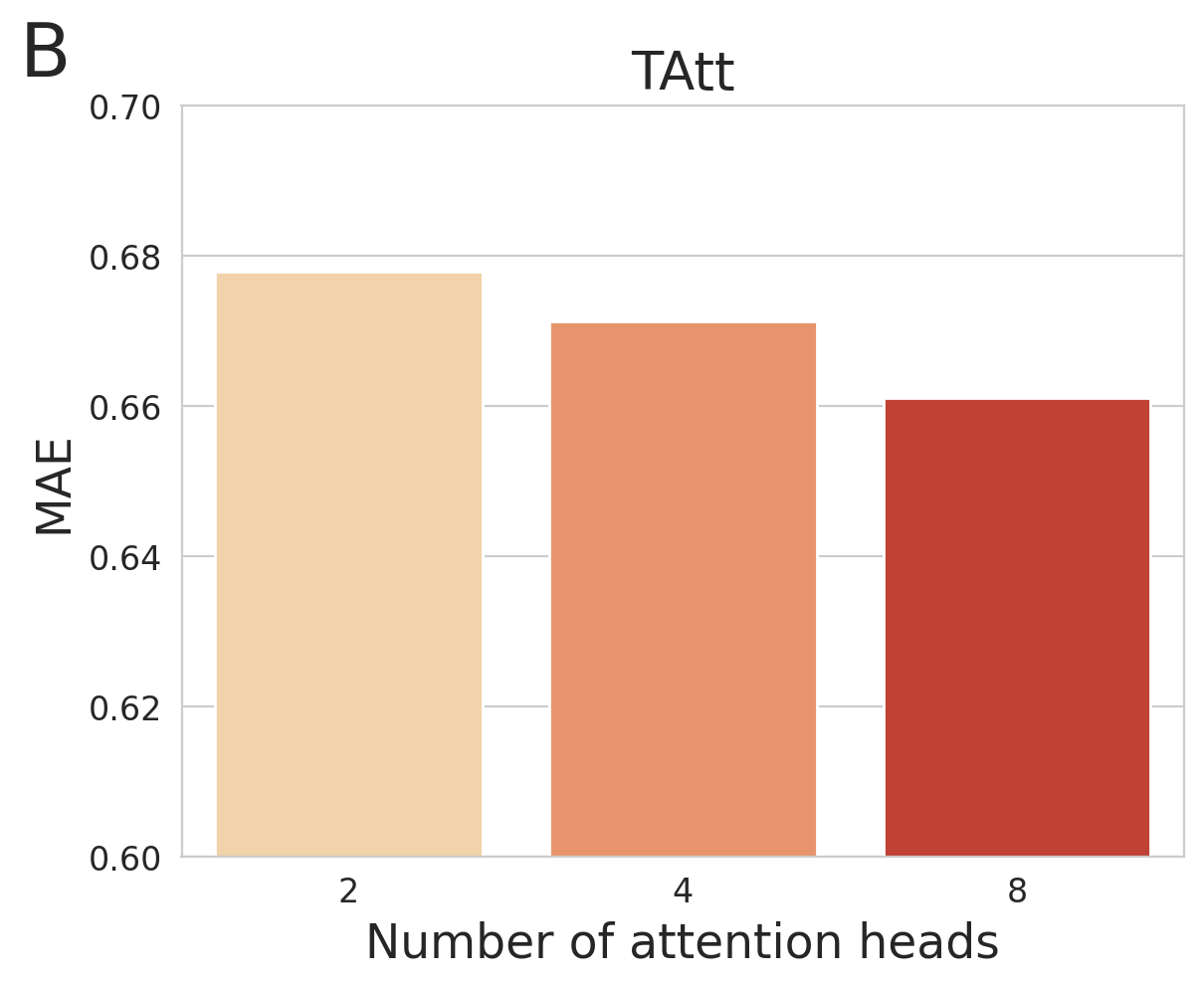}
  \end{minipage}   
}

\ec 
\caption{Here the influence of hyperparameters on the TAtt accuracy is illustrated. In (A) the test MAE in dependence of the number of neurons used in each TAtt mechanism is shown. In (B) the influence of the number of TAtt layers is depicted and (C) illustrates the impact of number of attention heads incorporated.}
\label{fig:hyperparameters_TAtt}
\end{figure}

In this supplement the influence of the model hyperparameters for the different neural network architectures is discussed. The hyperparameters are chosen as described in the \nameref{sec:model_training} section and held constant, while only the hyperparameter of interest is varied in the following evaluations. Figure \ref{fig:hyperparameters_DCRNN} and \ref{fig:hyperparameters_GWN} show that the performance of he DCRNN and GWN in general still could slightly improve with a larger number of model parameters. Because the computation time and memory requirements linearly grow with the number of parameters, we chose the model hyperparameters as described in the \nameref{sec:model_training} section to yield a good trade-off between model performance and computational requirements. Also the TAtt model in figure \ref{fig:hyperparameters_TAtt} shows some improvement with a larger number of parameters, however the MAE is still considerably higher compared to the RNN and WN based GNN architectures. 

\clearpage


\subsection*{Supplement II} \label{sec:supp_visual_ROIs}

\begin{table}[!ht]
\caption{List of ROIs involved in visual processing based on the multi-modal parcellation proposed by \cite{Glasser2016}. The table shows the index of the region in the atlas for the right/left hemisphere including the name of the region.}
\vspace{2mm}
\label{tab:list_ROIs}
\centering
\begin{tabular}{lccc}
\hline
  Index & Name \\
\hline

1/181 & V1 \\ 
2/182 & MST \\ 
3/183 & V6 \\ 
4/184 & V2 \\ 
5/185 & V3 \\ 
6/186 & V4 \\ 
7/187 & V8 \\ 
13/193 & V3A \\ 
16/196 & V7 \\ 
19/199 & V3B \\ 
20/200 & LO1 \\ 
21/201 & LO2 \\ 
22/202 & PIT \\ 
23/203 & MT \\ 
152/332 & V6A \\ 
153/333 & VMV1 \\ 
154/334 & VMV3 \\ 
156/336 & V4t \\ 
158/338 & V3CD \\ 
159/339 & LO3 \\ 
160/340 & VMV2 \\ 
163/343 & VVC \\  

\hline
\end{tabular}
\end{table}

\clearpage


\subsection*{Supplement III} \label{sec:supp_comparison_GNNs}

\begin{figure}[!htb]
\bc

\makebox[\textwidth][c]
{  
  \begin{minipage}[b]{0.255\textwidth}
  \includegraphics[width=1\textwidth]{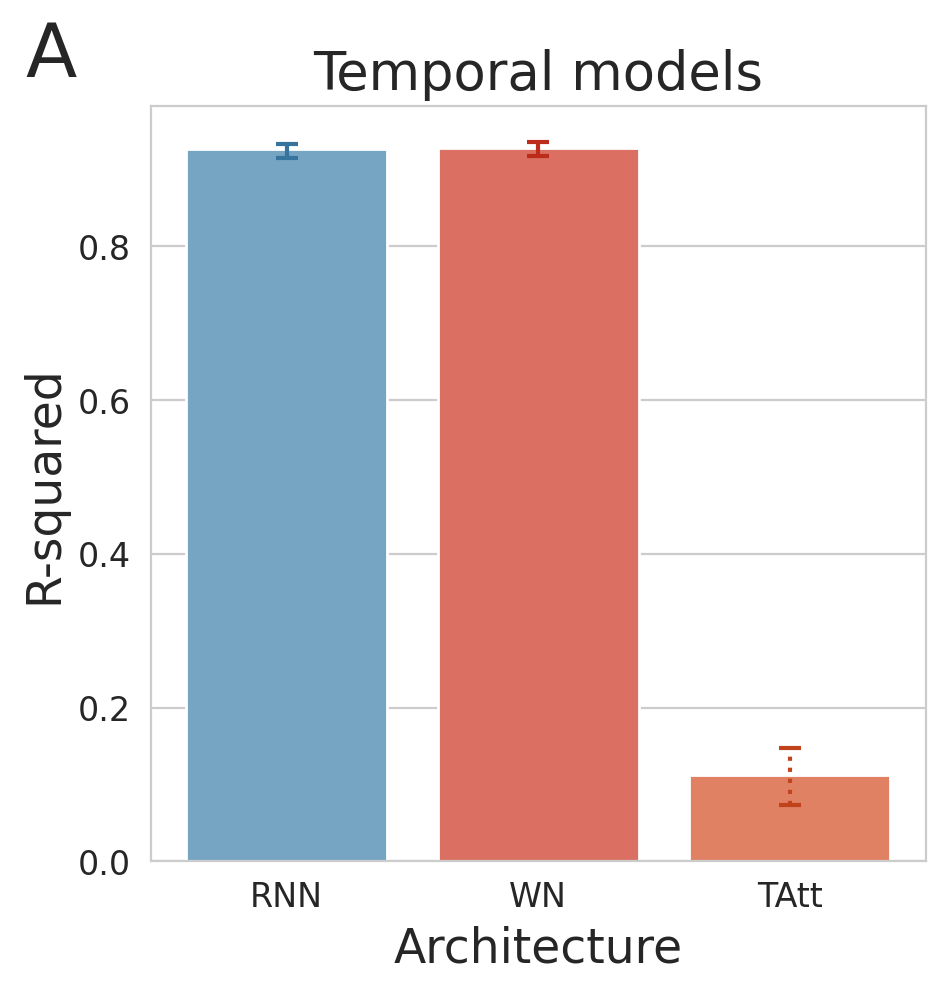}  
  \end{minipage}
        
  \begin{minipage}[b]{0.26\textwidth}
  \includegraphics[width=1\textwidth]{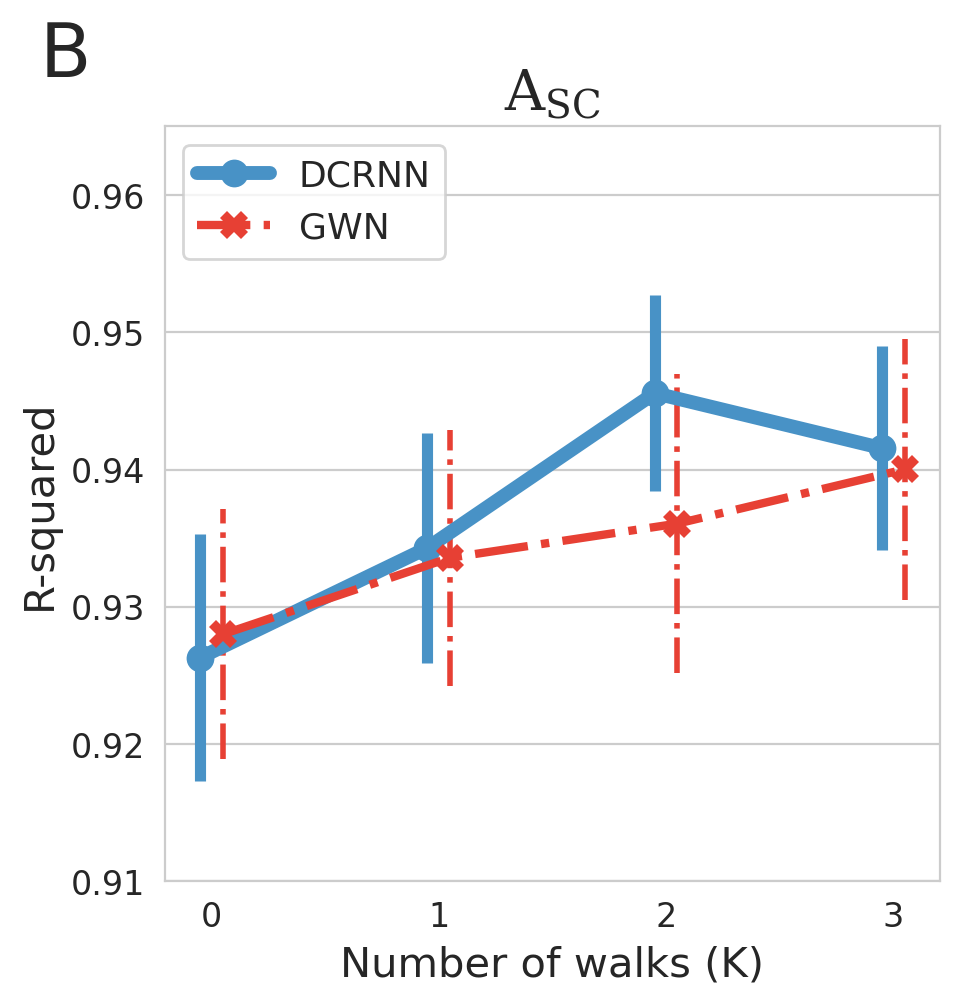}
  \end{minipage}

  \begin{minipage}[b]{0.26\textwidth}
  \includegraphics[width=1\textwidth]{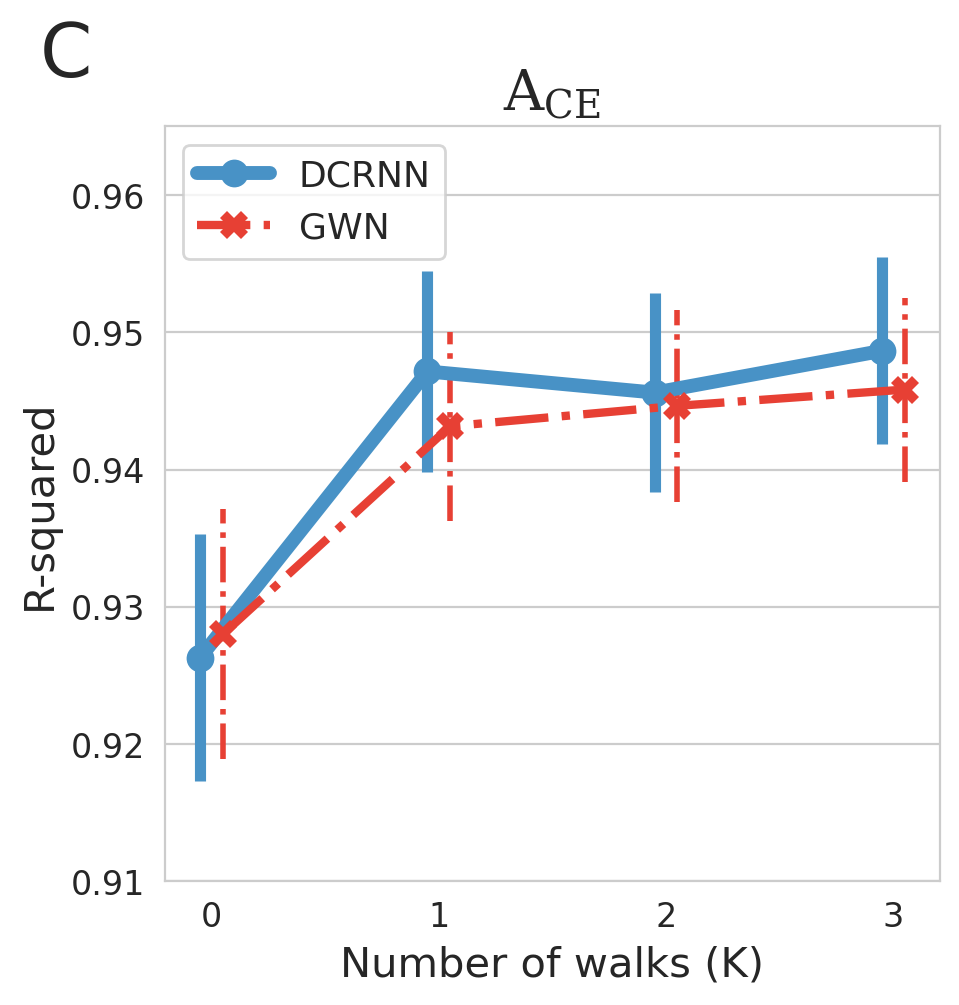}
  \end{minipage}  
    
  \begin{minipage}[b]{0.26\textwidth}
  \includegraphics[width=1\textwidth]{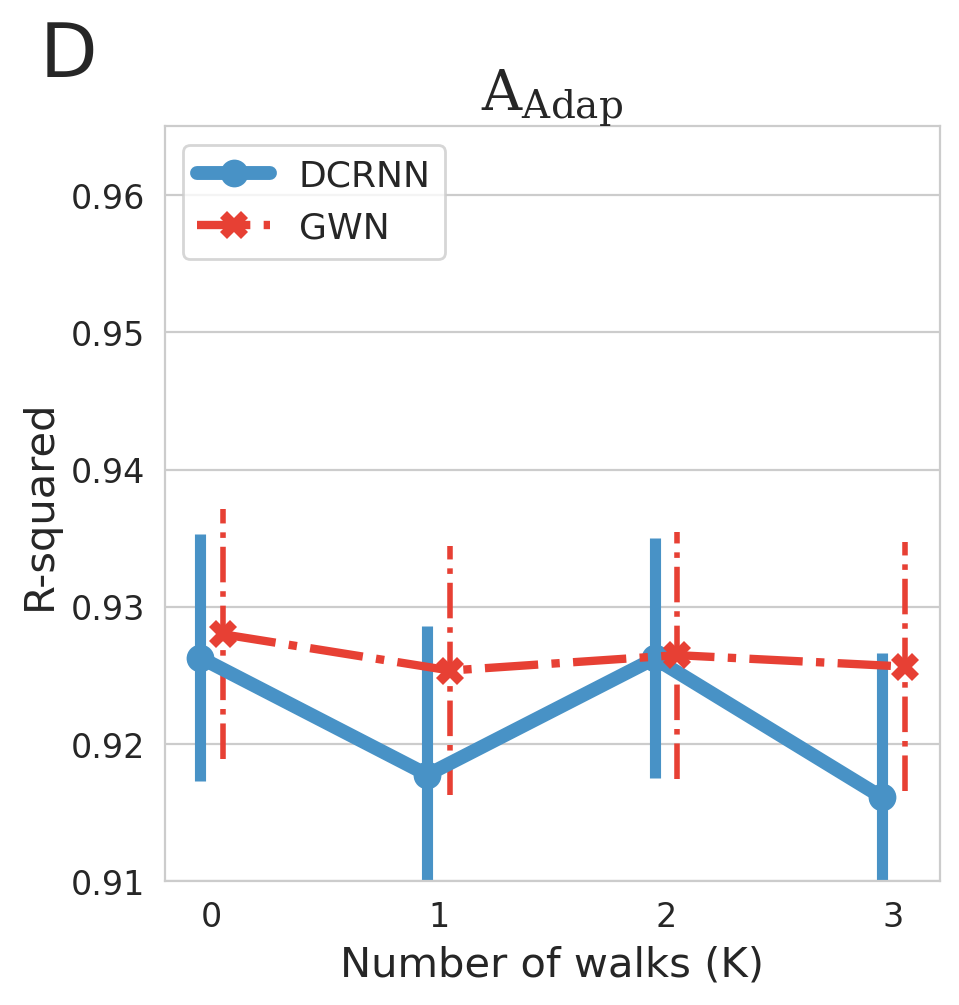}
  \end{minipage}  
}

\ec 
\caption{Figure (A) shows a comparison of different modeling strategies for temporal dynamics in the BOLD signal, comparing the $ \mb{R}^2 $ values of the recurrent neural network (RNN), the WaveNet (WN) and the temporal attention (TAtt) architecture. The errorbars represent the standard deviation of the values across subjects. Figure (B), (C) and (D) show the prediction accuracies of the DCRNN and GWN model in dependence of the walk order $ K $. In figure (B) the $ \mb{R}^2 $ values are shown when incorporating the SC as an adjacency matrix $ \mb{A}_{SC} $, figure (C) illustrates the test MAE when employing CEs in an adjacency matrix $ \mb{A}_{CE} $ to define spatial relationships, and (D) displays the case when using a self-adaptive weight matrix $ \mb{A}_{Adap} $.}
\label{fig:supp_comparison_temporal_spatial_R2}
\end{figure}

In this supplement the different spatial and temporal modeling approaches are evaluated using the scale-free R-squared ($ R^2 $) measure and the similarity of predicted connectivity states. The $ R^2 $ measure can be obtained with:

\beq 
	R^2 (\mb{x} , \hat{\mb{x}}) = 1 - \frac{ \sum_{n,t} ( x_n^{(t)} - \hat{x}_n^{(t)} )^2 }{ \sum_{n,t} ( x_n^{(t)} - \bar{x}_n^{(t)} )^2 }\label{eqn:R2}
\eeq 
where $ x_n^{(t)} $ represents the true BOLD signal in brain region $ n $ at timestep $ t $. Correspondingly $ \bar{x}_n^{(t)} $ denotes the average BOLD signal across all $ N $ regions and $ T_f $ timepoints, and $ \hat{x}_n^{(t)} $ indicates the predicted signal of the respective model. Using this scale-free measure the evaluations of the section \nameref{sec:comparison_GNNs} are replicated in figure \ref{fig:supp_comparison_temporal_spatial_R2}. To test the significance also based on this measure, the $ R^2 $ values were computed for the individual subjects, and by applying a paired t-test the RNN and WN model both outperformed the TAtt model with a $ p $-value of $ p \leq 0.0001 $ (Cohen's $ d \gg 1 $) based on this measure. Also the impact of structural modeling showed to be significant with $ p \leq 0.0001 $ (Cohen's $ d > 1 $) for both STGNN models.

\begin{figure}[!htb]
\bc

\makebox[\textwidth][c]
{  
  \begin{minipage}[b]{0.255\textwidth}
  \includegraphics[width=1\textwidth]{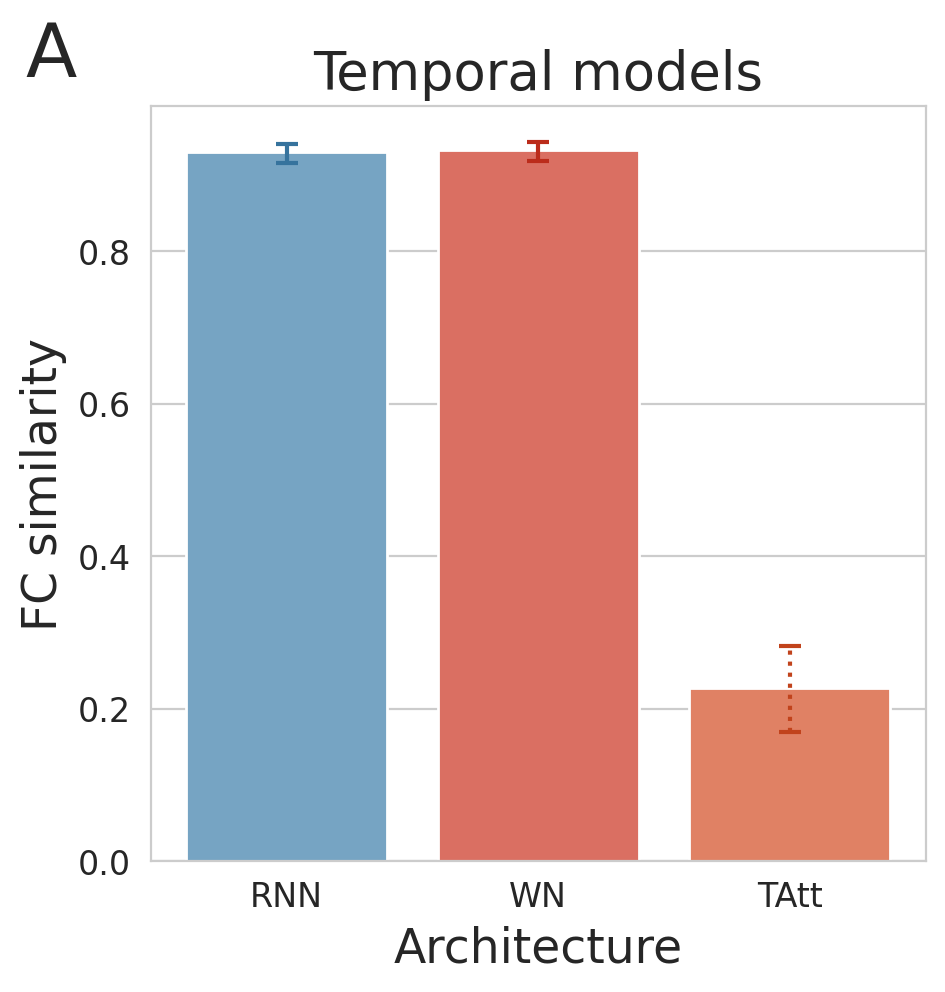}  
  \end{minipage}
        
  \begin{minipage}[b]{0.26\textwidth}
  \includegraphics[width=1\textwidth]{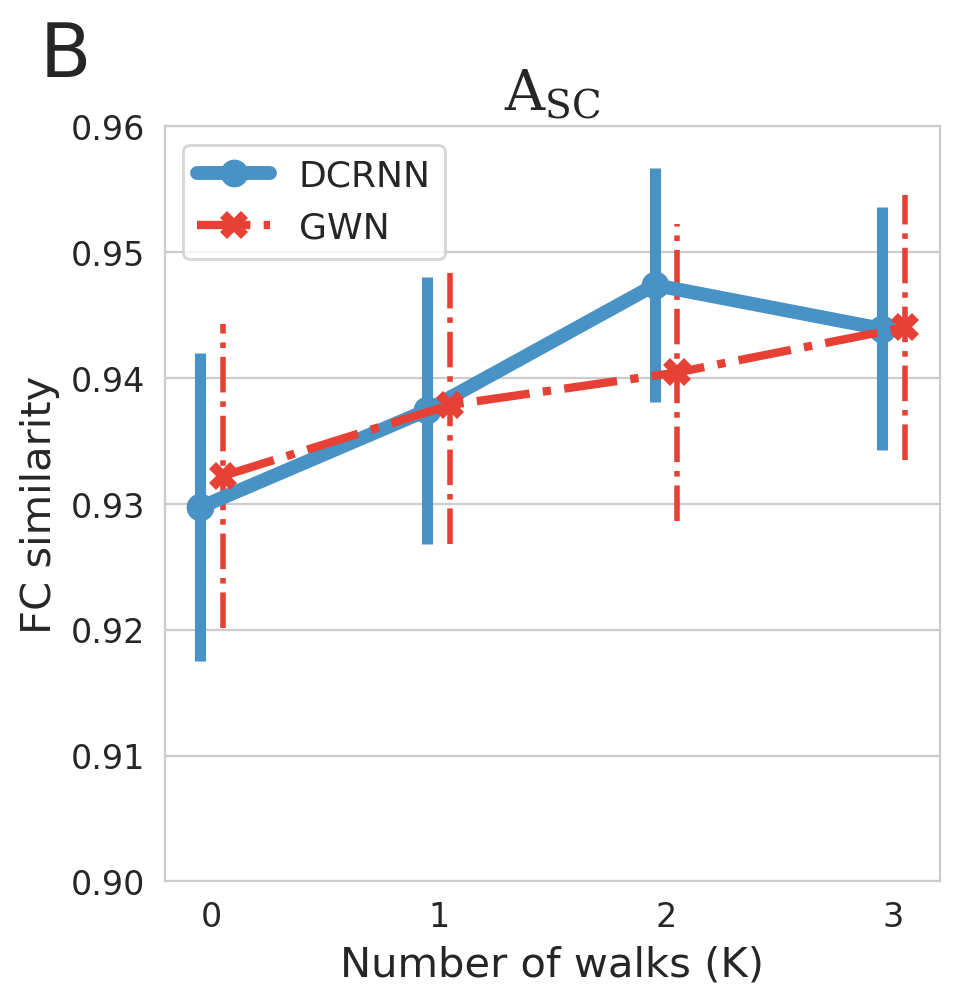}
  \end{minipage}

  \begin{minipage}[b]{0.26\textwidth}
  \includegraphics[width=1\textwidth]{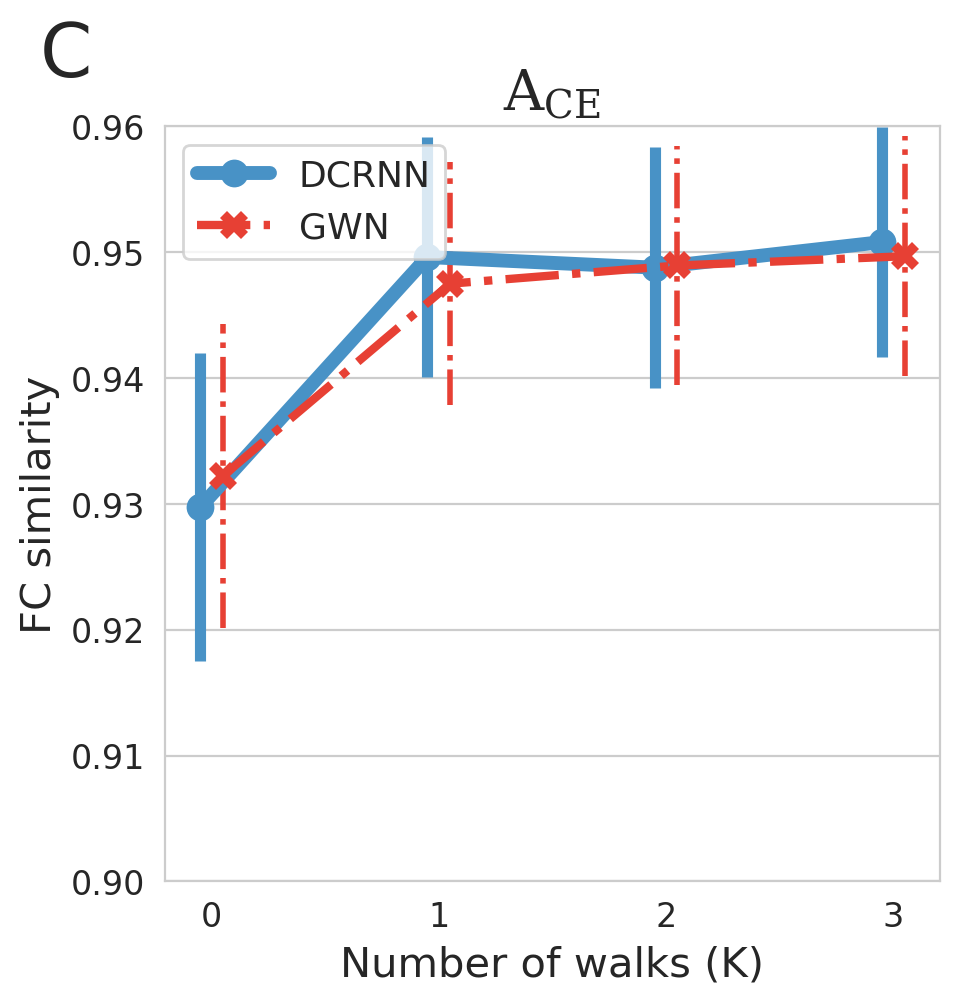}
  \end{minipage}  
    
  \begin{minipage}[b]{0.26\textwidth}
  \includegraphics[width=1\textwidth]{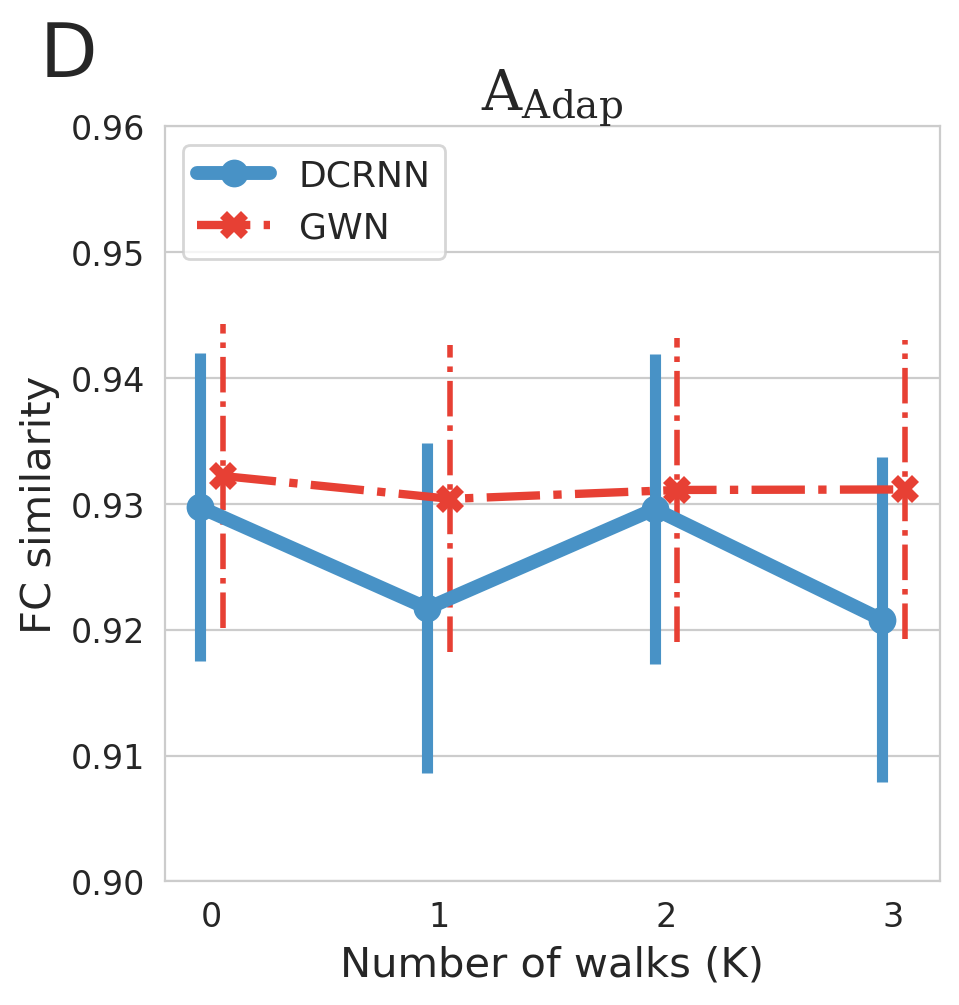}
  \end{minipage}  
}

\ec 
\caption{Figure (A) shows a comparison of different modeling strategies for temporal dynamics in the BOLD signal, comparing the FC similarities of the recurrent neural network (RNN), the WaveNet (WN) and the temporal attention (TAtt) architecture. The errorbars represent the standard deviation of the values across subjects. Figure (B), (C) and (D) show the prediction accuracies of the DCRNN and GWN model in dependence of the walk order $ K $. In figure (B) the similarity values are shown when incorporating the SC as an adjacency matrix $ \mb{A}_{SC} $, figure (C) illustrates the test MAE when employing CEs in an adjacency matrix $ \mb{A}_{CE} $ to define spatial relationships, and (D) displays the case when using a self-adaptive weight matrix $ \mb{A}_{Adap} $.}
\label{fig:supp_comparison_temporal_spatial_FC_simi}
\end{figure}

\begin{figure}[!htb]
\bc

\makebox[\textwidth][c]
{\includegraphics[width=0.75\textwidth]{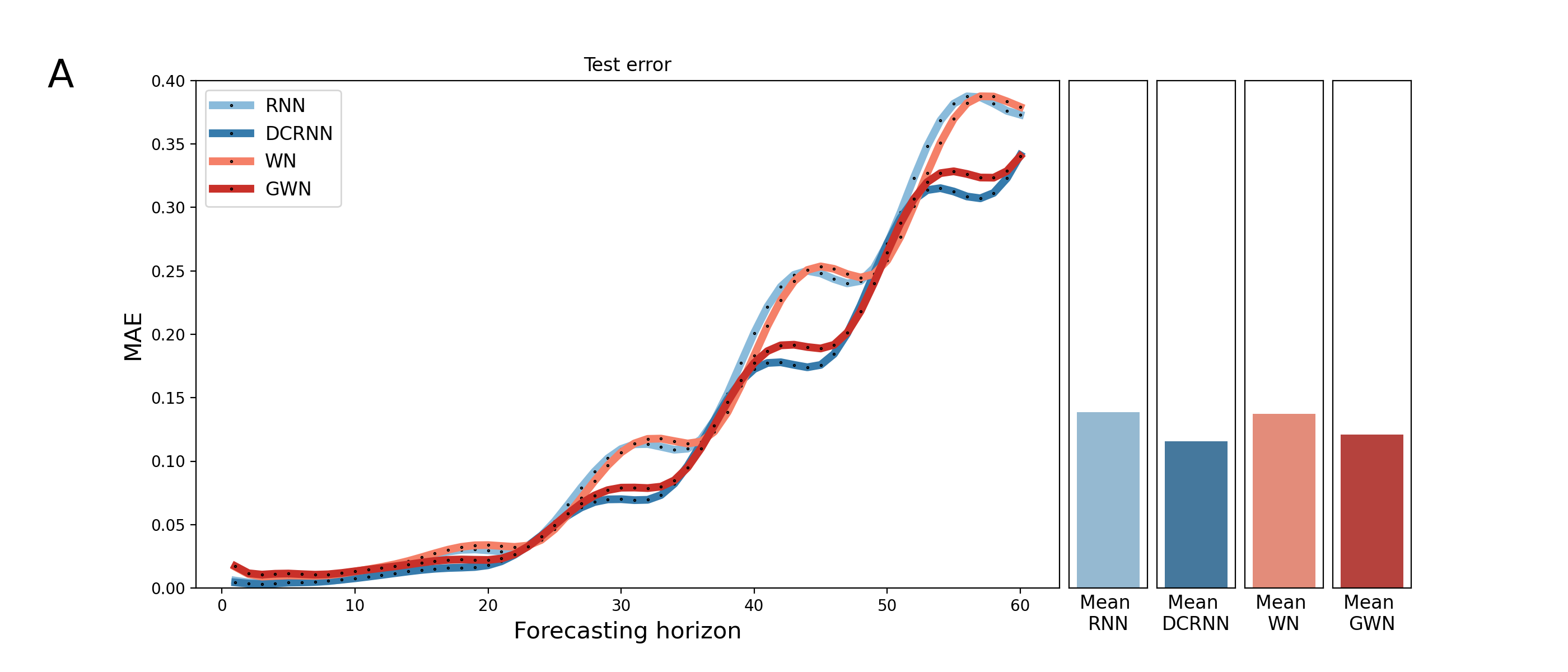}
}

\makebox[\textwidth][c]
{\includegraphics[width=0.79\textwidth]{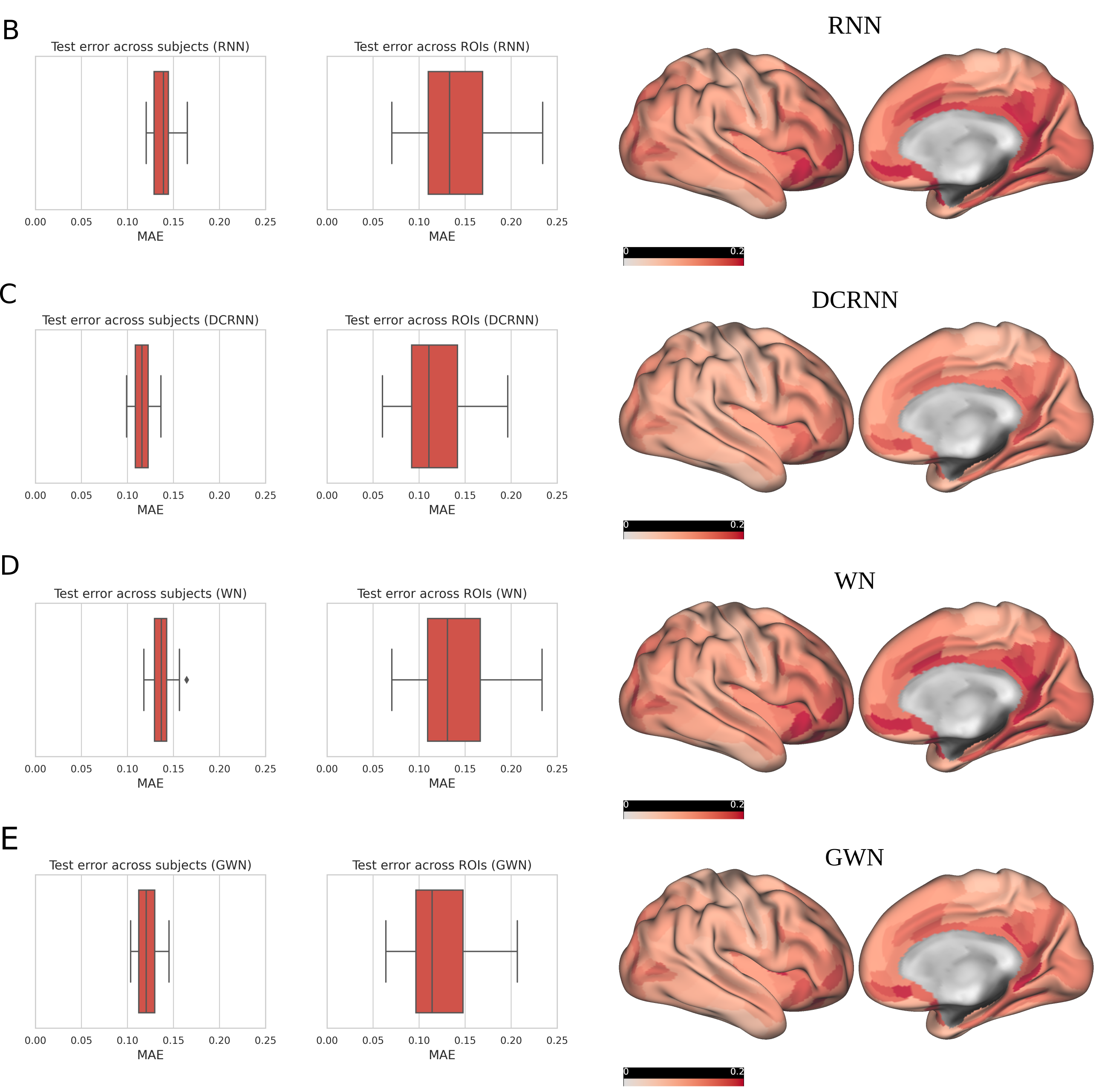}
}

\ec 
\caption{The distribution of the test MAE error across timepoins, subjects and ROIs is illustrated for the STGNNs, with and without incorporating spatial modeling. In (A) the prediction error in dependence of the forecasting horizon is depicted.
Further (B) shows the MAE across subjects and brain regions of the RNN, visualized in boxplots on the left side. On the right side the MAEs in dependence of the brain regions were projected onto the cortical surface. The same analysis was carried out for the DCRNN, WN and GWN model in figure (C), (D) and (E) respectively.}
\label{fig:supp_MAE_ROIs}
\end{figure}

In addition, we evaluated the similarity of the predicted FC states to the ground truth FC states. For this purpose we computed a FC matrix $ \mb{A}_{FC} \in \mbb{R}^{N \times N} $ from the true ROI timecourses $ \mb{x} \in \mbb{R}^{N \times T_f} $, and respectively the predicted FC matrix $  \mb{\hat{A}}_{FC} $ from predicted timecourses $ \mb{\hat{x}} $ based on Pearson correlation. Then we sorted the elements below the diagonal of the symmetric FC matrix $ \mb{A}_{FC} $ into a vector $ \mb{v} \in \mbb{R}^{\frac{1}{2} N (N-1)} $, and defined a correlation based measure of FC similarity as:

\beq 
	r(\mb{v}, \mb{\hat{v}}) = \frac{\sum_{c=1}^{\frac{1}{2} N (N-1)} ( \mb{v}_{c} - \mb{\bar{v}}_{c} ) ( \mb{\hat{v}}_{c} - \mb{\bar{\hat{v}}}_{c} )}{\sigma_{\mb{v}}\sigma_{\mb{\hat{v}}}}\label{eqn:FC_simi}
\eeq 
with $ \mb{\bar{v}}_{c} $ denoting the average and $ \sigma_{\mb{v}} $ the variance of FC values. Based on this metric we replicated the comparison of spatial and temporal GNN approaches in figure \ref{fig:supp_comparison_temporal_spatial_FC_simi}.  Also using this similarity measure, the RNN and WN model outperformed the TAtt model with $ p \leq 0.0001 $ (Cohen's $ d \gg 1 $) across subjects. The impact of structural modeling showed to be significant with $ p \leq 0.0001 $ (Cohen's $ d > 1 $) across subjects for both STGNN models. In addition the distributions of the test MAE across different subjects, timepoints and brain ROIs are visualized for the STGNNs with and without incorporating spatial modeling in figure \ref{fig:supp_MAE_ROIs}.

\clearpage


\subsection*{Supplement IV} \label{sec:supp_network_scaling}

\begin{figure}[!htb]
\bc

\makebox[\textwidth][c]
{
  \begin{minipage}[b]{0.31\textwidth}
  \includegraphics[width=1\textwidth]{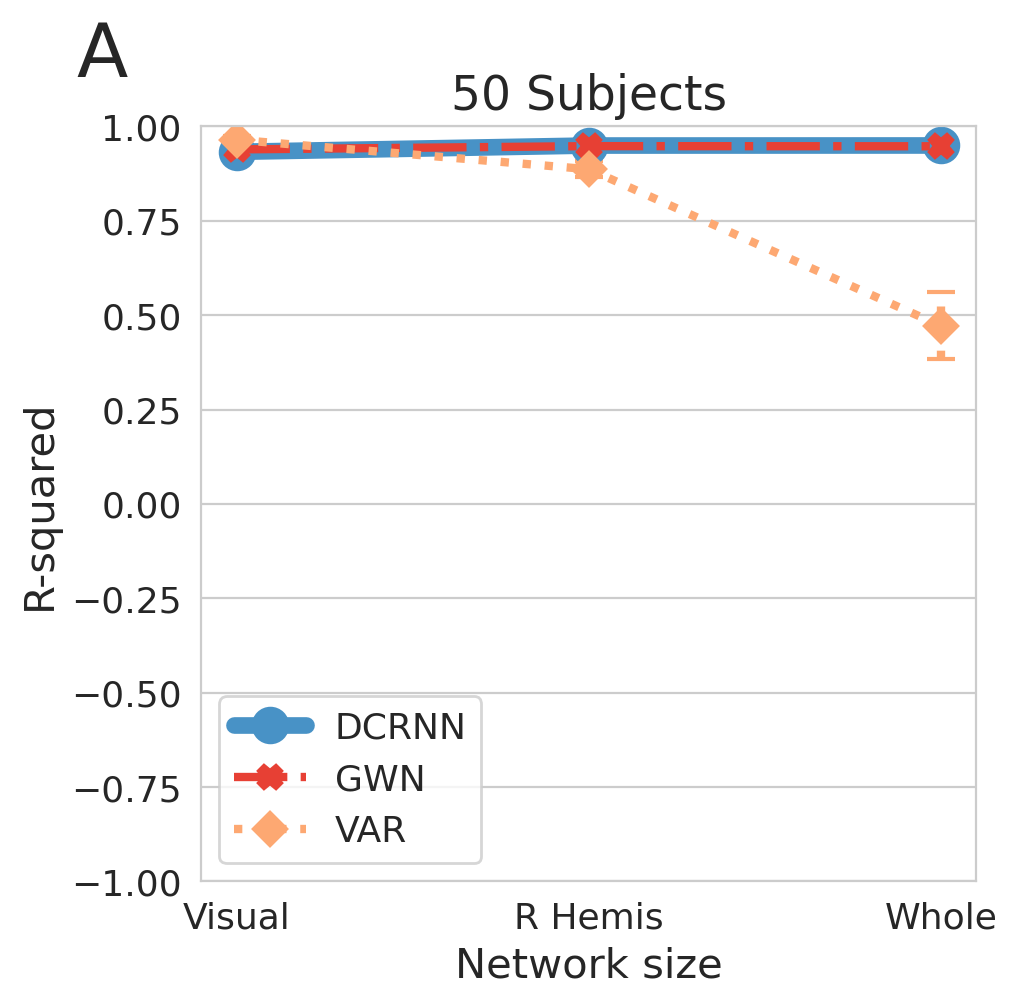}  
  \end{minipage}
    
  \begin{minipage}[b]{0.31\textwidth}
  \includegraphics[width=1\textwidth]{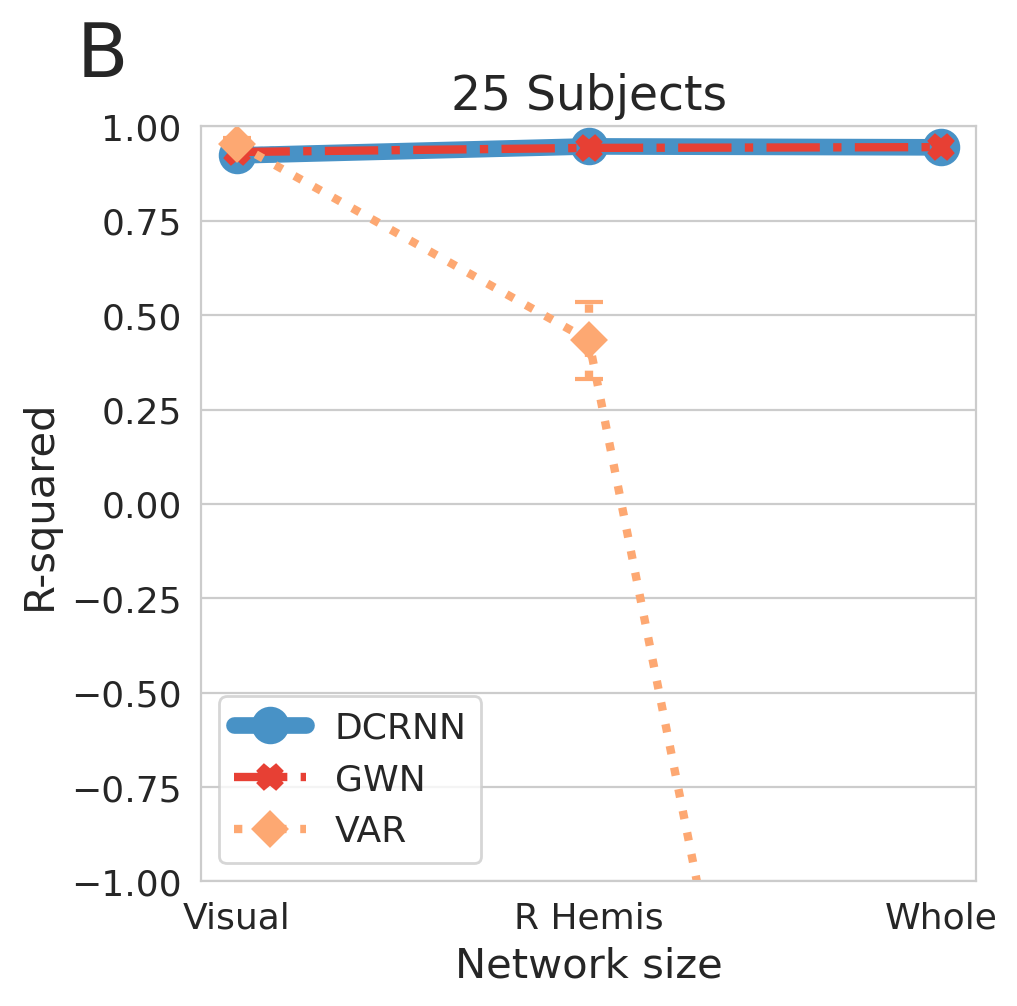}
  \end{minipage}  
  
  \begin{minipage}[b]{0.31\textwidth}
  \includegraphics[width=1\textwidth]{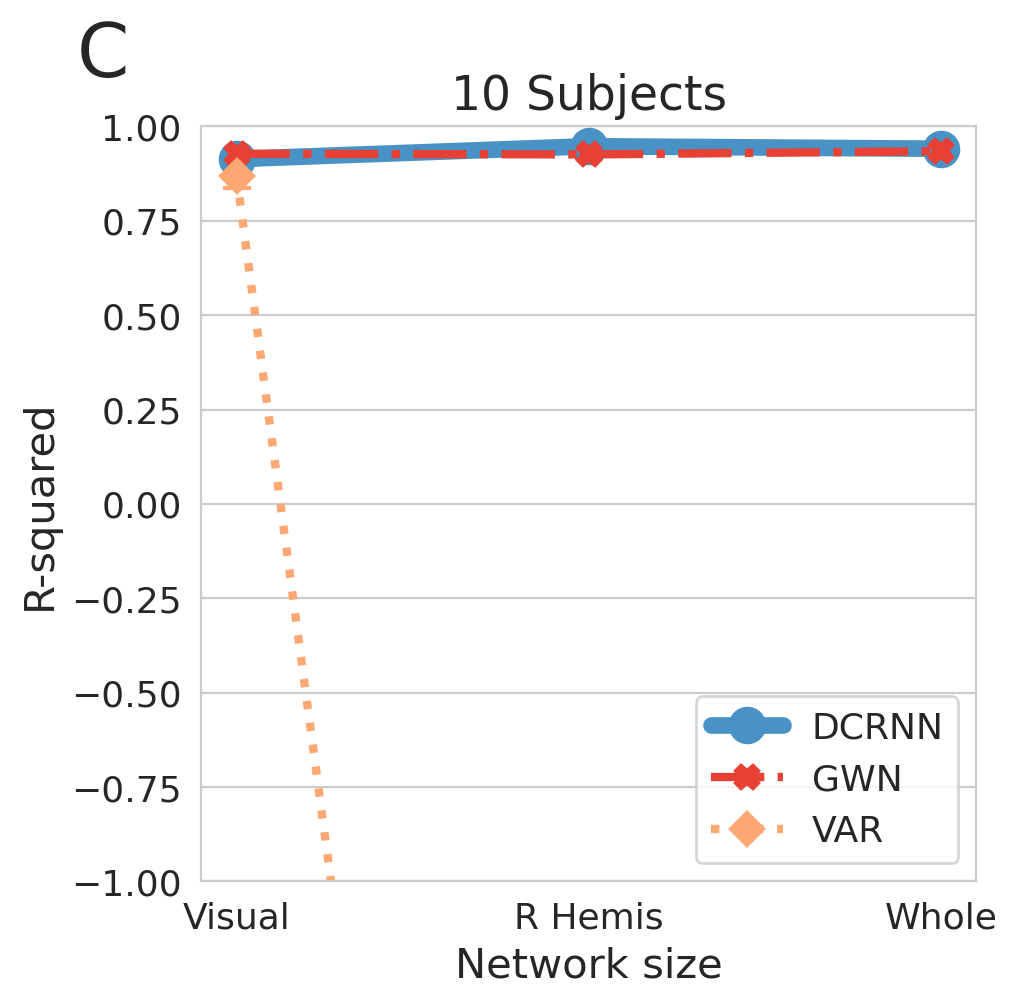}
  \end{minipage}  
}

\ec 
\caption{The figure shows a comparison of the model performances, based on the $ R^2 $ measure as defined in equation \ref{eqn:R2}. The errorbars represent the standard deviation of the $ R^2 $ values across subjects. In (A) the model accuracies when using a dataset of $ 50 $ subjects are shown for the visual network, the network within the right hemisphere and the whole brain network. Figure (B) and (C) show the test performances in dependence of the network size using the $ 25 $ and $ 10 $ subject dataset, respectively. Except for the single network dataset, the improvements of accuracy of the DCRNN and GWN in comparison to the VAR model became in all cases highly significant with $ p \leq 0.0001 $ (Cohen's $ d \gg 1 $).}
\label{fig:supp_comparison_models_R2}
\end{figure}

\begin{figure}[!htb]
\bc

\makebox[\textwidth][c]
{
  \begin{minipage}[b]{0.31\textwidth}
  \includegraphics[width=1\textwidth]{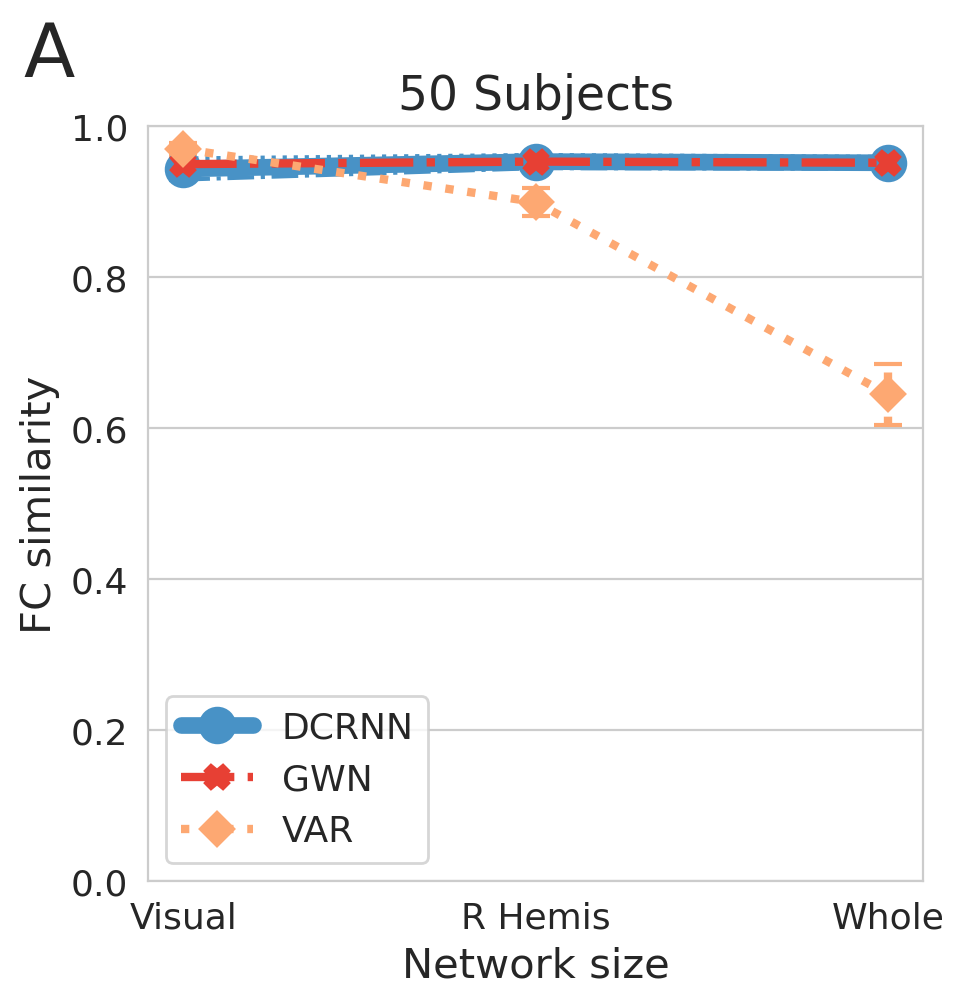}  
  \end{minipage}
    
  \begin{minipage}[b]{0.31\textwidth}
  \includegraphics[width=1\textwidth]{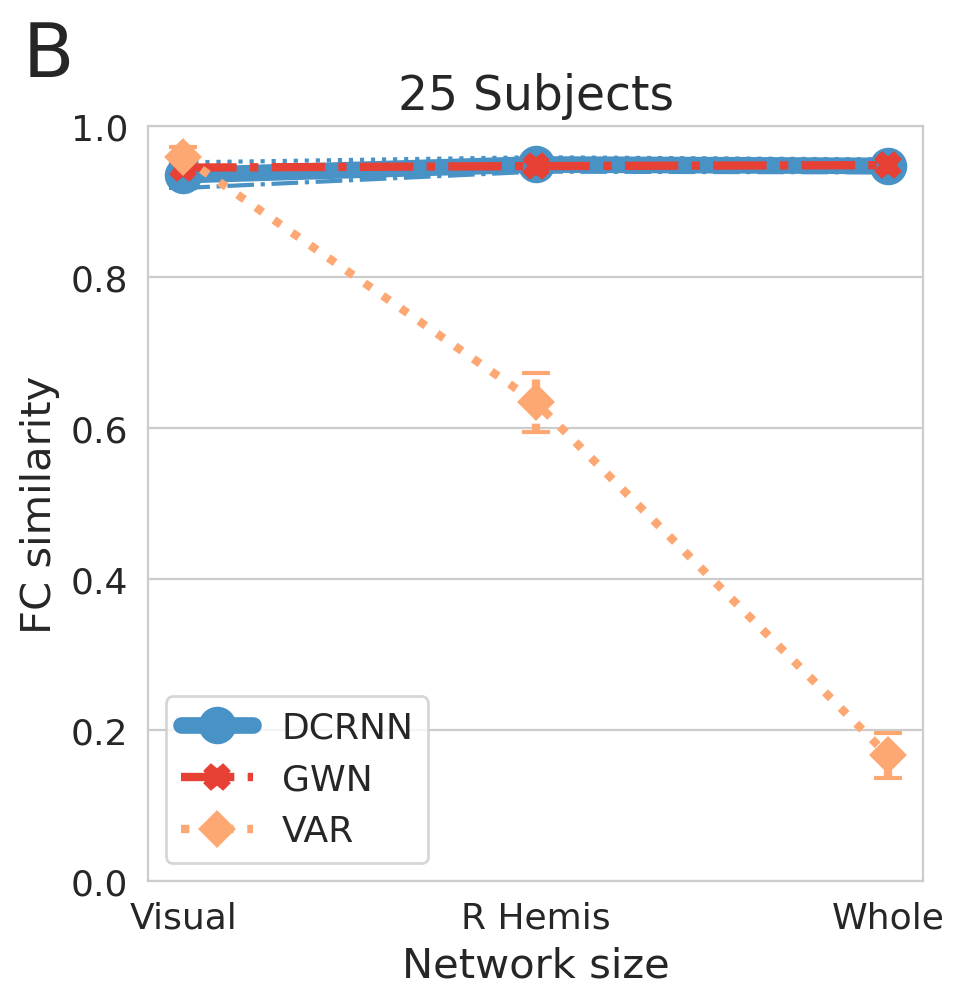}
  \end{minipage}  
  
  \begin{minipage}[b]{0.31\textwidth}
  \includegraphics[width=1\textwidth]{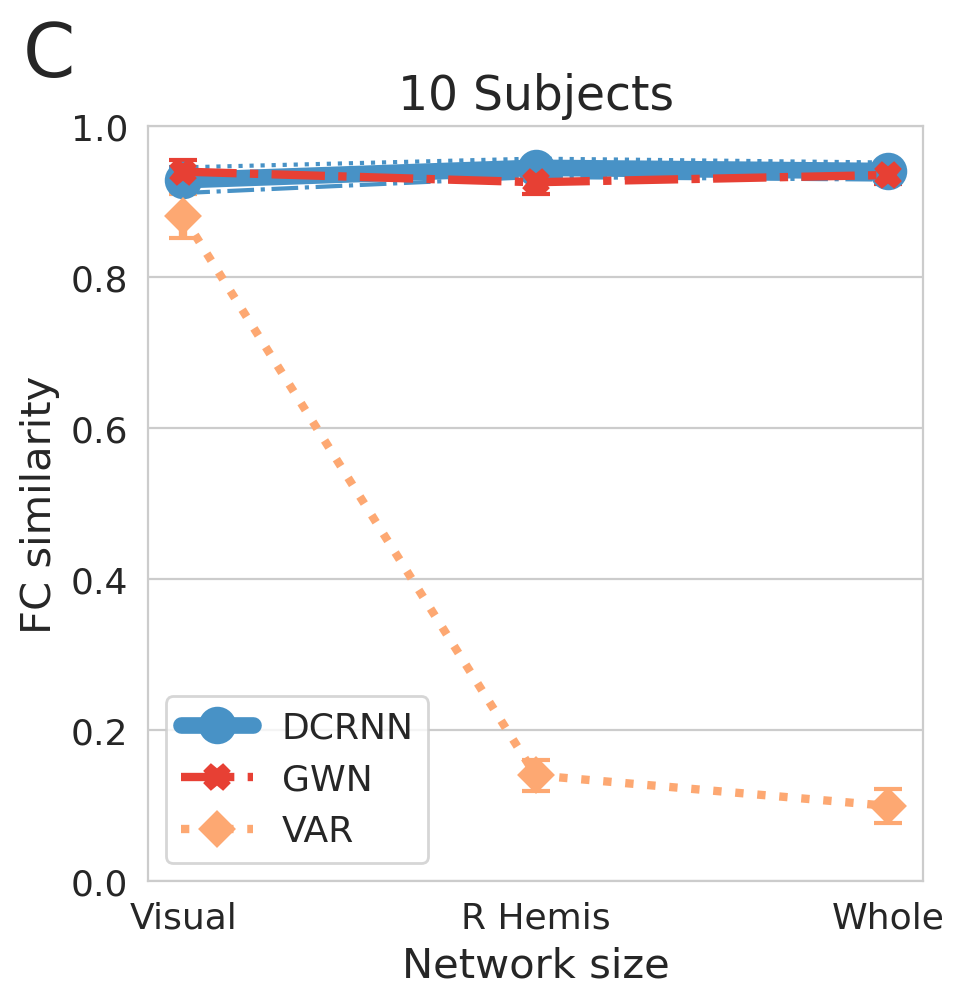}
  \end{minipage}  
}

\ec 
\caption{The figure shows a comparison of the model performances, based on the similarity of predicted FC states, as defined in equation \ref{eqn:FC_simi}. The errorbars represent the standard deviation of the values across subjects. In (A) the model accuracies when using a dataset of $ 50 $ subjects are shown for the visual network, the network within the right hemisphere and the whole brain network. Figure (B) and (C) show the test performances in dependence of the network size using the $ 25 $ and $ 10 $ subject dataset, respectively.  Except for the single network dataset, the improvements of accuracy of the DCRNN and GWN in comparison to the VAR model became in all cases highly significant with $ p \leq 0.0001 $ (Cohen's $ d \gg 1 $).
}
\label{fig:supp_comparison_models_FC_simi}
\end{figure}

\clearpage


\subsection*{Supplement V} \label{sec:supp_liberal_bandpass}

\begin{figure}[!htb]
\bc

\makebox[\textwidth][c]
{\includegraphics[width=0.8\textwidth]{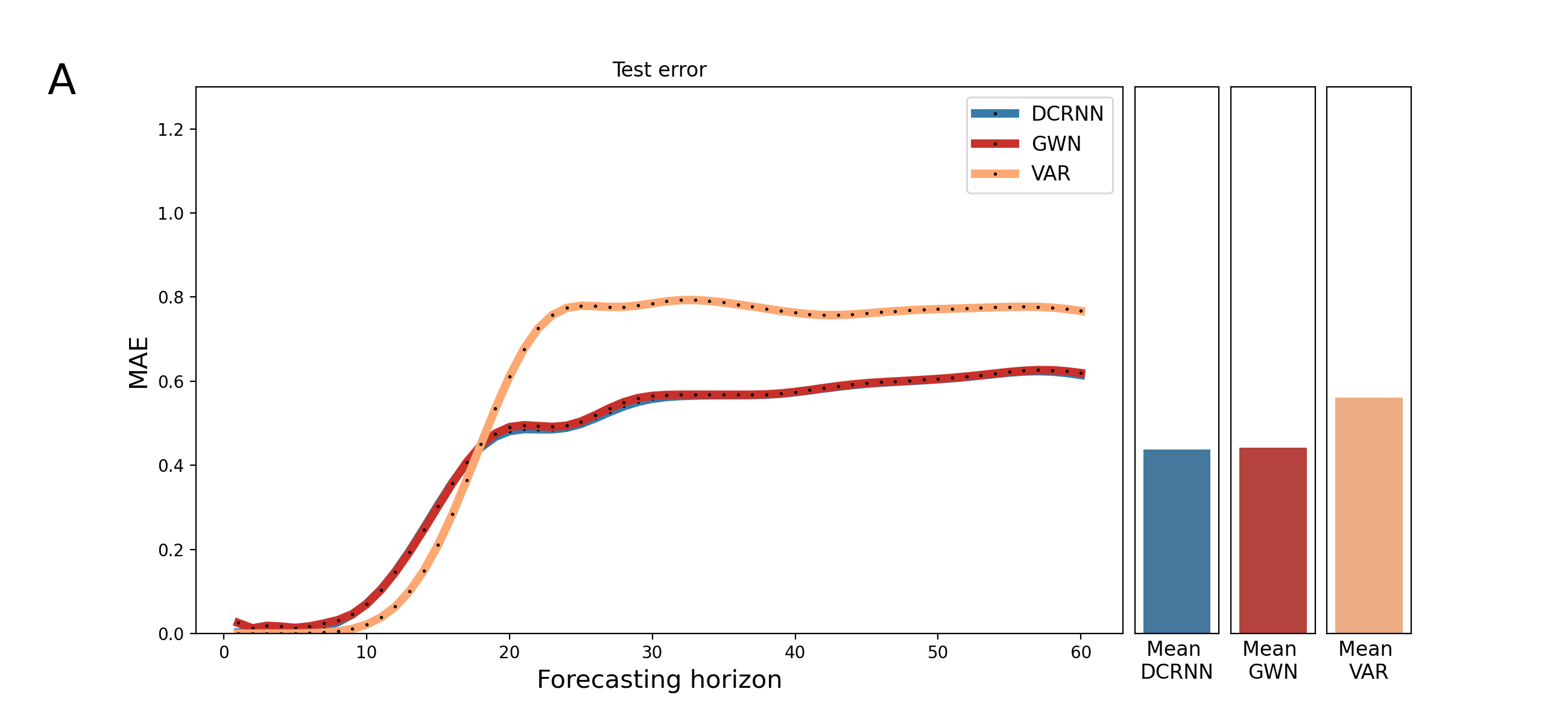}
}

\makebox[\textwidth][c]
{\includegraphics[width=0.85\textwidth]{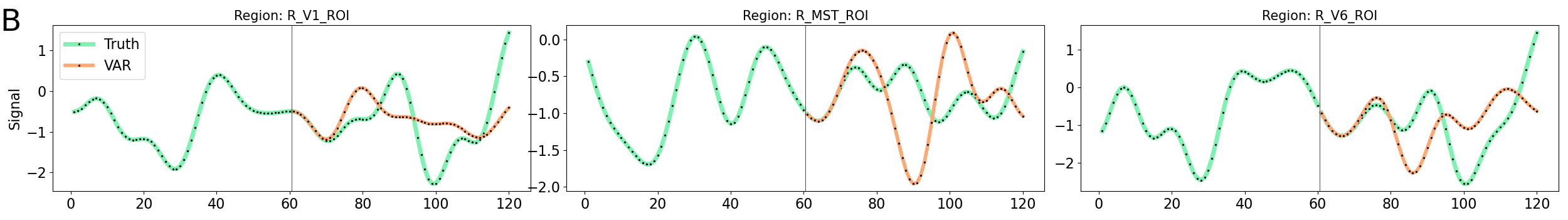}
}
\makebox[\textwidth][c]
{\includegraphics[width=0.85\textwidth]{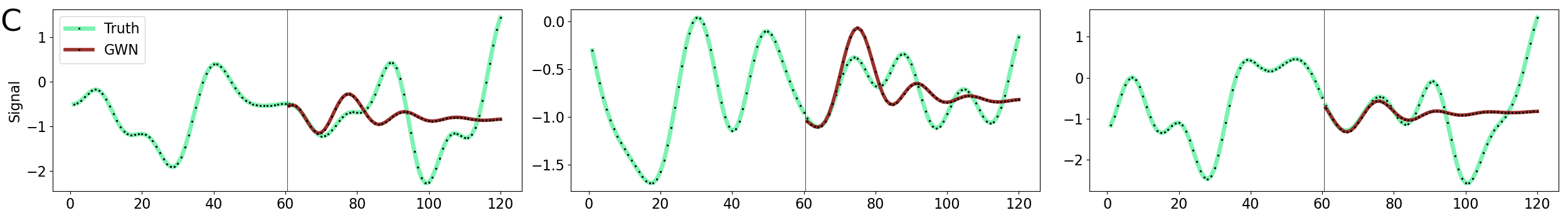}
}
\makebox[\textwidth][c]
{\includegraphics[width=0.85\textwidth]{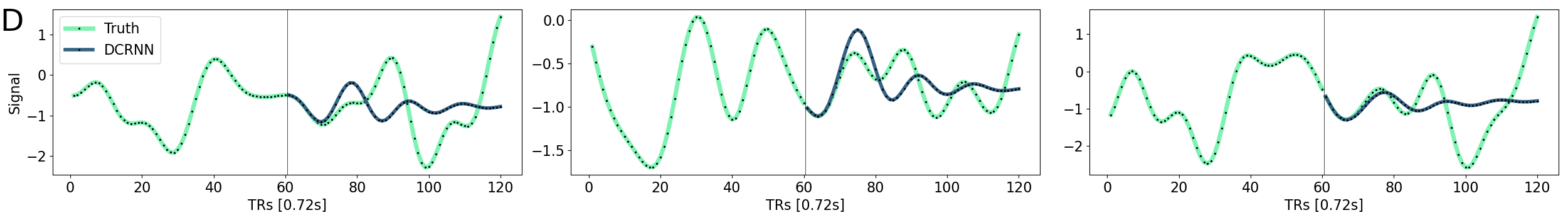}
}

\ec 
\caption{In this figure the prediction accuracy of the different models is presented when using a bandpass filter within the $ 0.01 - 0.1 Hz $ frequency range for the $ 25 $ subject dataset and the brain network including the ROIs within the right hemisphere \citep{Glasser2016}. In (A) the test MAE in dependence of the forecasting horizon is shown, computed as an average across test samples and brain regions. Figure (B), (C) and (D) show examples of predictions generated by the VAR, GWN and DCRNN model respectively. The examples in this figures were chosen to be representative for the whole test set, by selecting only examples which errors maximally deviate by $ 0.02 $ from the corresponding average test MAE of the models.}
\label{fig:supp_liberal_bandpass}
\end{figure} 

\clearpage


\subsection*{Supplement VI} \label{sec:supp_connectivity}

\begin{figure}[!htb]
\bc

\includegraphics[width=0.6\textwidth]{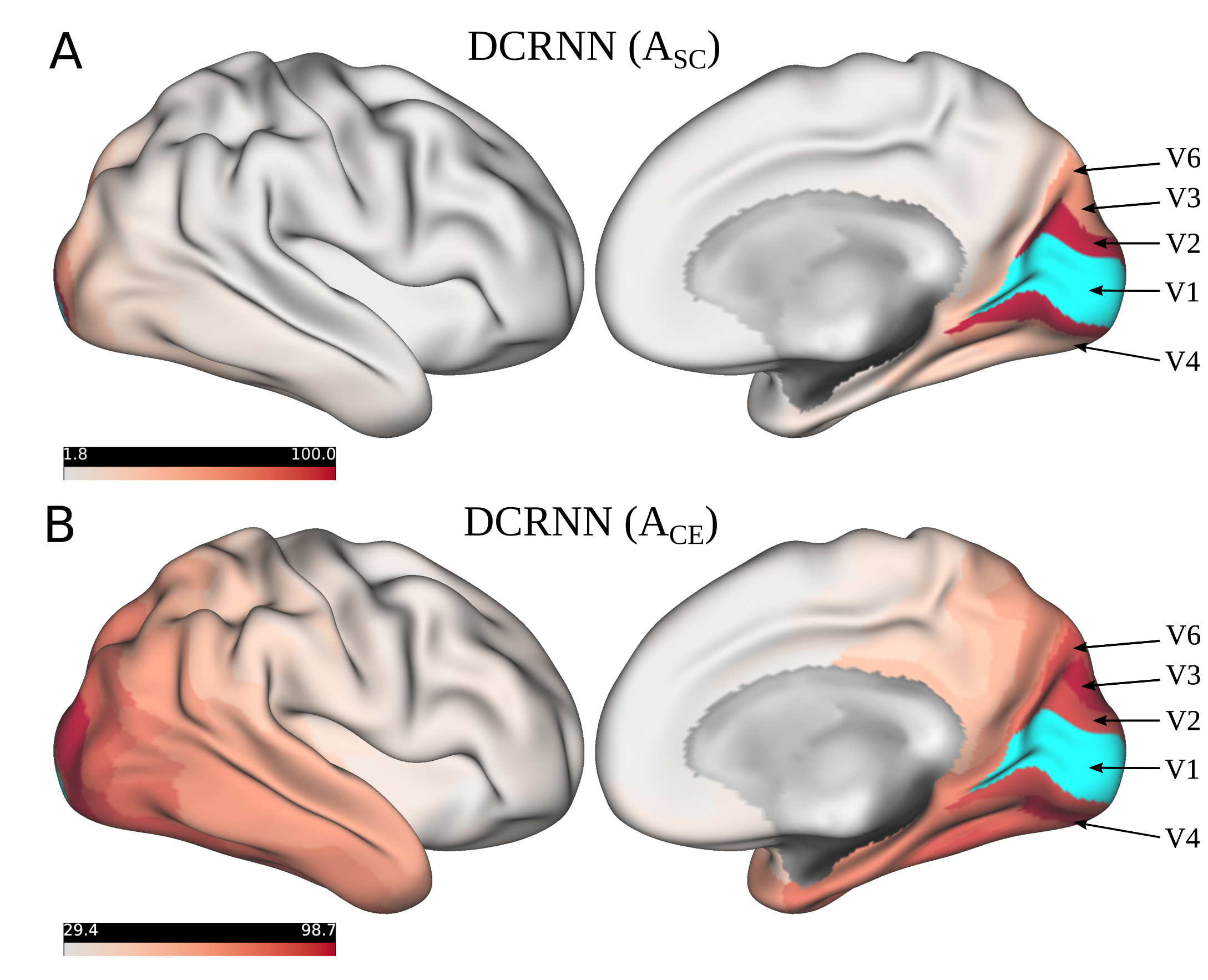}

\ec 
\caption{This figure illustrates directed spatial relations learned by the DCRNN model. Figure (A) shows the measures of influence $ \mb{I}(n') $, derived from the DCRNN model when using the SC for information propagation and figure (B) depicts the influence when incorporating CEs for the information exchange. The values of the connectivity measures were linearly mapped between $ 0 $ and $ 100 $ and the default scaling of the color values provided by the \textit{connectome workbench} (version 1.4.2) was used, adjusting the colormap between the $ 2th $ and $ 98th $ percentile of the values respectively.}
\label{fig:supp_connectivity_DCRNN}
\end{figure} 



\begin{figure}[!htb]
\bc

\includegraphics[width=1\textwidth]{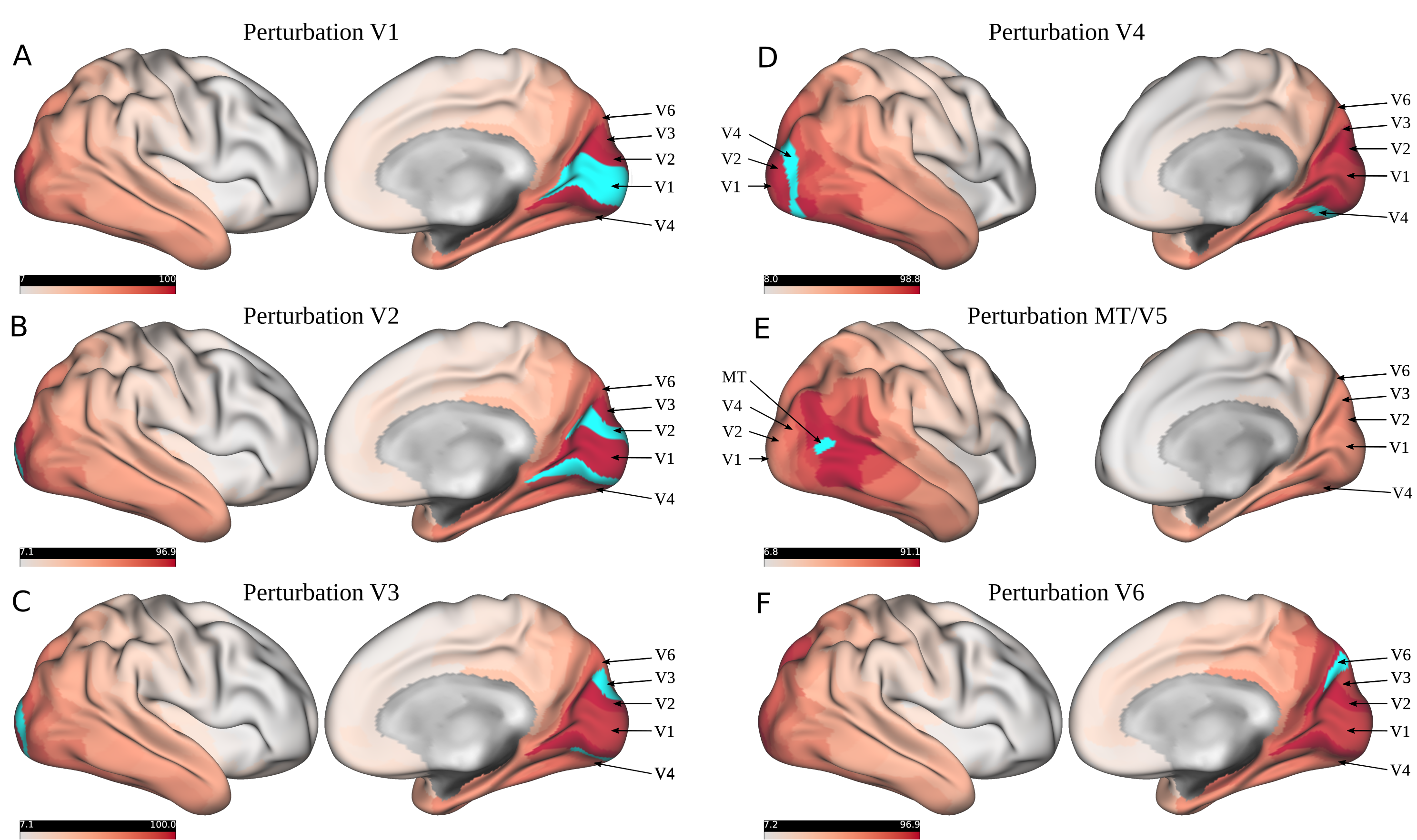}

\ec 
\caption{The figure illustrates the effect of a perturbation in a STGNN model systematically induced in target regions V1 (A), V2 (B), V3 (C), V4 (D), MT/V5 (E) and V6 (F). The connectivity values were linearly mapped between $ 0 $ and $ 100 $ and the default scaling of the color values provided by the \textit{connectome workbench} (version 1.4.2) was used, adjusting the colormap between the $ 2th $ and $ 98th $ percentile of the values respectively.}
\label{fig:supp_connectivity_visual}
\end{figure} 



\begin{figure}[!htb]
\bc

\makebox[\textwidth][c]
{
  \includegraphics[width=0.6\textwidth]{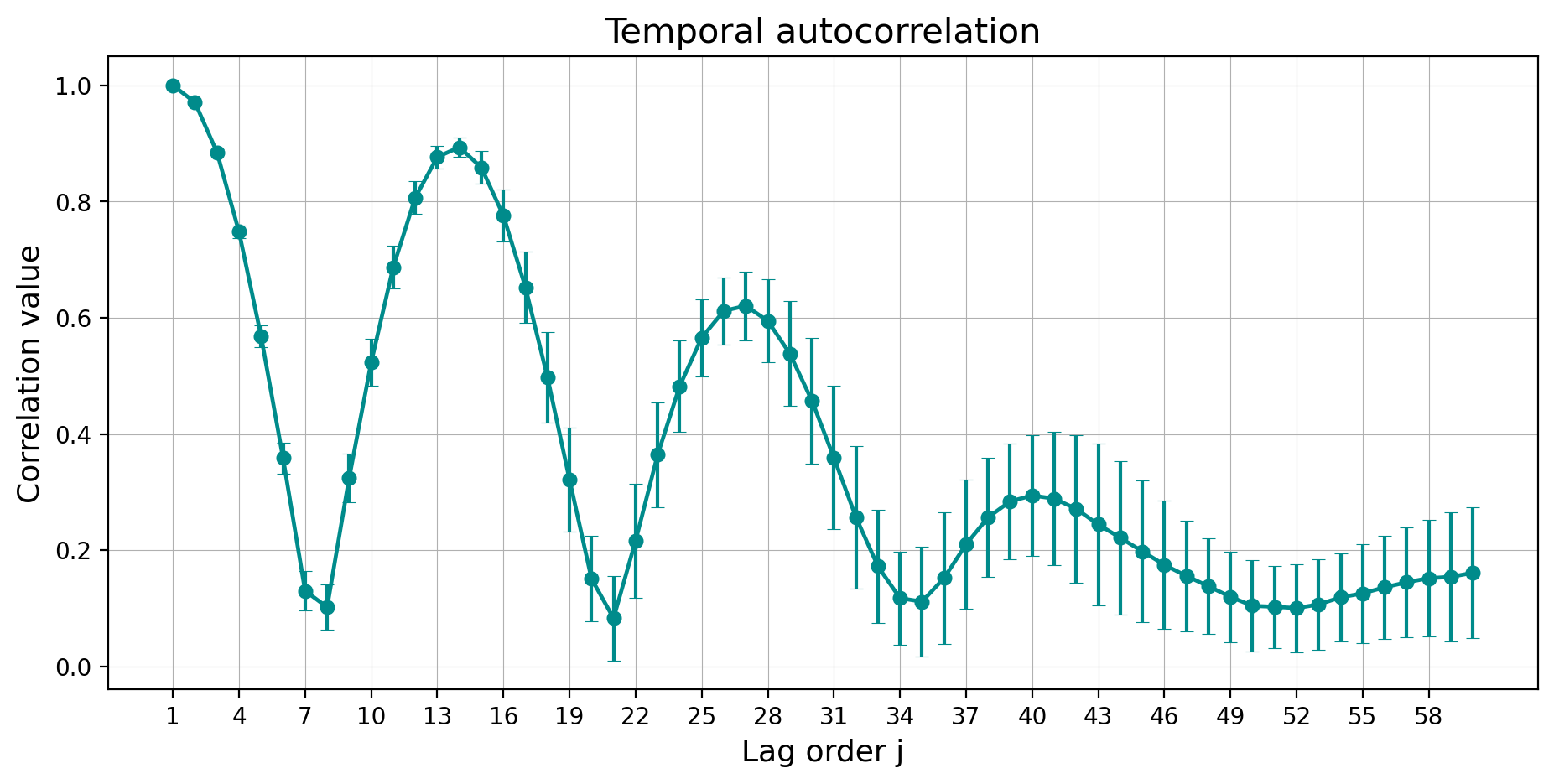}
}

\ec 
\caption{In this figure the temporal average autocorrelation values of fMRI timeseries in dependence of their lag order $ j $ are shown. The values were computed as the Pearson correlation between the original fMRI timeseries $ \mb{x}_n^{(t)} $ and its lagged values $ \mb{x}_n^{(t-j)} $, and then averaged across all subjects and brain regions. The errorbars represent the standard deviations across subjects.
}
\label{fig:supp_temp_autocorr}
\end{figure}


\begin{figure}[!htb]
\bc

\makebox[\textwidth][c]
{
\begin{minipage}[b]{0.165\textwidth}
\includegraphics[width=\textwidth]{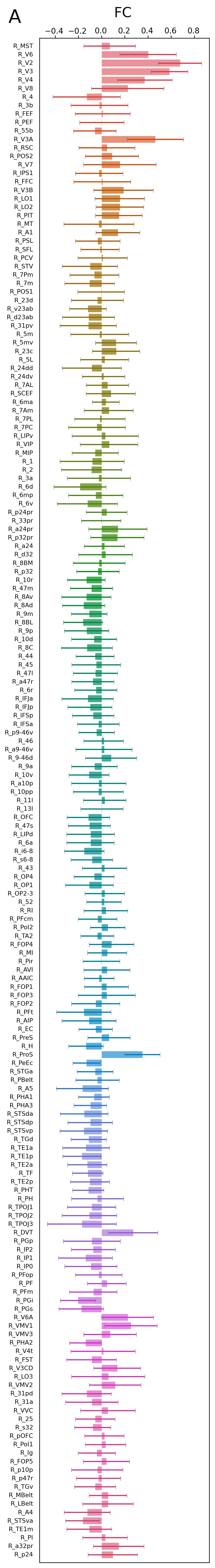}
\end{minipage}

\hspace{6mm}

\begin{minipage}[b]{0.165\textwidth}
\includegraphics[width=\textwidth]{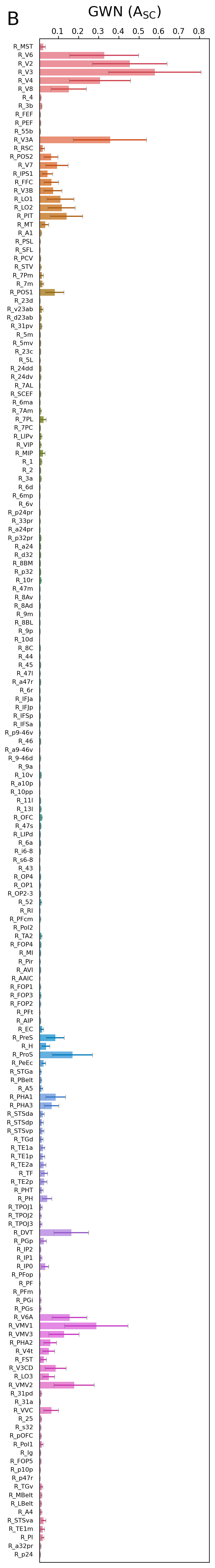}
\end{minipage}

\hspace{6mm}

\begin{minipage}[b]{0.165\textwidth}
\includegraphics[width=\textwidth]{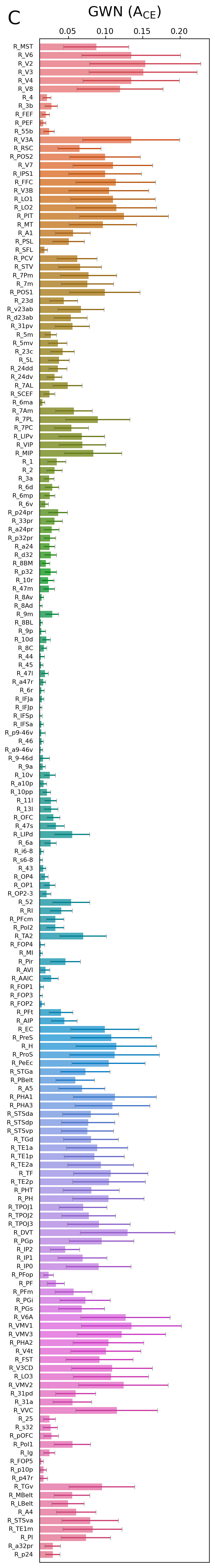}
\end{minipage}
}

\ec 
\caption{The FC connectivity of V1 to all other regions within the right hemisphere is shown in (A). In (B) and (C) the GWN based directed connectivity is illustrated when using the SC ($ \mb{A}_{SC} $) and CE similarity ($ \mb{A}_{CE} $) for the information propagation respectively.}
\label{fig:supp_connectivity_comparison}
\end{figure} 

\clearpage


\subsection*{Supplement VII} \label{sec:supp_respiratory_freq}

\begin{figure}[!htb]
\bc

\includegraphics[width=0.6\textwidth]{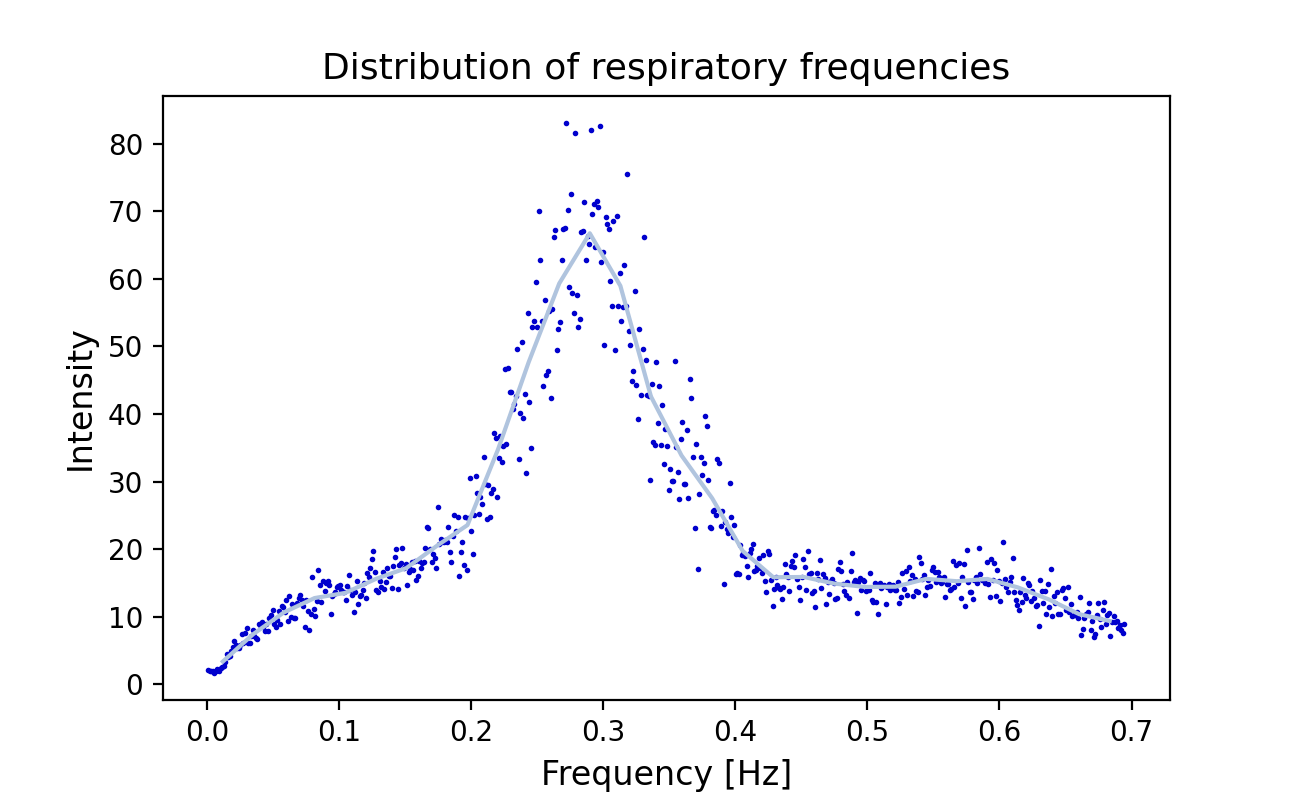}

\ec 
\caption{The distribution of respiratory frequencies is shown in this figure. For the 25 subjects resting-state fMRI dataset, physiological data was available from 22 subjects, and the depicted spectrum was computed as the average across the individual spectra of the 22 subjects.
}
\label{fig:supp_respiratory_freq}
\end{figure}

\end{document}